\documentclass[conference]{IEEEtran}

\usepackage{graphicx}
\usepackage{textcomp}
\usepackage{xcolor}
\def\BibTeX{{\rm B\kern-.05em{\sc i\kern-.025em b}\kern-.08em
    T\kern-.1667em\lower.7ex\hbox{E}\kern-.125emX}}
    
\usepackage{amsmath}
\usepackage{amsthm}
\usepackage{amssymb}

\usepackage{subcaption}
\usepackage{wrapfig}
\usepackage{multirow}
\usepackage{makecell}
\usepackage{url}
\usepackage{orcidlink}

\usepackage{fontawesome}
\usepackage{filecontents}
\usepackage{pgfplots}
\newsavebox{\imagebox}

\usepackage{soul}
\setstcolor{red}

\usepackage{tikz}
\usetikzlibrary{
positioning,
arrows,
automata,
shapes,
hobby,
backgrounds,
calc,
trees,
fit,
shapes.multipart,
fadings,
through,
patterns,
decorations.pathreplacing,
patterns,
shapes.geometric}
\usepackage{tikz-cd}
\pgfdeclarelayer{background}
\pgfsetlayers{background,main}

\definecolor{darkspringgreen}{rgb}{0.09, 0.45, 0.27}
\definecolor{my_grey}{RGB}{191, 191, 191}

\newcommand{\shortColRed}[1]{\textcolor{red}{#1}}

\begin{document}

\title{Order Fairness Evaluation of DAG-based ledgers}

\author{\IEEEauthorblockN{Erwan Mahe \orcidlink{0000-0002-5322-4337}}
\IEEEauthorblockA{\textit{Université Paris Saclay, CEA LIST} \\
Palaiseau, France}
\and
\IEEEauthorblockN{Sara Tucci-Piergiovanni \orcidlink{0000-0001-9738-9021}}
\IEEEauthorblockA{\textit{Université Paris Saclay, CEA LIST} \\
Palaiseau, France}
}

\maketitle

\begin{abstract}
In Distributed Ledgers (DL), Order Fairness (OF) refers to properties that relate the order in which transactions are sent or received to the order in which they are eventually finalized (totally ordered). 
The study of OF is relatively new and has been stimulated by the rise of Maximal Extractable Value attacks.
In classical Blockchain protocols, leaders are responsible for selecting the transactions to be included in blocks, which creates a vulnerability and opportunity for transaction order manipulation.
Unlike Blockchains, DAG-based DL allow participants in the network to independently propose blocks, which are then arranged as vertices of a directed acyclic graph. Interestingly, leaders in DAG-based ledgers are elected only \textit{après coup} (after the fact), once transactions are already part of the graph, to determine their total order. In other words, transactions are collectively validated by the nodes, and leaders are only elected to establish an ordering. This approach intuitively reduces the risk of transaction manipulation and enhances fairness.

In this paper, we aim to quantify the capability of DAG-based DL to achieve OF. To this end, we define new variants of OF adapted to DAG-based DL and evaluate the impact of an adversary capable of compromising a limited number of nodes (below the one-third threshold) to reorder transactions. We analyze how often our OF properties are violated under different network conditions and parameterizations of the DAG algorithm, depending on the adversary’s power.
Our study shows that DAG-based DL are still vulnerable to reordering attacks, as an adversary can coordinate a minority of Byzantine nodes to manipulate the DAG's structure.
\end{abstract}


\section{Introduction}

A Distributed Ledger (DL) is a set of replicated state machines. 
In cryptocurrencies and Web3, individual machines are called nodes and transitions in the state machine correspond to the finalization
of a transaction.
Clients may modify the ledger state by sending transactions to nodes.
Nodes maintain consensus on the current state by agreeing on the initial state and on the order with which transactions are finalized.

Distributed computing literature recognizes two main classes of properties of DL which are ``\textit{consistency}'' (i.e., consensus agreement and validity which are safety properties \cite{recognizing_safety_and_liveness}) and ``\textit{liveness}'' (an umbrella term that can refer to consensus wait-freedom, starvation-freedom or local progress) \cite{safety_liveness_exclusion_in_distributed_computing}.
However, as detailed in \cite{order_fairness_for_byzantine_consensus}, these properties have no say on the actual order of transactions that is agreed upon.
It is then possible, for a malicious adversary, to manipulate the order of transactions if given the opportunity.
In decentralized finance \cite{sok_preventing_transaction_reordering_manipulations_in_decentralized_finance}, front-running attacks \cite{flash_boys_frontrunning_in_decentralized_exchanges_miner_extractable_value_and_consensus_instability} consist in, having knowledge of a pending transaction $x$, placing a new transaction $x'$ in front of it (i.e., so that $x'$ is finalized before $x$).
Combining such attacks can be used to extract profits via manipulating the value of financial assets.
In practice, this is done by automated Maximum Extractable Value (MEV) bots \cite{flash_boys_frontrunning_in_decentralized_exchanges_miner_extractable_value_and_consensus_instability,sok_preventing_transaction_reordering_manipulations_in_decentralized_finance}. 
In Ethereum PoS \cite{exploiting_ethereum_after_the_merge_the_interplay_between_pos_and_mev_strategies}, front running can be performed by providing a higher gas reward for $x'$, thus providing an incentive to place $x'$ before $x$.
However, even without relying on reward mechanisms (gas), control over nodes still allows front running \cite{adversary_augmented_simulation_to_evaluate_client_fairness_on_hyperledger_fabric}.
According to Forbes, between January 2020 and September 2022, MEV bots have extracted $\sim675$ million \$ profits on the Ethereum blockchain alone.

Blockchains constitute a means to implement a DL.
In a blockchain, transactions are batched into successive blocks and their finalization is ordered according to the order in which they appear on and within blocks.
Each block is proposed by a unique node called a leader who gathers new transactions in that block.
In contrast to Blockchains, DAG-based DL \cite{sok_dag_based_blockchain_systems} such as \cite{all_you_need_is_dag,narwhal_and_tusk,bullshark,reducing_latency_of_dag_based_consensus_in_the_asynchronous_setting_via_the_utxo_model} rely on a Directed Acyclic Graph which vertices contain batches of transactions.
Each node can broadcast new vertices that are then added to the DAG. The definition of edges from more recent vertices to older vertices ensures consistency as it enables the nodes to agree on a shared view of their local copies of the DAG.
Similarly to Blockchains, transactions are then finalized in batches except that these batches rather correspond to sets of vertices, called ``waves'' in \cite{all_you_need_is_dag,bullshark,narwhal_and_tusk}.
With most current Blockchain algorithms, manipulations of transactions order may trivially occur if, e.g., the current leader has an incentive to do so.
\cite{exploiting_ethereum_after_the_merge_the_interplay_between_pos_and_mev_strategies} discusses it for the Ethereum PoS Blockchain and \cite{adversary_augmented_simulation_to_evaluate_client_fairness_on_hyperledger_fabric} experiments on such manipulations in HyperLedger Fabric.
The absence of block-proposing leaders in DAG-based DLT should, theoretically, make them less vulnerable to such attacks.
Yet, in the literature, such evaluations are lacking.

Transaction reordering attacks in DL may be mitigated using cryptography (commit \& reveal \cite{sok_preventing_transaction_reordering_manipulations_in_decentralized_finance,maximal_extractable_value_protection_on_a_DAG,fairness_notions_in_dag_based_dlts}). 
However, this solution comes with a high overhead.
As such, to alleviate these risks, ``\textit{order fairness}'' (OF) properties and Blockchain protocols that uphold them (called Algorithmic Committee Ordering in \cite{sok_preventing_transaction_reordering_manipulations_in_decentralized_finance}) have been designed \cite{order_fairness_for_byzantine_consensus,quick_order_fairness}.
OF properties relate various partial orders of events to one another.
These include the orders with which transactions are submitted (by clients), received (by nodes) and finalized.
These orders are not necessarily observable and their precise definitions may vary depending on the algorithms that are considered.
OF is a relatively new subject of study in traditional Blockchains and this is all the more true for DAG-based ledgers \cite{fairness_notions_in_dag_based_dlts}.

In this paper, we define several novel OF metrics that are particularly adapted to DAG algorithms such as DagRider \cite{all_you_need_is_dag}, Bullshark \cite{bullshark} and Tusk \cite{narwhal_and_tusk}.
Then, in the spirit of \cite{adversary_augmented_simulation_to_evaluate_client_fairness_on_hyperledger_fabric}, we devise DAG-specific attack scenarios aiming at manipulating transactions ordering and observe their effect on OF.
In particular, we consider the DagRider protocol \cite{all_you_need_is_dag}, complemented with mechanisms inspired by \cite{reducing_latency_of_dag_based_consensus_in_the_asynchronous_setting_via_the_utxo_model} and \cite{order_fairness_for_byzantine_consensus,themis_fast_strong_order_fairness_in_byzantine_consensus}.
We study the impact of attacks orchestrated by an adversary under various hypotheses on the network and algorithm parameterization.
The adversary may infect a number of nodes (so as to modify their behavior) bounded by the fault tolerance thresholds of the involved algorithms.
Our study suggests that, even though they are more robust than their Blockchain counterparts (e.g., HyperLedger Fabric in \cite{adversary_augmented_simulation_to_evaluate_client_fairness_on_hyperledger_fabric}), permissioned DAG-based ledgers such as DagRider \cite{all_you_need_is_dag} are still vulnerable to transaction reordering attacks.

This paper is organized as follows:
After preliminaries and related works in Sec.\ref{sec:related}, Sec.\ref{sec:prel} introduces our system model and application use case.
We then present DagRider and OF properties adapted to its specificities in Sec.\ref{sec:dag}.
Sec.\ref{sec:attacks} introduces attack scenarios before Sec.\ref{sec:experiments} details our experiments.
We conclude in Sec.\ref{sec:conc}.

\section{Preliminaries and related works\label{sec:related}}

Fairness, in the context of Blockchains \cite{on_fairness_in_commitee_based_blockchains,on_fairness_in_voting_consensus_protocols} and DAGs \cite{fairness_notions_in_dag_based_dlts} may refer to various notions, including the fairness in committee selection \cite{on_fairness_in_commitee_based_blockchains}, rewarding \cite{do_the_rich_get_richer_fairness_analysis_for_blockchain_incentives} or the ability to take decisions \cite{on_fairness_in_voting_consensus_protocols,fairledger_a_fair_blockchain_protocol_for_financial_institutions} w.r.t.~individual nodes' voting power. 
While \cite{fairness_and_efficiency_in_dag_based_cryptocurrencies} studies reward fairness in PoW DAGs, \cite{on_fairness_in_voting_consensus_protocols} deals with fairness to validators.
\cite{fairness_notions_in_dag_based_dlts} reviews notions of fairness applied to DAG-based DL.
In \cite{sok_preventing_transaction_reordering_manipulations_in_decentralized_finance} ``fairness'' is achieved whenever participants cannot include, exclude or front-run \cite{flash_boys_frontrunning_in_decentralized_exchanges_miner_extractable_value_and_consensus_instability} a transaction after having seen its content.

In this paper, we are specifically interested in {\em order-fairness} \cite{order_fairness_for_byzantine_consensus,quick_order_fairness,byzantine_ordered_consensus_without_byzantine_oligarchy,themis_fast_strong_order_fairness_in_byzantine_consensus} (OF) properties, that relate partial orders on transaction finalization and on communication events.
\cite{order_fairness_for_byzantine_consensus} defines ``\textit{receive-order fairness}'' as follows: if a majority of nodes receive a transaction $x$ before $x'$ then $x$ must be finalized before $x'$.
However, for three transactions $x_1$, $x_2$ and $x_3$, it may be so that a majority of nodes receive $x_1$ before $x_2$, $x_2$ before $x_3$ and $x_3$ before $x_1$. Thus, this property is impossible to achieve, as $\{x_1,x_2,x_3\}$ forms a Condorcet cycle \cite{condorcet_attack_against_fair_transaction_ordering}.
As a solution, \cite{quick_order_fairness} proposes the achievable ``\textit{differential-order fairness}'', which rather considers the difference between the number of honest nodes that receive $x$, and resp.~$x'$ first.
\cite{themis_fast_strong_order_fairness_in_byzantine_consensus} proposes ``\textit{$\gamma$-(all)-batch-order fairness}'' that reasons on pairs of transactions that are not in the same Condorcet cycle.


Because their premise involves the order with which nodes receive transactions, upholding such OF properties can only prevent an attacker from manipulating the finalization order, supposing the reception order remains unchanged (i.e., the attacker does not change it).
Yet, an attacker with a better network connection (than that of honest nodes and clients) may listen to an incoming transaction $x$ and front-run it via submitting $x'$ and ensuring that most nodes receive $x'$ before $x$ \cite{sok_preventing_transaction_reordering_manipulations_in_decentralized_finance}.
In that case, the attack is undetectable and even worse, upholding receive-order-fairness guarantees the attack to succeed.
\cite{condorcet_attack_against_fair_transaction_ordering} describes another attack, targeting protocols that uphold batch-order fairness.
This Condorcet attack consists in sending several transactions with high and small transmission delays to the nodes so as to artificially create Condorcet cycles. Its goal is to trap honest transactions in theses cycles so that e.g., they are easier to front-run (because they are all forced into the same batch).
As per \cite{sok_preventing_transaction_reordering_manipulations_in_decentralized_finance}, we refer to such attack as transaction reordering manipulations.
These attacks may rely on the power of individual clients and nodes (coordinated by the adversary) to order transactions.
\cite{condorcet_attack_against_fair_transaction_ordering} highlights that the existing Algorithmic Committee Ordering \cite{sok_preventing_transaction_reordering_manipulations_in_decentralized_finance} algorithms do not necessarily protect against transaction reordering.
On the other hand, cryptographic solutions such as commit \& reveal \cite{sok_preventing_transaction_reordering_manipulations_in_decentralized_finance,maximal_extractable_value_protection_on_a_DAG,fairness_notions_in_dag_based_dlts} are expensive.
DAG-based ledgers, which do not rely upon leaders in the same way as classical Blockchains do, could, theoretically, be more robust to transaction reordering and thus constitute a convenient alternative and a partial solution to that problem.
In this paper, we study the robustness of existing DAG-based algorithms to specific scenarios of transaction reordering.

{\em Send-order fairness} \cite{order_fairness_for_byzantine_consensus} relates the orders of transaction emission (by clients) and finalization. Upholding this property would prevent front-running by an adversary with a superior network connection.
However, in contrast to receive-order fairness, which involves locally observed reception orders on specific nodes, send-order fairness involves a global order of send events across all clients.
Distant machines having uncorrelated local clocks, and malicious clients being likely to falsify timestamps, maintaining that order would require consequent additional mechanisms and remains an unsolved problem \cite{order_fairness_for_byzantine_consensus}.
Yet, this notion of fairness remains a useful theoretical tool and it can actually be measured in a simulated environment (we have access to the simulator's clock).
Send-order fairness can be facilitated if one upholds ``fairness to clients'' \cite{fairness_notions_in_dag_based_dlts}.
The notion of {\em client-fairness} may refer to clients being treated fairly by the nodes. This includes how nodes should handle receiving multiple transactions from multiple clients \cite{byzid_byzantine_fault_tolerane_from_intrusion_detection,rbft_redundant_byzantine_fault_tolerance}. 
It can also correspond to nodes having ``fair access to transactions from all clients'', which can be modeled using Jain's fairness index \cite{a_quantitative_measure_of_fairness_and_discrimination_for_resource_allocation_in_shared_computer_systems}.

\section{Model\label{sec:prel}}

\paragraph{System Model}
We consider a permissioned system with $m$ clients and $n$ nodes.
These nodes are the replicated state machines of the DL and agree together on the order of transactions send by clients.
Once an honest node $\eta$ receives a new transaction $x$, $x$ is stored in the node's local mempool.
In a Blockchain network such as Cosmos \cite{dissecting_tendermint} or Ethereum PoS \cite{exploiting_ethereum_after_the_merge_the_interplay_between_pos_and_mev_strategies}, if $\eta$ emerges as the proposer of the next block, $x$ may be included in that block.
In DAG-based ledgers such as DagRider \cite{all_you_need_is_dag}, $\eta$ may include $x$ in its next vertex proposal.


\paragraph{Network Model}
Nodes and clients operate on a partially synchronous network \cite{consensus_in_the_presence_of_partial_synchrony} i.e., there exists a bound $\Delta$ that is not known by any node ($\Delta$ being unknown is equivalent to formulating partial synchrony using the Global Stabilization Time \cite{consensus_in_the_presence_of_partial_synchrony,themis_fast_strong_order_fairness_in_byzantine_consensus}) s.t., all messages send at time $t$ are received within $]t,t+\Delta]$.
A public key infrastructure is used to authenticate clients and nodes via digital signatures.

\paragraph{Application layer use-case}
The study of OF is notably motivated by the costs of MEV attacks \cite{flash_boys_frontrunning_in_decentralized_exchanges_miner_extractable_value_and_consensus_instability,sok_preventing_transaction_reordering_manipulations_in_decentralized_finance} in decentralized finance.
This usecase is characterized by \textbf{(1)} non commutative transactions and \textbf{(2)} incentives for an attacker to manipulate the order of transactions.
As in \cite{adversary_augmented_simulation_to_evaluate_client_fairness_on_hyperledger_fabric}, we consider a simple application layer which has both \textbf{(1)} and \textbf{(2)}. 
It consists in a game where puzzles are regularly revealed and clients compete to solve them.
Whenever a client $\chi$ solves a puzzle $k$, it submits a transaction $\chi:k$ that contains its solution.
For puzzle $k$, the winner is the first client $\chi$ s.t., $\chi:k$ is finalized.
For a given execution of the system, \%$g(\chi)$ denotes the percentage of games a client $\chi$ has won.
Supposing every client has knowledge of the puzzles at the same time and has the same ability, the game is {\em client-fair} iff \%$g(\chi)$ converges towards $1/m$.
By defining $\mathtt{score}(\chi) = \%g(\chi) * m$ we have a metric to evaluate client-fairness that is protocol agnostic and independent from the parameterization (numbers $m$ of clients, $n$ of nodes, etc.).
Unlike Jain's fairness index \cite{a_quantitative_measure_of_fairness_and_discrimination_for_resource_allocation_in_shared_computer_systems}, which is global, our $\mathtt{score}(\chi)$ is ``client-specific'' (an interesting metric in cases where specific clients are targeted). It also measures {\em send-order fairness} because, if the latter is upheld, then puzzle solutions of clients are finalized in the order in which they are send. Thus their success likelihood only depends upon their ability (and not on e.g., their network connection).

\paragraph{Adversary}
Our adversary may control $b \in [0,f]$ nodes (where $f = (n-1)/3$) and force them to deviate from the protocol. However, it does not have direct control over the network (e.g., it can only delay messages emitted by the nodes it controls).
The goal of the adversary is to reduce $\mathtt{score}(\chi)$ for a target client $\chi$, reducing its chances of winning puzzles.
We also evaluate the impact of the adversary by measuring numbers of violations of OF properties.
This amounts to comparing, for each pair of finalized transactions $x$ and $x'$, their finalization order in the ledger and either the order of their initial emission (by the corresponding clients) or the order of their reception in each of the nodes.
Given $X$ the total number of finalized transactions at one moment in time, there are $X*(X-1)/2$ such pairs. 
In order to have statistically significant results, we need a high number $p$ of puzzle games, which yields a large number of transactions $X = m*p$ having $m$ transactions per game.
Thus, comparing reception orders on every pair of transactions on every node amounts to $n*m*p*(m*p -1)/2$ comparisons, which quickly become intractable.
In our usecase application, the order between two transactions $\chi_1:k_1$ and $\chi_2:k_2$ only matters if they concern the same puzzle game i.e., iff $k_1 = k_2$ (otherwise they can commute).
We leverage this by only comparing transactions involving the same puzzle, which only requires $n*p*m*(m-1)/2$ comparisons.
Then, counting the numbers of times $z$ that transaction $x$ is received before $x'$, we have that, if $z > n/2$, then receive-order fairness is violated if $x$ is finalized after $x'$.
By counting the numbers of times each property is violated (for all properties and all relevant pairs of transactions) during the run of a system, we can evaluate their respect and the effect that the adversarial attacks have on them.

\section{DAG-based ledgers and DagRider\label{sec:dag}}

\subsection{DAGs \& Blockchains\label{ssec:dags_and_blockchains}}

\begin{wrapfigure}{r}{.25\textwidth}
\vspace*{-.9cm}
    \centering
    \scalebox{.75}{\begin{tikzpicture}
\node[fill=violet,fill opacity=.25,draw=black,text opacity=1,minimum height=1.75cm, minimum width=.8cm,inner sep=0cm] (v_0_A) at (0,0) {
$
\begin{array}{l}
v_0^A
\\
~
\\
~
\\
\end{array}
$
};
\node[fill=cyan,fill opacity=.25,draw=black,text opacity=1,minimum height=1.75cm, minimum width=.8cm,right=.5cm of v_0_A,inner sep=0cm] (v_1_A) {
$
\begin{array}{l}
v_1^A
\\
\hline 
x_1
\\
~
\\
\end{array}
$
};
\node[fill=red,fill opacity=.25,draw=black,text opacity=1,minimum height=1.75cm, minimum width=.8cm,right=.5cm of v_1_A,inner sep=0cm] (v_2_A) {
$
\begin{array}{l}
v_2^A
\\
\hline 
x_3
\\
~
\\
\end{array}
$
};
\node[fill=red,fill opacity=.25,draw=black,text opacity=1,minimum height=1.75cm, minimum width=.8cm,right=.5cm of v_2_A,inner sep=0cm] (v_3_A) {
$
\begin{array}{l}
v_3^A
\\
\hline 
x_2
\\
x_4
\\
\end{array}
$
};
\node[fill=red,fill opacity=.25,draw=black,text opacity=1,minimum height=1.75cm, minimum width=.8cm,right=.5cm of v_3_A,inner sep=0cm] (v_4_A) {
$
\begin{array}{l}
v_4^A
\\
\hline 
x_5
\\
~
\\
\end{array}
$
};
%
%
\node[fill=violet,fill opacity=.25,draw=black,text opacity=1,minimum height=1.75cm, minimum width=.8cm,inner sep=0cm] (v_0_B) at (0,-2.25) {
$
\begin{array}{l}
v_0^B
\\
~
\\
~
\\
\end{array}
$
};
\node[fill=cyan,fill opacity=.25,draw=black,text opacity=1,minimum height=1.75cm, minimum width=.8cm,right=.5cm of v_0_B,inner sep=0cm] (v_1_B) {
$
\begin{array}{l}
v_1^B
\\
\hline 
x_1
\\
x_2
\\
\end{array}
$
};
\node[fill=cyan,fill opacity=.25,draw=black,text opacity=1,minimum height=1.75cm, minimum width=.8cm,right=.5cm of v_1_B,inner sep=0cm] (v_2_B) {
$
\begin{array}{l}
v_2^B
\\
\hline 
x_3
\\
~
\\
\end{array}
$
};
\node[fill=red,fill opacity=.25,draw=black,text opacity=1,minimum height=1.75cm, minimum width=.8cm,right=.5cm of v_2_B,inner sep=0cm] (v_3_B) {
$
\begin{array}{l}
v_3^B
\\
\hline 
x_4
\\
~
\\
\end{array}
$
};
\node[draw=black,minimum height=1.75cm, minimum width=.8cm,right=.5cm of v_3_B,inner sep=0cm] (v_4_B) {
$
\begin{array}{l}
v_4^B
\\
\hline 
x_5
\\
~
\\
\end{array}
$
};
%
%
\node[fill=violet,fill opacity=.25,draw=black,text opacity=1,minimum height=1.75cm, minimum width=.8cm,inner sep=0cm] (v_0_C) at (0,-4.5) {
$
\begin{array}{l}
v_0^C
\\
~
\\
~
\\
\end{array}
$
};
\node[fill=red,fill opacity=.25,draw=black,text opacity=1,minimum height=1.75cm, minimum width=.8cm,right=.5cm of v_0_C,inner sep=0cm] (v_1_C) {
$
\begin{array}{l}
v_1^C
\\
\hline 
x_2
\\
~
\\
\end{array}
$
};
%
%
%
\draw[->] (v_1_A.150) -- (v_0_A.east);
\draw[->] (v_1_A.150) -- (v_0_B.east);
\draw[->] (v_2_A.150) -- (v_1_A.east);
\draw[->] (v_2_A.150) -- (v_1_B.east);
\draw[->] (v_3_A.150) -- (v_2_A.east);
\draw[->] (v_3_A.150) -- (v_2_B.east);
\draw[->] (v_4_A.150) -- (v_3_A.east);
\draw[->] (v_4_A.150) -- (v_3_B.east);
\draw[->] (v_1_B.150) -- (v_0_B.east);
\draw[->] (v_1_B.150) -- (v_0_C.east);
\draw[->] (v_2_B.150) -- (v_1_B.east);
\draw[->] (v_2_B.150) -- (v_1_A.east);
\draw[->] (v_3_B.150) -- (v_2_A.east);
\draw[->] (v_3_B.150) -- (v_2_B.east);
\draw[->] (v_4_B.150) -- (v_3_A.east);
\draw[->] (v_4_B.150) -- (v_3_B.east);
\draw[->] (v_1_C.150) -- (v_0_C.east);
\draw[->] (v_1_C.150) -- (v_0_B.east);
\node[inner sep=0] (anchor1) at ($(v_2_B.south east) + (.1,-.1)$) {}; 
\node[inner sep=0] (anchor2) at ($(v_2_B.south west) + (-.2,-.1)$) {}; 
\draw (v_3_B.235) edge[bend left=10,dashed] (anchor1);
\draw (anchor1) edge[dashed] (anchor2);
\draw (anchor2) edge[->,bend right=5,dashed] (v_1_C);
%
%
%
\node[draw,inner sep=.5] at (4.15,-4.35) {
\scalebox{.9}{
\input{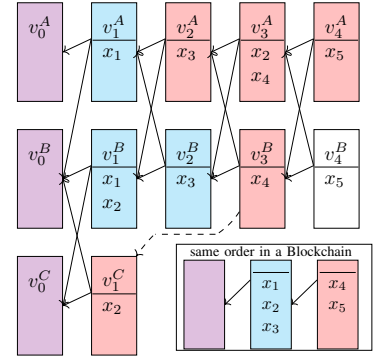}
}
};
\end{tikzpicture}}
    \caption{Simplified example}
    \label{fig:distributed_ledger_dag_and_blockchains}
\vspace*{-.4cm}
\end{wrapfigure}

DAG-based DL such as \cite{all_you_need_is_dag,narwhal_and_tusk,bullshark} rely on a Directed Acyclic Graph which vertices contain batches of transactions.
These graphs are structured in rows, corresponding to individual nodes that author vertices, and columns, representing the passing of time.
Fig.\ref{fig:distributed_ledger_dag_and_blockchains} represents a snapshot of such a DAG, built by a ledger that contains 3 nodes $A$, $B$ and $C$.
In the following we may use such simplified diagrammatic examples (with $n=3$) for didactic purposes.
The top/middle/bottom row represents vertices added by node $A$/$B$/$C$. 
We denote by $v_c^r$ the vertex at column $c \in \mathbb{N}$ and row $r \in \{A,B,C\}$.
Each node adds new vertices sequentially.
As in Blockchains, all nodes start with a common genesis state. 
In DAG-based ledgers this corresponds to a set of genesis vertices (in purple on Fig.\ref{fig:distributed_ledger_dag_and_blockchains}).
Then, so as to ensure consistency, new vertices contain hash references to previous vertices (found in previous columns).
Unlike Blockchains, in which there is a unique hash reference in each block, here, a vertex contains multiple references that are called {\em edges}.
\textbf{Strong} (resp.~\textbf{weak}) edges relate vertices of column $c$ to vertices of column $c-1$ (resp.~$c' < c-1$).
On Fig.\ref{fig:distributed_ledger_dag_and_blockchains} the only weak edge corresponds to the dashed arrow from $v_3^A$ to $v_1^C$.
To be able to broadcast a new vertex at column $c+1$, a node must have knowledge of a sufficient number of vertices at a column $c$, a well formed vertex having a minimum number of strong edges.
Weak edges are not required but can be used to allow vertices from ``slow'' nodes to be finalized.
The order with which vertices are ordered is inferred deterministically from the structure of the DAG.
So-called {\em leader vertices} are selected regularly (at fixed intervals of columns, a node/row is selected).
The causal sub-graphs of leader vertices are then ordered deterministically and finalized in \textit{waves}.

In \cite{all_you_need_is_dag}, columns are called rounds and a wave has $4$ rounds.
Also, the leader of a wave $w$ is defined as a vertex on the first round of the wave, that is elected retrospectively once the wave is complete (i.e., at the fourth round of the wave). It is therefore the leader of wave $w+1$ which determines the content of wave $w$ as its causal sub-graph.
In classical Blockchains, the notion of ``round'' is prevalent and refers to steps that can be repeated multiple times inside a cryptographic or consensus algorithm.
To avoid potential confusions, in the following, we will use the term ``column'' exclusively.
Also, it is generally the leader at a height $h$ that determines the content of the $h^{th}$ block.
In order to more easily juxtapose notions of waves and blocks\footnote{and to highlight that if the height $h$ leader is Byzantine, the adversary may more easily reorder transactions in the $h^{th}$ block}, in the following, we consider that ``the leader of wave $w$'' ($w$ starting at $0$) refers to the vertex at column $1 + 4*w$, that determines the wave's content as its causal sub-graph.

On Fig.\ref{fig:distributed_ledger_dag_and_blockchains}, we identify 2 waves of size $2$ (simplified illustrations use smaller waves), the first in cyan and the second in red .
Here, vertex $v_2^B$ (the right-most cyan vertex) is selected as leader for column $2$. 
The first wave, in cyan, therefore correspond to the causal sub-graph of $v_2^B$. 
In this simplified example, a wave's vertices are ordered from top to bottom and left to right. The final order then ignores duplicated transactions.
This yields $\lbrack x_1,x_2,x_3 \rbrack$ being finalized. 
Then, $v_4^A$ is selected as leader of column $4$, determining the second wave in red, which yields $\lbrack x_4,x_5 \rbrack$ being finalized.
Matching waves of vertices to blocks, we may obtain the same ordering of transactions in a traditional Blockchain as is illustrated on the bottom right of Fig.\ref{fig:distributed_ledger_dag_and_blockchains}.
In both cases, $\lbrack x_1, x_2, x_3 \rbrack$ and then $\lbrack x_4,x_5 \rbrack$ are finalized.
Even though Blockchains can provide the same service as DAG-based ledgers, the latter claim a better decentralization (due to a less obvious dependency on selected leader nodes) and a higher throughput at the price of a higher latency.
DAG-based ledgers may also arguably be less vulnerable to order-fairness attacks, which evaluation is the object of this present paper.

\subsection{DagRider\label{ssec:dagrider}}

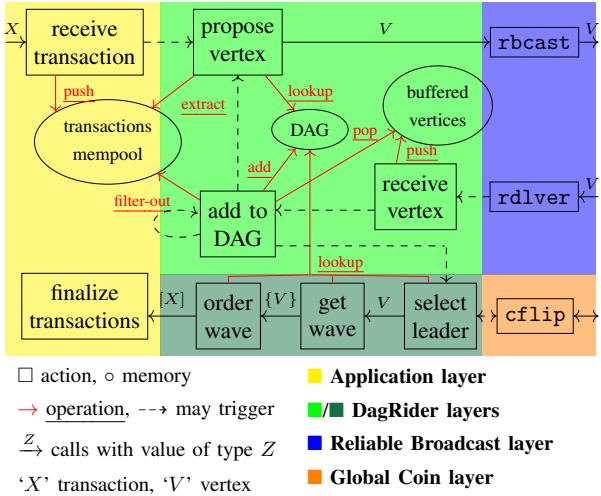
\begin{figure}[h]
\vspace*{-.4cm}
    \centering
\scalebox{.875}{\begin{tikzpicture}
\tikzset{
    protocol_action/.style={
        draw,align=center,minimum height=1cm
        },
    buffer_memory/.style={
        draw,align=center,ellipse
        },
}
%
%
\node[protocol_action] (rec_tr) at (0,0) {receive\\transaction};
\node[protocol_action,right=.75cm of rec_tr] (prop_vert) {propose\\vertex};
\node[protocol_action,below=1.75cm of prop_vert] (add_todag) {add to\\DAG};
\node[protocol_action,above right=-.6cm and 1.5cm of add_todag] (rec_vert) {receive\\vertex};
\node[draw, right=.5cm of rec_vert] (r_deliver) {$\mathtt{rdlver}$};
\node[draw, right=3.15cm of prop_vert] (r_bcast) {$\mathtt{rbcast}$};
\node[protocol_action,below right=.8cm and -.8cm of rec_vert] (sel_leader) {select\\leader};
\node[protocol_action,left=.6cm of sel_leader] (get_wave) {get\\wave};
\node[protocol_action,left=.6cm of get_wave] (order_vert) {order\\wave}; 
\node[protocol_action,below=3cm of rec_tr] (del_tr) {finalize\\transactions};
\node[draw, right=.3cm of sel_leader] (c_flip) {$\mathtt{cflip}$};
%
%
%
\node[buffer_memory,below left=.55cm and .5cm of prop_vert] (bf_tr) {\footnotesize transactions\\\footnotesize mempool};
\node[buffer_memory,below left=.3cm and .2cm of r_bcast] (bf_vert) {\footnotesize buffered\\\footnotesize vertices};
\node[buffer_memory,below right=.6cm and 0cm of prop_vert] (bf_dag) {\footnotesize DAG};
%
%
\draw (bf_tr.150 |- rec_tr.south) edge[->,red]  node[right] {\scriptsize \ul{push}} (bf_tr.150);
\draw (prop_vert) edge[->,red] node[below right] {\scriptsize \ul{extract}} (bf_tr);
\draw (prop_vert) edge[->,red] node[right] {\scriptsize \ul{lookup}} (bf_dag);
\draw (add_todag) edge[->,red]  node[below left] {\scriptsize \ul{filter-out}} (bf_tr);
\draw (rec_vert.120) edge[->,red] node[right,pos=.4] {\scriptsize \ul{push}} (bf_vert.220);
\draw (add_todag) edge[->,red] node[above,pos=.75] {\scriptsize \ul{pop}} (bf_vert);
\draw (add_todag) edge[->,red] node[left] {\scriptsize \ul{add}} (bf_dag);
\node[inner sep=0pt] (anchor_lookup_dag_split) at ($(bf_dag.south) + (0,-1.9)$) {};
\draw (bf_dag.south) edge[<-,red] node[right,pos=.9] {\scriptsize \ul{lookup}} (anchor_lookup_dag_split.south);
\draw (anchor_lookup_dag_split) edge[red] (sel_leader.110 |- anchor_lookup_dag_split.south);
\draw (sel_leader.110) edge[red] (sel_leader.110 |- anchor_lookup_dag_split.south);
\draw (anchor_lookup_dag_split) edge[red] (order_vert.north |- anchor_lookup_dag_split.south);
\draw (order_vert.north) edge[red] (order_vert.north |- anchor_lookup_dag_split.south);
\draw (get_wave.north) edge[red] (get_wave.north |- anchor_lookup_dag_split.south);
%
%
\draw (rec_tr) edge[dashed,->] (prop_vert);
\draw (add_todag) edge[dashed,->] (prop_vert);
\draw (prop_vert) edge[->] node[above] {\scriptsize $V$} (r_bcast);
\draw (rec_vert.west |- add_todag.20) edge[dashed,->] (add_todag.20);
\draw (add_todag) edge[loop left,dashed,->] (add_todag);
\draw (sel_leader) edge[->] node[above] {\scriptsize $V$} (get_wave);
\draw (get_wave) edge[->] node[above] {\scriptsize $\{V\}$} (order_vert);
\draw (order_vert.west) edge[->] node[above] {\scriptsize $[X]$} (del_tr.east |- order_vert.west);
\node[inner sep=0pt] (anchor_add_to_dag_right) at (sel_leader.80 |- add_todag.330) {};
\draw (add_todag.330) edge[dashed] (anchor_add_to_dag_right);
\draw (anchor_add_to_dag_right) edge[->,dashed] (sel_leader.80);
%
%
%
%
\draw (r_deliver) edge[->,dashed] (rec_vert);
\draw (sel_leader) edge[<->,dashed] (c_flip);
%
%
\draw ($(r_deliver.east) + (.3,0)$) edge[->] node[above,pos=.3] {\scriptsize $V$} (r_deliver);
\draw (r_bcast.east) edge[->] node[above,pos=.7] {\scriptsize $V$} ($(r_deliver.east |- r_bcast.east) + (.3,0)$);
\draw ($(rec_tr.west) + (-.3,0)$) edge[->] node[above,pos=.3] {\scriptsize $X$} (rec_tr);
\draw (c_flip.east) edge[<->] ($(r_deliver.east |- c_flip.east) + (.3,0)$);
%
%
\node[inner sep=0pt] (anchor_full_top) at ($(rec_tr.north) + (0,.1)$) {};
\node[inner sep=0pt] (anchor_full_left) at ($(del_tr.west) + (-.25,0)$) {};
\node[inner sep=0pt] (anchor_full_right) at ($(r_bcast.east) + (.25,0)$) {};
\node[inner sep=0pt] (anchor_full_bottom) at ($(sel_leader.south) + (0,-.1)$) {};
\node[inner sep=0pt] (anchor_application_right) at ($(rec_tr.east) + (.25,0)$) {};
\node[inner sep=0pt] (anchor_broadcast_left) at ($(r_bcast.west) + (-.125,0)$) {};
\node[inner sep=0pt] (anchor_ordering_top) at ($(sel_leader.north) + (0,.125)$) {};

\begin{scope}[on background layer]
\path[fill=yellow,opacity=.5] 
    (anchor_full_left |- anchor_full_top) 
    -- (anchor_application_right |- anchor_full_top) 
    -- (anchor_application_right |- anchor_full_bottom) 
    -- (anchor_full_left |- anchor_full_bottom) ;
\path[fill=green,opacity=.5] 
    (anchor_application_right |- anchor_full_top) 
    -- (anchor_broadcast_left |- anchor_full_top) 
    -- (anchor_broadcast_left |- anchor_ordering_top) 
    -- (anchor_application_right |- anchor_ordering_top)  ;
\path[fill=darkspringgreen,opacity=.5] 
    (anchor_application_right |- anchor_ordering_top) 
    -- (anchor_broadcast_left |- anchor_ordering_top) 
    -- (anchor_broadcast_left |- anchor_full_bottom) 
    -- (anchor_application_right |- anchor_full_bottom)  ;
\path[fill=blue,opacity=.5] 
    (anchor_broadcast_left |- anchor_full_top) 
    -- (anchor_full_right |- anchor_full_top) 
    -- (anchor_full_right |- anchor_ordering_top) 
    -- (anchor_broadcast_left |- anchor_ordering_top)  ;
\path[fill=orange,opacity=.5] 
    (anchor_broadcast_left |- anchor_ordering_top) 
    -- (anchor_full_right |- anchor_ordering_top) 
    -- (anchor_full_right |- anchor_full_bottom) 
    -- (anchor_broadcast_left |- anchor_full_bottom)  ;
\end{scope}
%
%
\node[align=center] (leg1) at (.3,-5.1) {\small$\square$ action, $\circ$ memory}; 
\node[below=.25cm of leg1.south west,anchor=west,align=center] (leg2) {\small\shortColRed{$\rightarrow$} \ul{operation}, $\dashrightarrow$ may trigger}; 
\node[below=.25cm of leg2.south west,anchor=west,align=center] (leg3) {\small$\xrightarrow{Z}$ calls with value of type $Z$}; 
\node[below=.25cm of leg3.south west,anchor=west,align=center] (leg4) {\small `$X$' transaction, `$V$' vertex}; 
\node[right=1.5cm of leg1,align=center] (dr_leg0) {\small\textcolor{yellow}{$\blacksquare$} \textbf{Application layer}}; 
\node[below=.25cm of dr_leg0.south west,anchor=west,align=center] (dr_leg) {\small\textcolor{green}{$\blacksquare$}/\textcolor{darkspringgreen}{$\blacksquare$} \textbf{DagRider layers}};
\node[below=.25cm of dr_leg.south west,anchor=west,align=center] (rb_leg) {\small\textcolor{blue}{$\blacksquare$} \textbf{Reliable Broadcast layer}};  
\node[below=.25cm of rb_leg.south west,anchor=west,align=center] (gc_leg) {\small\textcolor{orange}{$\blacksquare$} \textbf{Global Coin layer}};
\end{tikzpicture}}
    \caption{Description of DagRider}
    \label{fig:layers_dagrider}
\vspace*{-.4cm}
\end{figure}

In \cite{all_you_need_is_dag}, DagRider is described via pseudocode.
For concision, we rather describe it via the atomic actions a DagRider node may perform and how these actions are triggered by each other or by external stimuli (c.f.~actor model \cite{a_universal_modular_actor_formalism_for_artificial_intelligence}).
Fig.\ref{fig:layers_dagrider} describes the behavior of any individual node. 
The rectangles and ellipses resp.~correspond to atomic actions it may execute and to persistent information stored in its memory.
The red arrows describe the effects of the actions on the memory.
Plain arrows carrying a symbol $Z$ correspond to calling an action on a value of type $Z$. 
Dotted arrows signify that the completion of an action may trigger the execution of another.
The colored areas correspond to layers of abstraction.

In the application layer, ``receive transaction'' is triggered whenever a node receives a new transaction $x$ of type $X$. 
This results in $x$ (if not duplicated) being \ul{pushed} in a ``transactions mempool'' buffer.
This may in turn trigger the proposal of a new vertex by the node $r$ if the following condition is met: the node is at column $c$ (i.e, it has already broadcast vertices for all $c'<c$ and has yet to do so for $c$) and there are at least $2*f+1$ distinct vertices at column $c-1$ in its local copy of the DAG (which can be verified via \ul{lookup} of the ``DAG''). 
In that case, ``propose vertex'' creates a new vertex $v_c^r$ that contains transactions \ul{extracted} from ``buffered transactions'', has at least $2*f+1$ strong edges and at most $f$ weak edges.
Details in Appendix \ref{anx:bug_dagrider}.

In the reliable broadcast layer (in blue on Fig.\ref{fig:layers_dagrider}), ``$\mathtt{rbcast}$'' and ``$\mathtt{rdlver}$'' resp.~correspond to triggering the broadcast and delivering the vertex.
As per \cite{all_you_need_is_dag}, the broadcast must guarantee: \textbf{agreement} i.e., if a correct node delivers a vertex $v$ then, eventually, all the other correct nodes will also deliver $v$, \textbf{integrity} i.e., for each round, only one vertex can be delivered per node and \textbf{validity} i.e., if a correct node broadcast a correct vertex $v$ then, eventually, every correct node will deliver $v$.
These properties are s.t.~we may safely ignore issues related to duplicated or equivocated vertices.

Once ``propose vertex'' outputs a vertex $v$, it is reliably broadcast to the $n$ nodes of the network via $\mathtt{rbcast}$.
Then, eventually, $\mathtt{rdlver}$ triggers ``receive vertex'' which \ul{pushes} $v$ in a set of ``buffered vertices''.
If all vertices $v'$ that are targets of the edges of $v$ are already present in the node's local copy of the DAG, then ``add to DAG'' is triggered, which causes the node to remove vertices from ``buffered vertices'' and to insert them at their correct position in its local copy of the DAG.
The reception of a single vertex may have cascading effects and cause the addition of a large number of vertices to the DAG.
Whenever a vertex is added to the DAG, the transactions it contains are \ul{filtered-out} from ``transactions mempool'', so that they won't be included in subsequent vertex proposals (thus purging the mempool).

As the local copy of the DAG grows, columns are progressively filled, which allows constructing the waves.
The first step is ``select leader'', which outputs the leader of wave $w \geq 0$ (at column $1 + w*4$). 
Its determination corresponds to the selection of a node/row.
So that the adversary cannot predict it and tamper with the broadcast of the leader vertex (via e.g., Denial of Service \cite{a_framework_for_classifying_denial_of_service_attacks}), this is done via a global perfect coin. 
This protocol (orange layer on Fig.\ref{fig:layers_dagrider}) provides a ``$\mathtt{cflip}$'' primitive that eventually returns the identity of the leader once at least $f+1$ distinct nodes called $\mathtt{cflip}$. 
This may be implemented via threshold signatures \cite{practical_threshold_signatures}.

Once a leader is selected at column $c = 1 + w*4$, the algorithm first verifies that there are at least $2*f+1$ strong paths between it and vertices at column $4 + w*4$. 
If this is not the case, the leader is skipped.
Otherwise, a new wave is formed, containing all the not-yet ordered vertices in the causal sub-graph of the leader.
This corresponds to the output of ``get wave'' on Fig.\ref{fig:layers_dagrider}.
As per \cite{all_you_need_is_dag}, the vertices of the wave are then ordered using a certain deterministic order ``order wave''.
This results in a total ordering of the transactions, which are then finalized.

\subsection{Reliable Broadcast and Global Coin}

In the following, we consider that the reliable broadcast layer of Fig.\ref{fig:layers_dagrider} is implemented via Bracha's Byzantine Reliable Broadcast algorithm \cite{asynchronous_byzantine_agreement_protocols} (as suggested in \cite{bullshark}, see Appendix \ref{anx:bracha}).
As for the global coin layer, for the sake of simplicity, we abstract it away via PRNG with a seed that depends on the wave number so that its ``agreement'' property \cite{all_you_need_is_dag} is ensured.

\subsection{Deterministic order for wave finalization\label{ssec:deterministic_order}}

In \cite{all_you_need_is_dag,bullshark,narwhal_and_tusk}, the deterministic process used to order vertices within a wave is not described beyond the requirement that it is deterministic (to guarantee consistency).
However, as we are interested in OF, implementation details matter.
In this paper, we consider three variants: ``FullShuffle'', ``PerColumnShuffle'' and ``VoteCount''.

\paragraph{FullShuffle} This simply consists in randomly shuffling the vertices. 
To ensure determinism, we can fix a PRNG seed. So that it is shared without communication overhead and cannot be predicted in advance, this seed may be computed from e.g., the leader vertex hash value.

\paragraph{PerColumnShuffle} In this variant: \textbf{(1)} the causal sub-graph is split by columns, \textbf{(2)} for each column, vertices are sorted according to a permutation of the rows that depends on a PNRG which seeds depends on the corresponding vertices hash values (ensuring fairness and consistency) and \textbf{(3)} vertices are delivered in ascending order of columns.

\paragraph{VoteCount} This variant is inspired by \cite{reducing_latency_of_dag_based_consensus_in_the_asynchronous_setting_via_the_utxo_model}, \cite{themis_fast_strong_order_fairness_in_byzantine_consensus} and \cite{diversified_top_k_graph_pattern_matching}.
\cite{reducing_latency_of_dag_based_consensus_in_the_asynchronous_setting_via_the_utxo_model} considers that a node $r$ ``votes'' for a vertex $v$ at column $c$ if $c$ is the earliest column such that $v$ is in the causal sub-graph of $v_c^r$.

\begin{figure}[h]
\vspace*{-.4cm}
    \centering

\begin{subfigure}{.24\textwidth}
    \centering
    \scalebox{.85}{\begin{tikzpicture}
\node[draw] at (0,0)                       (vA1) {$v^A_{1}$};
\node[draw=none,        below=.3cm of vA1] (vB1) {\phantom{$v^B_{1}$}};
\node[draw,             below=.3cm of vB1] (vC1) {$v^C_{1}$};
\node[draw,             below=.3cm of vC1] (vD1) {$v^D_{1}$};
\node[draw,        right=.6cm of vA1] (vA2) {$v^A_{2}$};
\node[draw,        right=.6cm of vA2] (vA3) {$v^A_{3}$};
\node[draw,        right=.6cm of vB1] (vB2) {$v^B_{2}$};
\node[draw,        right=.6cm of vC1] (vC2) {$v^C_{2}$};
\node[draw,        right=.6cm of vC2] (vC3) {$v^C_{3}$};
\node[draw,        right=.6cm of vC3] (vC4) {$v^C_{4}$};
\node[draw,        right=.6cm of vD1] (vD2) {$v^D_{2}$};
\node[draw,        right=.6cm of vD2] (vD3) {$v^D_{3}$};
\draw[->] (vA2) -- (vA1);
\draw[->] (vA2) -- (vC1);
\draw[->] (vB2) -- (vA1);
\draw[->] (vB2) -- (vC1);
\draw[->] (vC2) -- (vA1);
\draw[->] (vC2) -- (vC1);
\draw[->] (vD2) -- (vA1);
\draw[->] (vD2) -- (vC1);
\draw[->] (vA3) -- (vA2);
\draw[->] (vA3) -- (vB2);
\draw[->] (vA3) -- (vC2);
\draw[->] (vC3) -- (vB2);
\draw[->] (vC3) -- (vC2);
\draw[->] (vC3) -- (vD2);
\draw[->] (vD3) -- (vA2);
\draw[->] (vD3) -- (vB2);
\draw[->] (vD3) -- (vD2);
\draw[->] (vC4) -- (vA3);
\draw[->] (vC4) -- (vC3);
\draw[->] (vC4) -- (vD3);
\draw (vD3) edge[dashed,->,bend left=45] (vD1);
\draw (vC3) edge[dashed,->,bend left=55] (vD1);
\end{tikzpicture}}
    \caption{Wave}
    \label{fig:ex_full_wave}
\end{subfigure}
\begin{subfigure}{.23\textwidth}
    \centering
    \scalebox{.7}{\begin{tabular}{|c|c|c|c|c|}
\hline
& $A$ & $B$ & $C$ & $D$ \\
\hline 
$v_A^1$ & $1$ & $2$ & $2$ & $2$ \\
\hline 
$v_A^2$ & $2$ & $\infty$ & $4$ & $3$ \\
\hline 
$v_A^3$ & $3$ & $\infty$ & $4$ & $\infty$ \\
\hline 
$v_B^2$ & $3$ & $2$ & $3$ & $3$ \\
\hline 
$v_C^1$ & $2$ & $2$ & $1$ & $2$ \\
\hline 
$v_C^2$ & $3$ & $\infty$ & $2$ & $\infty$ \\
\hline 
$v_C^3$ & $\infty$ & $\infty$ & $3$ & $\infty$ \\
\hline 
$v_C^4$ & $\infty$ & $\infty$ & $4$ & $\infty$ \\
\hline 
$v_D^1$ & $\infty$ & $\infty$ & $3$ & $1$ \\
\hline 
$v_D^2$ & $\infty$ & $\infty$ & $3$ & $2$ \\
\hline 
$v_D^3$ & $\infty$ & $\infty$ & $4$ & $3$ \\
\hline 
\end{tabular}}
    \caption{Vote table}
    \label{fig:ex_full_table}
\end{subfigure}
    \caption{Example Wave}
    \label{fig:do_example_1}
\vspace*{-.4cm}
\end{figure}

Fig.\ref{fig:ex_full_wave} displays a wave which leader is $v^C_4$.
By interpreting the structure of the DAG, one can infer the vote table on Fig.\ref{fig:ex_full_table}.
Let us focus on $v_2^A$. Node $A$ votes for vertex $v_2^A$ at column $2$ because it is its own proposal. 
In the considered wave, $B$ never vote for $v_2^A$ because there is no edge between $v_2^B$ and $v_2^A$ (hence the $\infty$ on Fig.\ref{fig:ex_full_table}). 
$D$ votes for $v_2^A$ at column $3$ because there is an edge between $v_3^D$ and $v_2^A$. 
$C$ votes for $v_2^A$ at column $4$ because there is a path between $v_4^C$ and $v_2^A$ (via either $v_3^A$ or $v_3^D$) but no path between $v_3^C$ and $v_2^A$.

From the vote table, a partial order $\geq_{\text{\faEnvelopeO}}$ is inferred.
In \cite{reducing_latency_of_dag_based_consensus_in_the_asynchronous_setting_via_the_utxo_model} it corresponds to, $x <_{\text{\faEnvelopeO}} x'$ iff $|\{\text{nodes vote}~x~\text{first}\}| > |\{\text{nodes vote}~x'~\text{first}\}|$.
The reader may notice the similarity with receive-order-fairness \cite{order_fairness_for_byzantine_consensus}.
In light of \cite{order_fairness_for_byzantine_consensus,quick_order_fairness,themis_fast_strong_order_fairness_in_byzantine_consensus}, many variants of $\geq_{\text{\faEnvelopeO}}$ may be considered.
Then, $\geq_{\text{\faEnvelopeO}}$ can be used to order the vertices of the wave.
However, \cite{reducing_latency_of_dag_based_consensus_in_the_asynchronous_setting_via_the_utxo_model} does not acknowledge the issue of Condorcet cycles (see Appendix \ref{anx:bug_board_and_clerk}).
The mechanism which we retain is inspired by \cite{order_fairness_for_byzantine_consensus,themis_fast_strong_order_fairness_in_byzantine_consensus}, using an additional notion from \cite{diversified_top_k_graph_pattern_matching}.
For editorial constraints, we discuss it in Appendix \ref{anx:vote_count_mechanism}.

\section{Attacks and their effect\label{sec:attacks}}

The goal of the adversary is to prevent a specific target client $\chi$ to win puzzle games.
To do so, puzzle solutions send by $\chi$ must be ordered after solutions of the same puzzle send by other clients.
We call such attacks transaction reordering attacks \cite{sok_preventing_transaction_reordering_manipulations_in_decentralized_finance} as the adversary changes the natural order of transactions which would otherwise have occurred.

\begin{wrapfigure}{r}{.19\textwidth}
    \centering
    \scalebox{.75}{\begin{tikzpicture}
\node[fill=violet,fill opacity=.25,draw=black,text opacity=1,minimum height=1.75cm, minimum width=.8cm] (b0) at (0,0) {
$
\begin{array}{l}
~
\\
~
\\
~
\\
\end{array}
$
};
\node[fill=cyan,fill opacity=.25,draw=black,text opacity=1,minimum height=1.75cm, minimum width=.8cm,right=.5cm of b0] (b1) {
$
\begin{array}{l}
\hline 
x_1
\\
x_2
\\
x_3
\\
\end{array}
$
};
\draw[->] (b1.130) -- (b0.east);
\node[fill=red,fill opacity=.25,draw=black,text opacity=1,minimum height=1.75cm, minimum width=.8cm,right=1cm of b1] (b2) {
$
\begin{array}{l}
\hline 
x_4
\\
x_5
\\
~
\\
\end{array}
$
};
\draw[->] (b2.130) -- (b1.east);
\node[red] (anchor) at (1.37,.42) {\Huge$\times$};
\draw (anchor.center) edge[bend left=35,->,red,dashed] ($(anchor) + (.25,-1)$);
\draw (anchor.center) edge[bend left=25,->,red,dashed] ($(anchor) + (1.7,-.1)$);
\node[draw=red,fill=white] (mal) at ($(anchor) + (.7,-.3)$) {\shortColRed{or}};
\node[draw=red,fill=white,circle,inner sep=.05cm] (var1) at ($(anchor) + (.5,-1.1)$) {\small\shortColRed{1}};
\node[draw=red,fill=white,circle,inner sep=.05cm] (var1) at ($(anchor) + (1.45,-.25)$) {\small\shortColRed{2}};
\end{tikzpicture}}
    \caption{Desired effect on a Blockchain}
    \label{fig:blockchain_attacked}
\vspace*{-.5cm}
\end{wrapfigure}
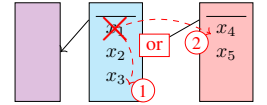

In some Blockchains such attacks may succeed if the leader of the next block is infected by the adversary.
Fig.\ref{fig:blockchain_attacked} illustrates this, supposing that the adversary wishes to delay the finalization of $x_1$ and has infected the leader node for the cyan block.
In this configuration, the adversary may (as represented by the red dashed arrows) either \shortColRed{\textcircled{1}} put $x_1$ at the end of the block or \shortColRed{\textcircled{2}} ignore $x_1$, which may then be added in the next block by the leader for the next height.
In doing so, the adversary accomplishes its goal without violating neither consistency nor liveness.
In both cases, it is likely (provided the honest leader would otherwise have respected it) that receive-order fairness is, as a result, violated.

In DAG-based DL, we have waves instead of blocks.
The equivalent of variant \shortColRed{\textcircled{1}} may amount to influencing the outcome of the deterministic order used to order vertices of a wave.
Variant \shortColRed{\textcircled{2}} in contrast, may amount to changing which vertices are included in the wave.

To achieve its goal, the adversary may infect a limited number of nodes and modify their behavior (on either or both the DagRider and Reliable Broadcast layers). This number is bound by the fault tolerance thresholds of the involved algorithms, here the same $f = (n-1)/3$ for Bracha and DagRider.
Honest nodes faithfully participate in both Bracha and DagRider.
This signifies that \textbf{(1)} they emit ECHO and READY messages as soon as possible, \textbf{(2)} they do not exclude received transactions from their vertex proposals, \textbf{(3)} they propose non-empty vertices as soon as possible and \textbf{(4)} upon proposing a vertex, they include all possible strong (between $2*f+1$ and $n$) and weak edges (between $0$ and $f$).
In the following, we present how infected nodes can be coordinated to achieve effect analogous to those in Fig.\ref{fig:blockchain_attacked}.

\subsection{Byzantine behavior on the DagRider layer\label{ssec:byz_dagrider_layer}}

Infected nodes may propose vertices differently.
In addition to not including transactions from the target client $\chi$, infected nodes will also avoid including edges that point towards other vertices that contain transactions from $\chi$.

\begin{wrapfigure}{l}{.25\textwidth}
\vspace*{-.3cm}
    \centering
    \scalebox{.75}{\input{figures/attack_goal/dag_attacked_dagrider_layer}}
    \caption{DagRider layer attack}
    \label{fig:dag_attacked_dagrider_layer}
\vspace*{-.4cm}
\end{wrapfigure}

Fig.\ref{fig:dag_attacked_dagrider_layer} illustrates this.
Here node B is controlled by the adversary.
If B were to behave honestly, the resulting DAG would be the one from Fig.\ref{fig:distributed_ledger_dag_and_blockchains}.
By contrast, Fig.\ref{fig:dag_attacked_dagrider_layer} results from B not including $x_1$ in $v_1^B$ and favoring a strong edge to $v_1^C$ rather than to $v_1^A$ in its $v_2^B$ vertex proposal.
Overall, this has the effect of excluding transaction $x_1$ from the first wave (in cyan), delaying its delivery to the second wave (in red). As a result, this has the same effect as variant \shortColRed{\textcircled{2}} of Fig.\ref{fig:blockchain_attacked}.
See Appendix \ref{anx:detail_dag_rider_layer_attack} for details.

\subsection{Byzantine behavior on the Bracha layer\label{ssec:attack_bracha}}

The delay between $\mathtt{rbcast}$ and $\mathtt{rdlver}$ depends on the termination of an instance of (in our case) Bracha's BRB algorithm \cite{on_the_versatility_of_bracha_byzantine_reliable_broadcast_algorithm}.
Because this algorithm relies on collecting (and thus waiting for) a number of messages from other nodes, the adversary can statistically delay it on all nodes via infecting a minority of nodes (which will not emit these messages for vertices that contain transactions from the target client).
See Appendix \ref{anx:detail_bracha_layer_attack} for details.

\begin{wrapfigure}{l}{.25\textwidth}
\vspace*{-.3cm}
    \centering
    \scalebox{.75}{\input{figures/attack_goal/dag_attacked_bracha_layer}}
    \caption{Bracha layer attack}
    \label{fig:dag_attacked_bracha_layer}
\vspace*{-.4cm}
\end{wrapfigure}

This ability to slow down the delivery of specific vertices allows colluding Byzantine nodes to manipulate the DAG edges.
In our example from Fig.\ref{fig:distributed_ledger_dag_and_blockchains}, node A being Byzantine could yield what is depicted on Fig.\ref{fig:dag_attacked_bracha_layer}.
Here, A not faithfully participating in the reliable broadcast of $v_1^B$ may slow its delivery and make it happen later than that of $v_1^A$ on node B.
In that case, once the honest B is ready to propose, it will produce a vertex $v_2^B$ that targets $v_1^A$ rather than $v_1^B$ (which is not available at the time of the proposal).

Both the attacks from Fig.\ref{fig:dag_attacked_dagrider_layer} and Fig.\ref{fig:dag_attacked_bracha_layer} amount to variant \shortColRed{\textcircled{2}} of Fig.\ref{fig:blockchain_attacked}
Implementing an attack on a DAG having an effect analogous to the one of variant \shortColRed{\textcircled{1}} of Fig.\ref{fig:blockchain_attacked} requires leveraging the deterministic order that is used to order vertices within a wave.
``FullShuffle'' is entirely random and as such cannot easily be leveraged.
However, ``VoteCount'' takes into consideration the structure of the DAG when formulating the finalization order. 
In particular, we remark that both attacks tend to delay the rounds at which nodes vote for the vertices that are targeted.
For instance, in our example if one consider the round at which $A$ votes for $v_1^B$, it is in round $2$ on Fig.\ref{fig:distributed_ledger_dag_and_blockchains} and $3$ on Fig.\ref{fig:dag_attacked_bracha_layer}.
As a result, even if the targeted vertex stays in the same wave, the attack may cause it to be ordered later, thus causing the analogous of variant \shortColRed{\textcircled{1}} of Fig.\ref{fig:blockchain_attacked}.

\subsection{Evaluation metrics\label{ssec:metrics}}

To asses the vulnerability of DAG-based DL to transaction reordering attacks, beyond reasoning on examples such as Fig.\ref{fig:dag_attacked_dagrider_layer} and Fig.\ref{fig:dag_attacked_bracha_layer}, we need concrete metrics.
As explained in Sec.\ref{sec:prel}, we can monitor the number of pairs of transactions $(x,x')$ for which OF properties are violated. We now introduce 8 DAG-specific OF properties (details in Appendix \ref{anx:enumeration_order_fairness_props}).

In Sec.\ref{ssec:dagrider} we characterized the lifecycle of a transaction $x$ on a specific DagRider node by four instants:
\textbf{(1)} reception of $x$ from a client,
\textbf{(2)} triggering the broadcast of the first vertex that contains $x$,
\textbf{(3)} delivery of the first vertex that contains $x$ 
and
\textbf{(4)} finalization of $x$ (the first occurrence of $x$, given that transactions may be duplicated in the DAG).
The definition of OF properties depend on these instants as OF relates the order of transaction reception on each individual node (``reception'', which can be understood as instant \textbf{(1)}, \textbf{(2)} or \textbf{(3)}) and the order of their eventual finalization (instant \textbf{(4)}).
Instants \textbf{(2)} and \textbf{(3)} are guaranteed to exist in all cases.
However, for instant \textbf{(1)}, it is true iff clients broadcast their transactions to all the nodes.

We consider properties $OF^\beta_\alpha$ of the form: for any transactions $x$ and $x'$, \textbf{if} $\alpha(x,x')$ \textbf{then} $\beta(x,x')$.

The $\beta$ predicate concerns transaction finalization, with two cases: $F_{IN}$ (for ``finalization'') and $W_{AV}$ (for ``wave''), so that $\beta \in \{F_{IN},~W_{AV}\}$.
For any two transactions $x$ and $x'$, $F_{IN}(x,x')$ signifies that all honest nodes must finalize $x$ before $x'$ while $W_{AV}(x,x')$ implies that no honest node can finalize $x$ in a wave after that in which $x'$ is finalized.
While $F_{IN}$ corresponds to {\em receive-order fairness}, $W_{AV}$ coincides with adapting a certain interpretation of the {\em block-order fairness} from \cite{order_fairness_for_byzantine_consensus} to DAGs.
$W_{AV}$ also provides an interesting metric. Indeed, violations of $OF^{W_{AV}}_*$ properties can be linked to variant variant \shortColRed{\textcircled{2}} of Fig.\ref{fig:blockchain_attacked} while those of $OF^{F_{IN}}_*$ properties may correspond to either \shortColRed{\textcircled{1}} or \shortColRed{\textcircled{2}}.

The $\alpha$ predicate involves relations between communication events. 
We consider four cases with $\alpha \in \{S_{ND},R_{EC},I_{NI},D_{LV}\}$.
$S_{ND}(x,x')$ signifies that the initial emission of $x$ by a certain client precedes that of $x'$.
$R_{EC}(x,x')$ (and resp.~$I_{NI}$ and $D_{LV}$) signifies that a majority of nodes receive $x$ from a client (and resp.~ begin their participation in the reliable broadcast of a vertex that contains $x$ and deliver a vertex containing $x$) before they do so for $x'$ i.e., instant \textbf{(1)} (and resp.~\textbf{(2)} and \textbf{(3)}).


\section{Parameterization and experiments\label{sec:experiments}}

\subsection{System parameterization\label{ssec:network_param}}

We consider $m=3$ clients and we simulate Peer To Peer communications with no loss.
Delays for the transmission of individual messages are sampled from hypoexponential probability distributions (as recommended in \cite{models_of_network_delay}), which we bound with an upper value $\Delta$.
This reflects the non-determinism in network communications while staying within the conservative hypothesis of a partially synchronous communication model \cite{consensus_in_the_presence_of_partial_synchrony}.
New puzzles are revealed at regular intervals.
The time required for a client to solve a puzzle is sampled from a Poisson distribution.
In addition of the transactions send by the clients, which contain puzzle solutions, the nodes also regularly receive third party transactions that are not part of the puzzle game (to simulate the DL being used concurrently by other applications at the application layer).

Our aim is to evaluate the robustness of the system w.r.t.~the proportion of nodes that are infected by the adversary (i.e., the Byzantine nodes)
We also vary three additional parameters:
\textbf{(1)} the deterministic order,
\textbf{(2)} the \textcolor{black}{\faClockO} distribution of delays between nodes (in this way, we vary the network ratio \cite{themis_fast_strong_order_fairness_in_byzantine_consensus} i.e., how the propagation time of transactions relates to the rate at which they are created),
and \textbf{(3)} the fanout from clients to nodes.
Details in Appendix \ref{anx:max_sim}.

In the following, we present three sets of experiments.
For the first two, we fix the value of $n$ to $13$, i.e., we have up to $f=4$ byzantine nodes.
Then, in the third set of experiments, we vary the value of $n$ to show that the results of the first two experiments do not depend on $n$.
All the materials required to reproduce them are available at \cite{max_dagrider_order_fairness_exp}.

\subsection{Experiments on the delay distribution\label{ssec:exp1}}

\begin{figure}[h]
    \centering

\setlength\tabcolsep{1.5pt}
\begin{tabular}{|c|c|c|}
\hline
{\scriptsize score}
&
{\scriptsize$OF_{S_{ND}}^{F_{IN}}$ violations}
&
{\scriptsize Legend}
\\
\hline 
\includegraphics[scale=.25]{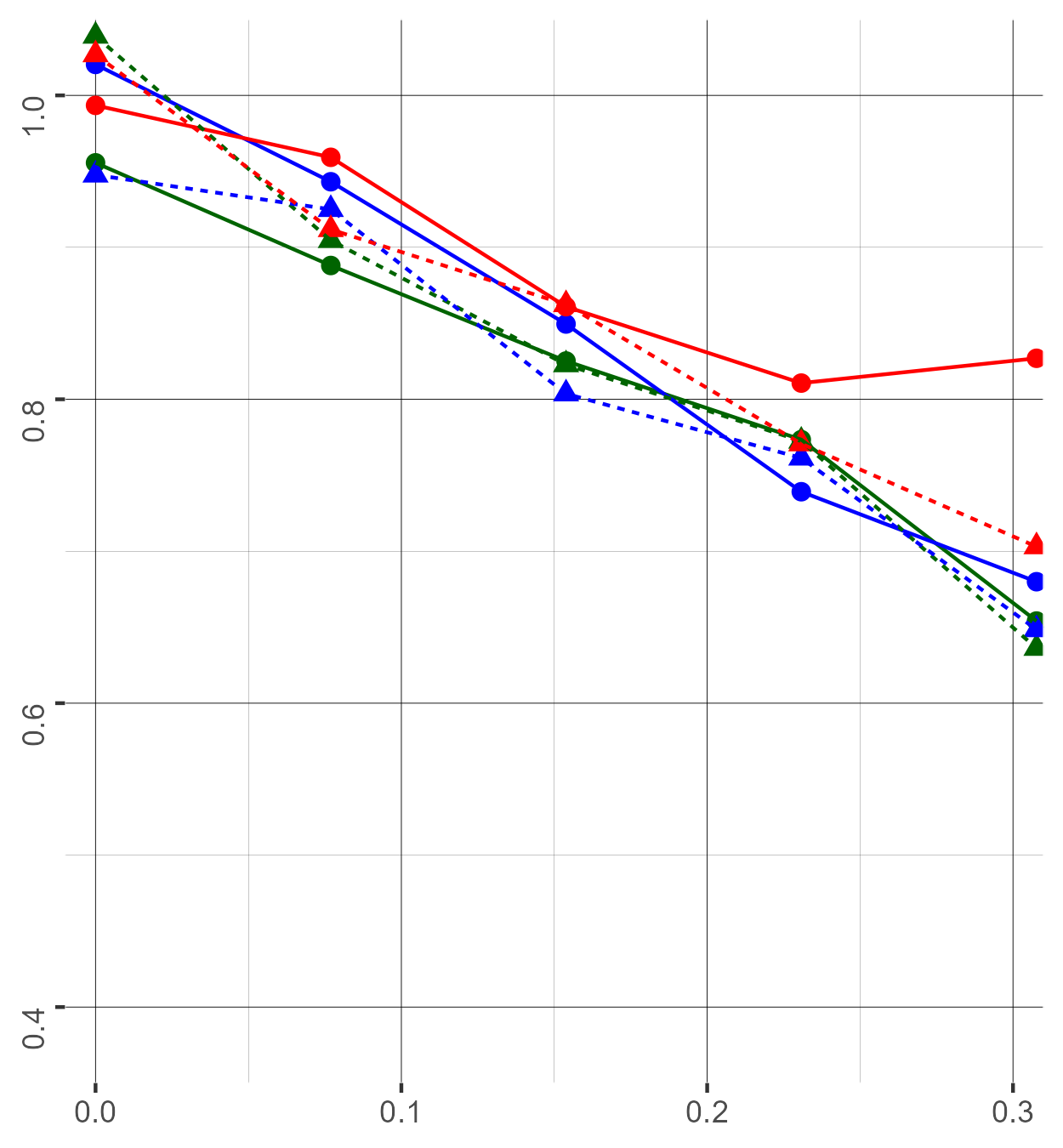}
&
\includegraphics[scale=.25]{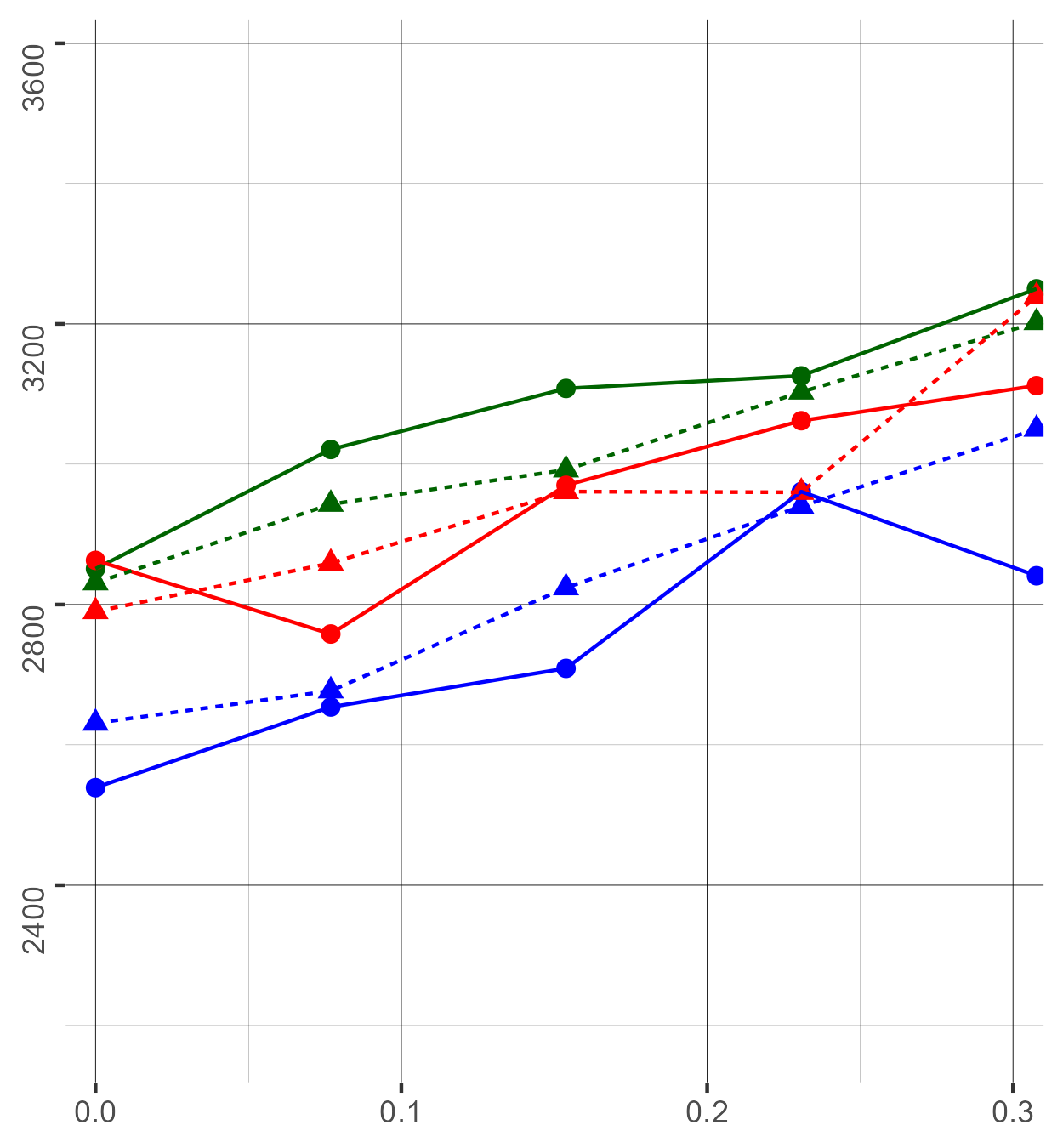}
&
\scalebox{.7}{
\begin{tikzpicture}
\node[align=center] (leg1) at (0,0) {\textcolor{darkspringgreen}{$\blacksquare$} {\footnotesize FullShuffle}}; 
\node[below=.25cm of leg1.south west,anchor=west,align=center] (leg2) {\textcolor{blue}{$\blacksquare$} {\footnotesize PerColumnShuffle}};
\node[below=.25cm of leg2.south west,anchor=west,align=center] (leg3) {\textcolor{red}{$\blacksquare$} {\footnotesize VoteCount}};
\node[below right=.3cm and -.9cm of leg3] (leg4) {\footnotesize Smaller Delays};
\node[draw=black,fill=black,left=.2cm of leg4,circle,inner sep=2pt] (leg4s) {};
\draw[thick] ($(leg4s) + (-.3,0) $) -- ($ (leg4s) + (.3,0) $);
\node[below=.1cm of leg4.south east,anchor=east,align=right] (leg4b) {\scriptsize (quick network)};
\node[below=.75cm of leg4.south west,anchor=west,align=center] (leg5) {\footnotesize Larger Delays};
\node[left=.2cm of leg5,inner sep=0pt] (leg5s) {$\blacktriangle$};
\draw[dotted,thick] ($(leg5s) + (-.3,0) $) -- ($ (leg5s) + (.3,0) $);
\node[below=.1cm of leg5.south east,anchor=east,align=right] (leg5b) {\scriptsize (slow network)};
\end{tikzpicture}
}
\\
\hline 
\hline
{\scriptsize$OF_{S_{ND}}^{W_{AV}}$ violations}
&
{\scriptsize$OF_{R_{EC}}^{W_{AV}}$ violations}
&
{\scriptsize$OF_{D_{LV}}^{W_{AV}}$ violations}
\\
\hline 
\includegraphics[scale=.25]{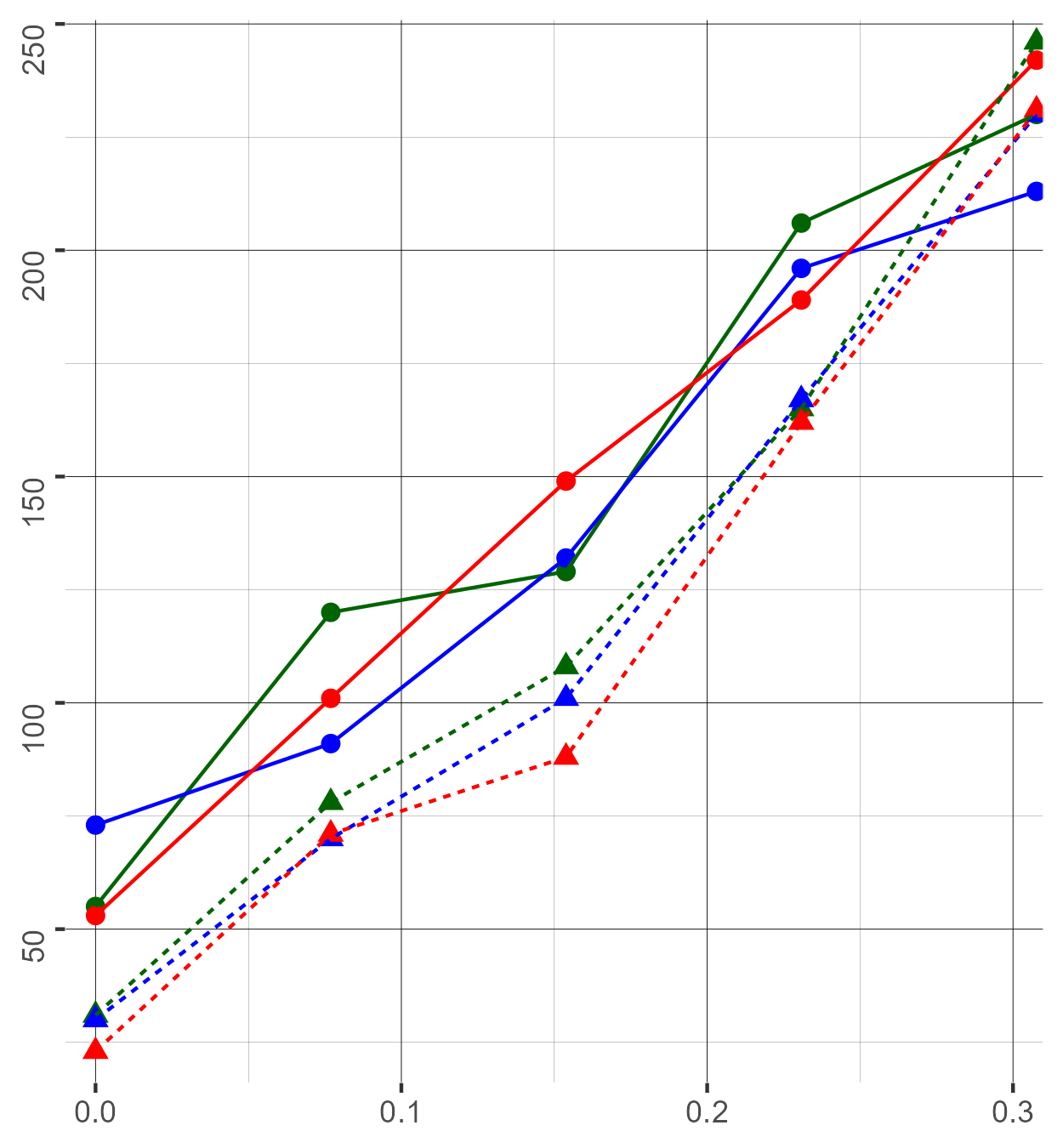}
&
\includegraphics[scale=.25]{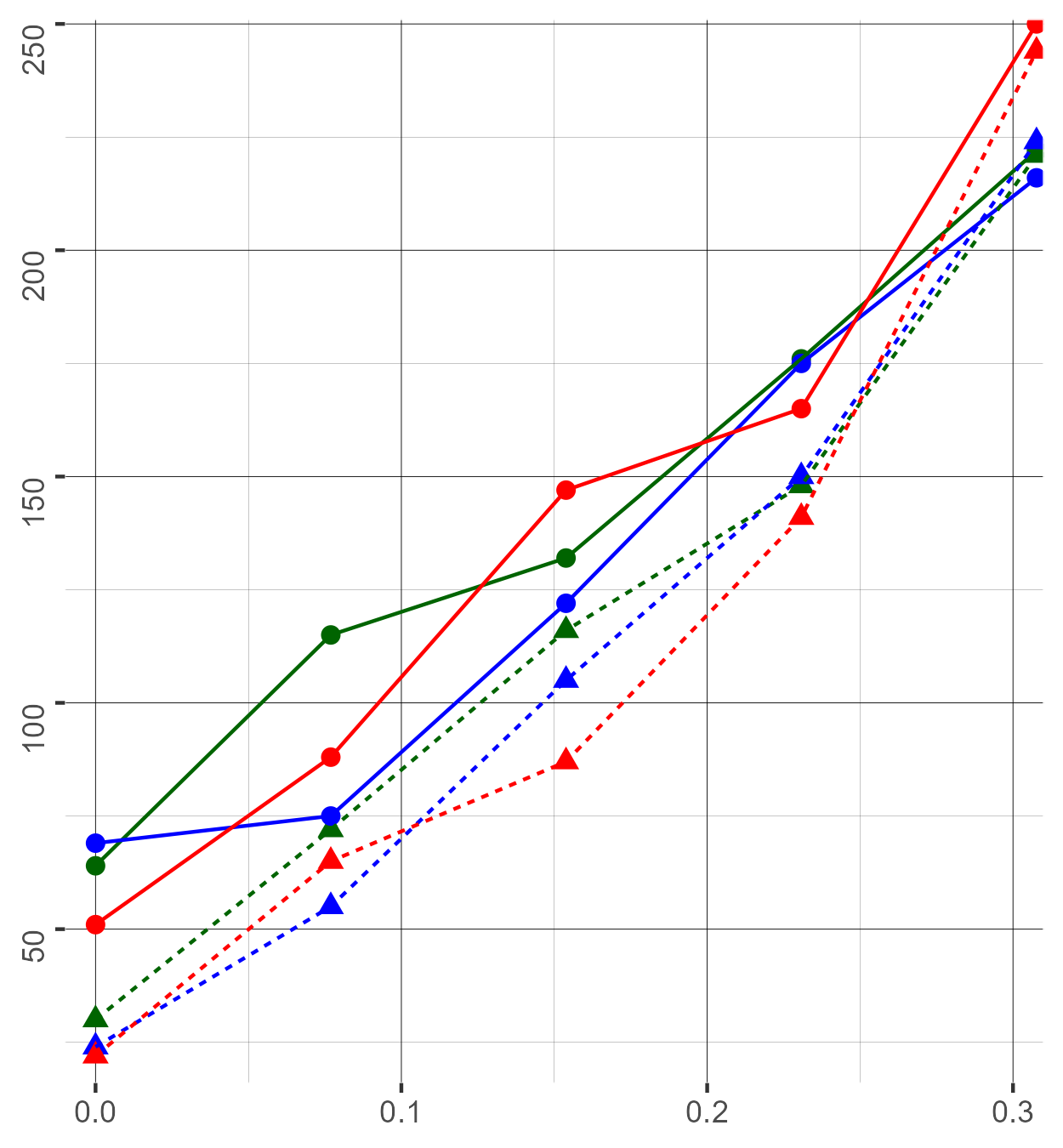}
&
\includegraphics[scale=.25]{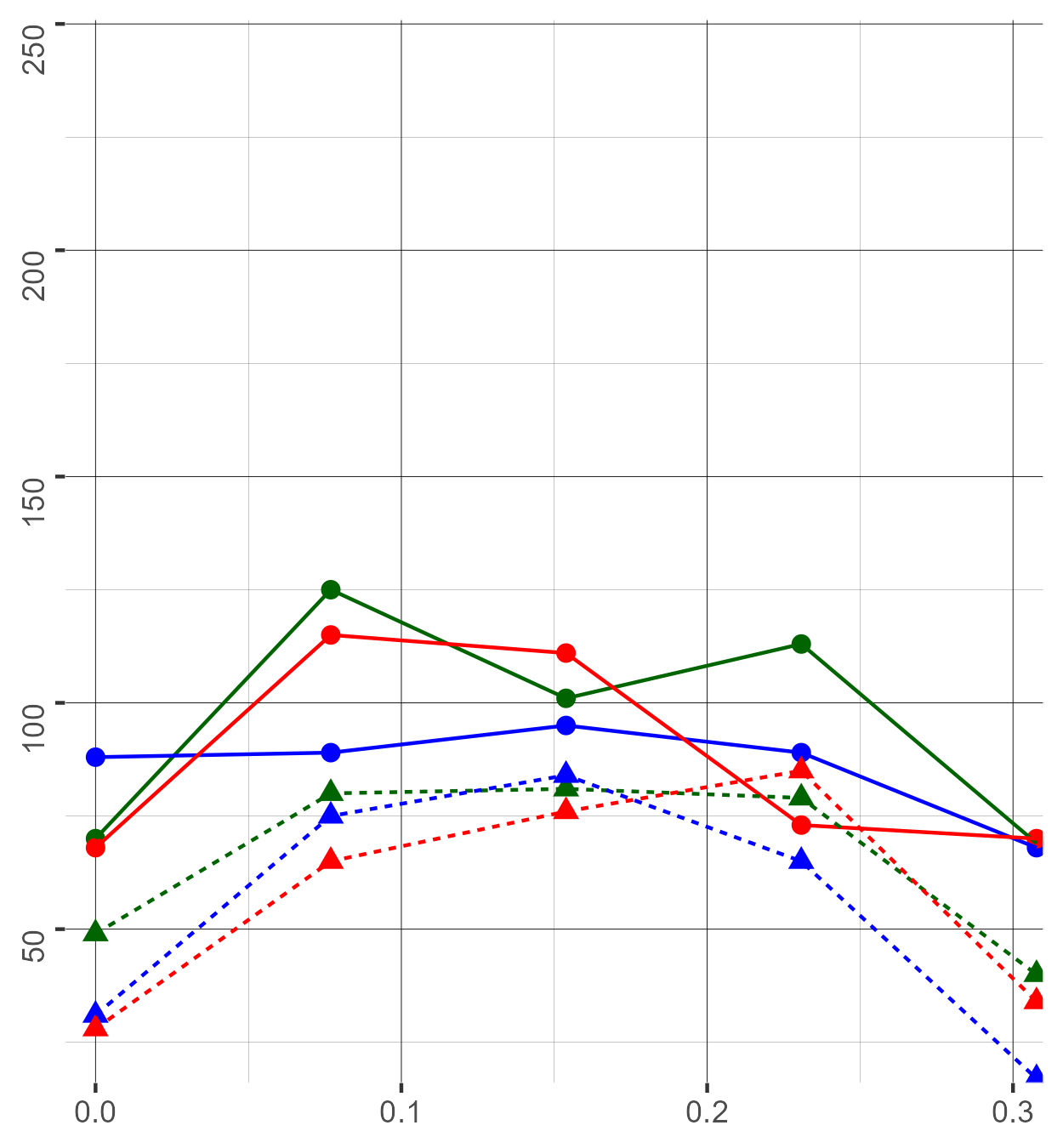}
\\
\hline 
\end{tabular}
\setlength\tabcolsep{6pt}
    
    \caption{Experiments w.r.t.~the deterministic order and \faClockO}
    \label{fig:exp1}
\end{figure}
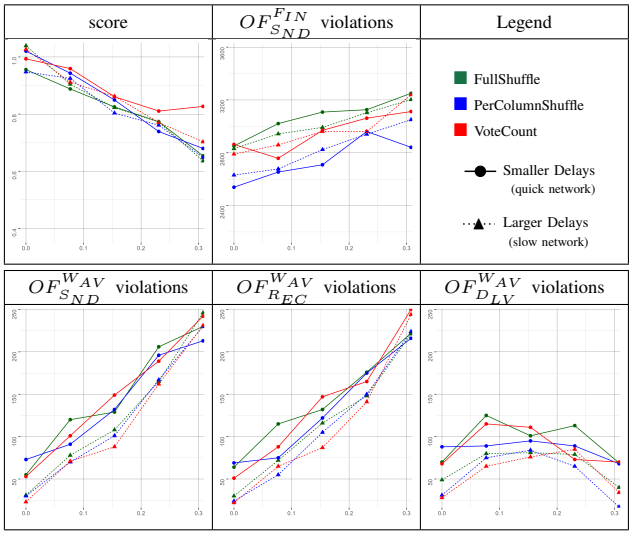

This first set of experiments investigates the effect of parameters \textbf{(1)} (the deterministic order) and \textbf{(2)} (the \faClockO~delay distribution).
Parameter \textbf{(3)} is fixed : clients send their transactions to all the nodes.
Fig.\ref{fig:exp1} summarizes experimental results.
On each diagram, the proportion $b/n$ of Byzantine nodes (with $b \in [0,f]$) corresponds to the $x$-axis (it represents the power of the adversary). 
The colors and styles of the different curves resp.~correspond parameters \textbf{(1)} and \textbf{(2)}.
The first diagram, on the top left of Fig.\ref{fig:exp1}, gives the $\mathtt{score}$ of the client targeted by the attack on the $y$ axis.
As for the other four diagrams, they plot the number of pairs of transactions for which the corresponding property is violated.

In all cases, the adversary succeeds in reducing $\mathtt{score}(\chi)$.
Still, we remark that VoteCount has a protective effect.
However, and this is especially true for VoteCount, higher delays in the network may cause a higher vulnerability of the system.
Indeed, this gives the adversary more leeway to manipulate the order of delivery of the vertices, which has a stronger effect on VoteCount because it determines the in-wave order based on the columns at which nodes include delivered vertices in their latest vertices' causal sub-graph.

The decrease in the score is correlated to an increase in the number of $OF_{S_{ND}}^{F_{IN}}$ violations.
As expected, the FullShuffle deterministic order is more susceptible to violate $OF_{S_{ND}}^{F_{IN}}$ because the in-wave order is entirely random.
However, we observe that VoteCount yields more violations than PerColumnShuffle.
This can be explained by the fact that VoteCount determines the order based on the order of delivery of the vertices and not on the order of reception of individual transactions.
On the other hand, PerColumnShuffle always order a vertex at column $c+1$ after a vertex at column $c$.
Then, because honest nodes always include transactions as early as possible, the number of violations only depends on the client to node delay distribution and the randomness introduced by PerColumnShuffle when shuffling vertices at the same column.
On the other hand, because VoteCount takes into account the delivery order, the node to node \faClockO delay distribution also introduces noise, which explains the higher number of violations.

When considering the four diagrams on the left, what we observe is the combined effect of the DagRider layer and Bracha layer attacks (Sec.\ref{ssec:byz_dagrider_layer} and Sec.\ref{ssec:attack_bracha}).
Overall, we observe that the increase in wave-order-fairness violations ($OF_{S_{ND}}^{W_{AV}}$ and $OF_{R_{EC}}^{W_{AV}}$) is sharper than that for finalization-order-fairness ($OF_{S_{ND}}^{F_{IN}}$).
Indeed, the attack may exclude vertices from certain waves (especially whenever the wave leader is Byzantine).

Because the Bracha-layer attack's effect is to manipulate the order of delivery, and because $OF_{D_{LV}}^{W_{AV}}$ only relates this vertex delivery order to the finalization order, this means that the initial increase in $OF_{D_{LV}}^{W_{AV}}$ violations on the bottom right diagram is only due to the DagRider layer attack. Then, the decrease at higher adversarial power can be explained by the fact that the Bracha layer attack already successfully reorders the relevant pairs of transactions (and thus, new violations are less likely to be added at the DagRider layer).

See Appendix \ref{anx:exp1} for further details.

\subsection{Experiments on the fanout\label{ssec:exp2}}

\definecolor{fanout1}{RGB}{165,42,42}
\definecolor{fanoutfp1}{RGB}{255,165,0}
\definecolor{fanout2fp1}{RGB}{160,32,240}
\definecolor{fanout3fp1}{RGB}{255,192,203}

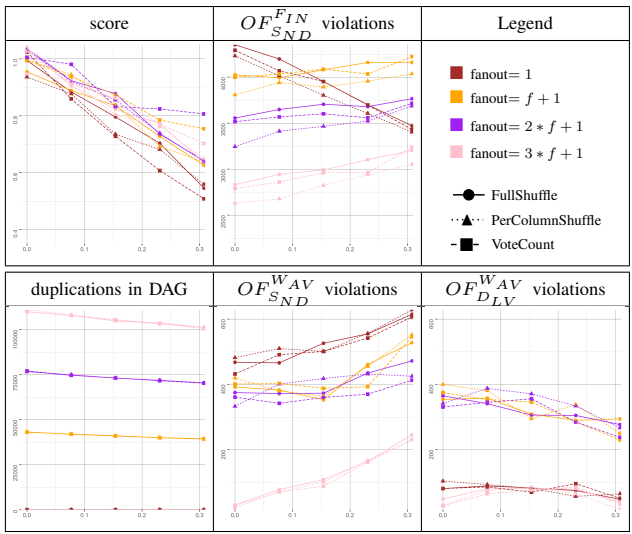
\begin{figure}[h]
    \centering

\setlength\tabcolsep{1.5pt}
\begin{tabular}{|c|c|c|}
\hline
{\scriptsize score}
&
{\scriptsize$OF_{S_{ND}}^{F_{IN}}$ violations}
&
{\scriptsize Legend}
\\
\hline 
\includegraphics[scale=.25]{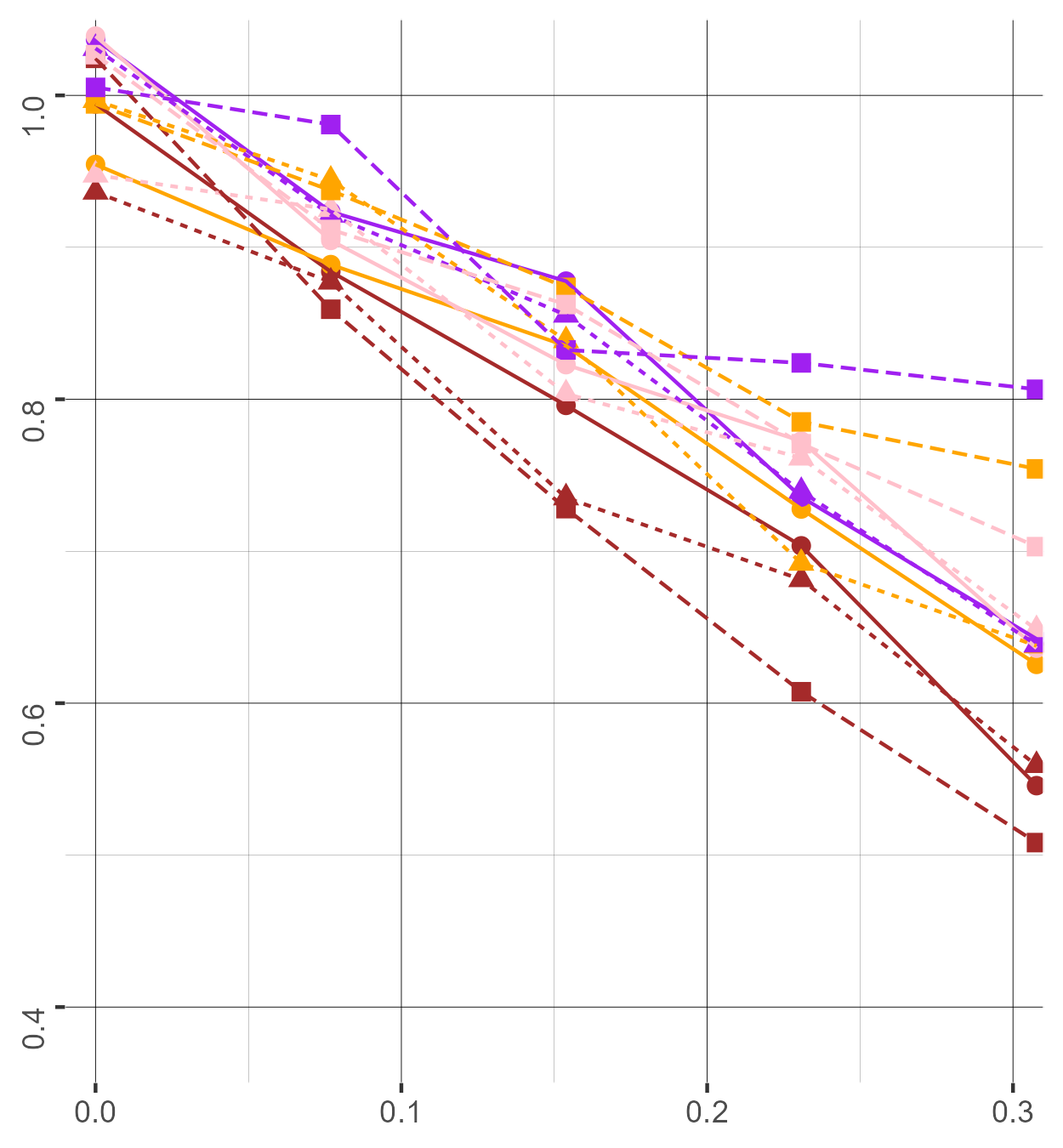}
&
\includegraphics[scale=.25]{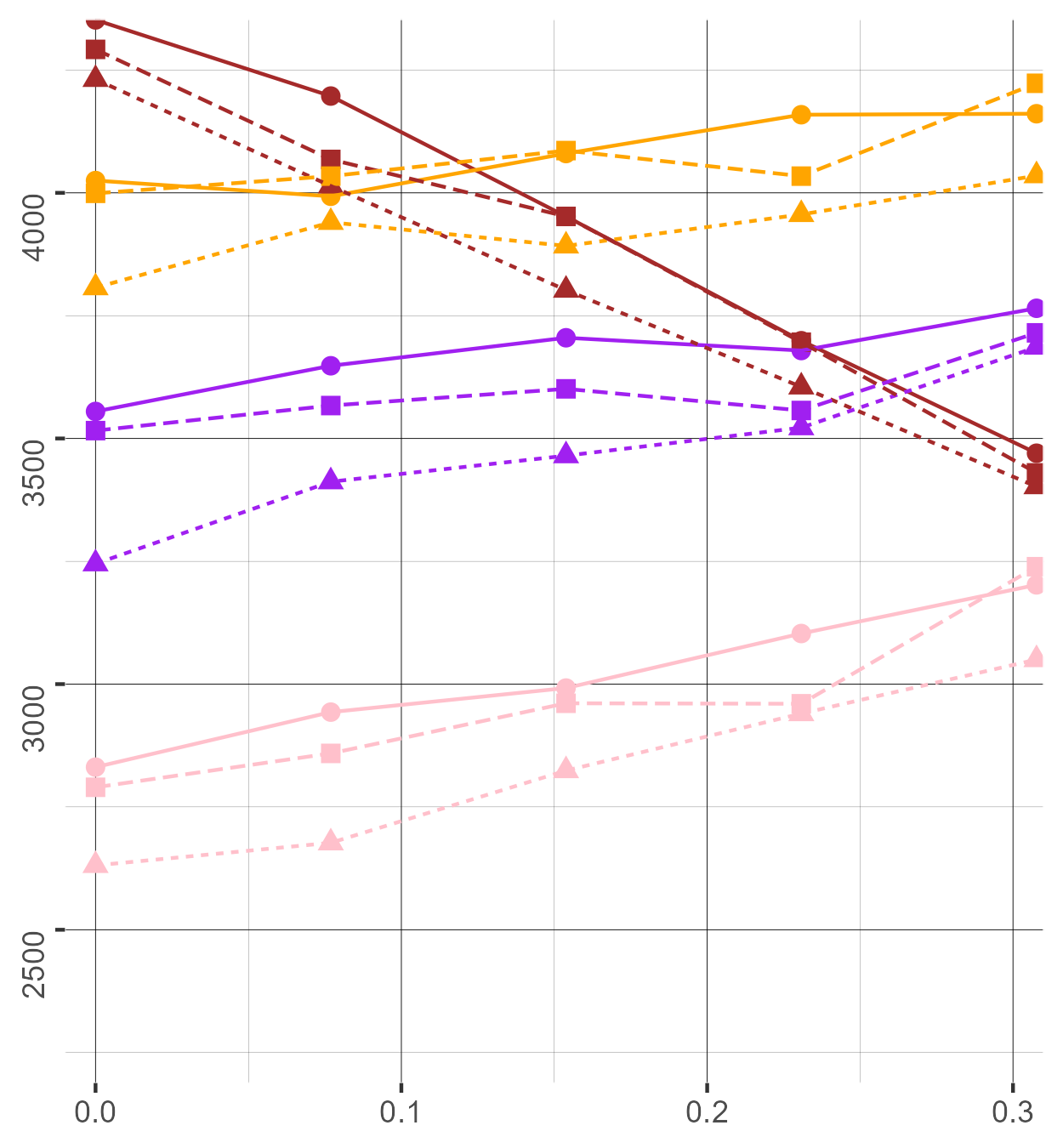}
&
\scalebox{.7}{
\begin{tikzpicture}
\node[align=center] (leg1) at (0,0) {\textcolor{fanout1}{$\blacksquare$} {\footnotesize fanout$=1$}}; 
\node[below=.25cm of leg1.south west,anchor=west,align=center] (leg2) {\textcolor{fanoutfp1}{$\blacksquare$} {\footnotesize fanout$=f+1$}};
\node[below=.25cm of leg2.south west,anchor=west,align=center] (leg3) {\textcolor{fanout2fp1}{$\blacksquare$} {\footnotesize fanout$=2*f+1$}};
\node[below=.25cm of leg3.south west,anchor=west,align=center] (leg4) {\textcolor{fanout3fp1}{$\blacksquare$} {\footnotesize fanout$=3*f+1$}};
\node[below right=.3cm and -2cm of leg4] (leg5) {\footnotesize FullShuffle};
\node[draw=black,fill=black,left=.2cm of leg5,circle,inner sep=2pt] (leg5s) {};
\draw[thick] ($(leg5s) + (-.3,0) $) -- ($ (leg5s) + (.3,0) $);
\node[below=.25cm of leg5.south west,anchor=west,align=center] (leg6) {\footnotesize PerColumnShuffle};
\node[left=.2cm of leg6,inner sep=0pt] (leg6s) {$\blacktriangle$};
\draw[thick,dotted] ($(leg6s) + (-.3,0) $) -- ($ (leg6s) + (.3,0) $);
\node[below=.25cm of leg6.south west,anchor=west,align=center] (leg7) {\footnotesize VoteCount};
\node[left=.2cm of leg7,inner sep=0pt] (leg7s) {$\blacksquare$};
\draw[thick,dashed] ($(leg7s) + (-.3,0) $) -- ($ (leg7s) + (.3,0) $);
\end{tikzpicture}
}
\\
\hline 
\hline
{\scriptsize duplications in DAG}
&
{\scriptsize$OF_{S_{ND}}^{W_{AV}}$ violations}
&
{\scriptsize$OF_{D_{LV}}^{W_{AV}}$ violations}
\\
\hline 
\includegraphics[scale=.25]{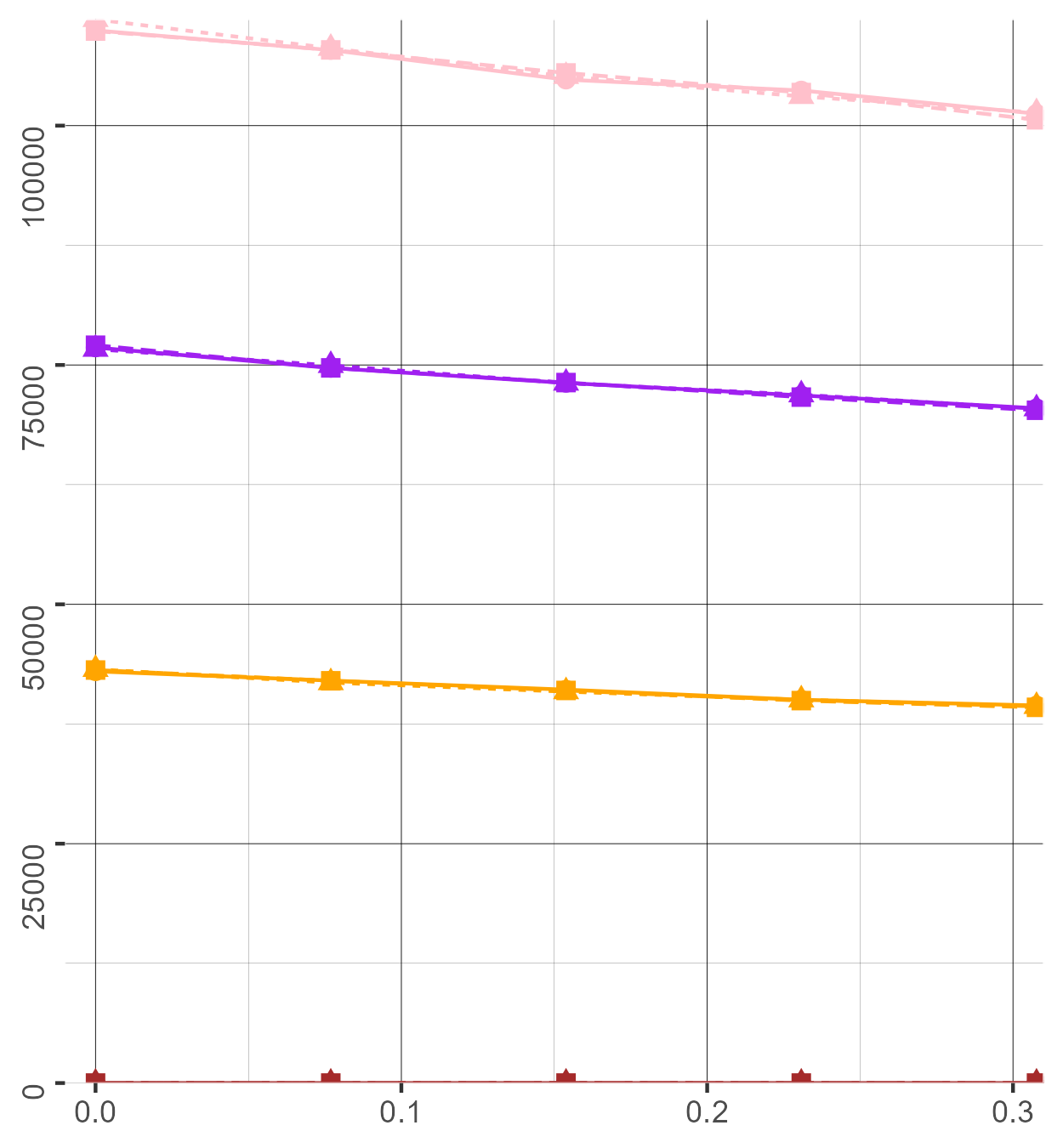}
&
\includegraphics[scale=.25]{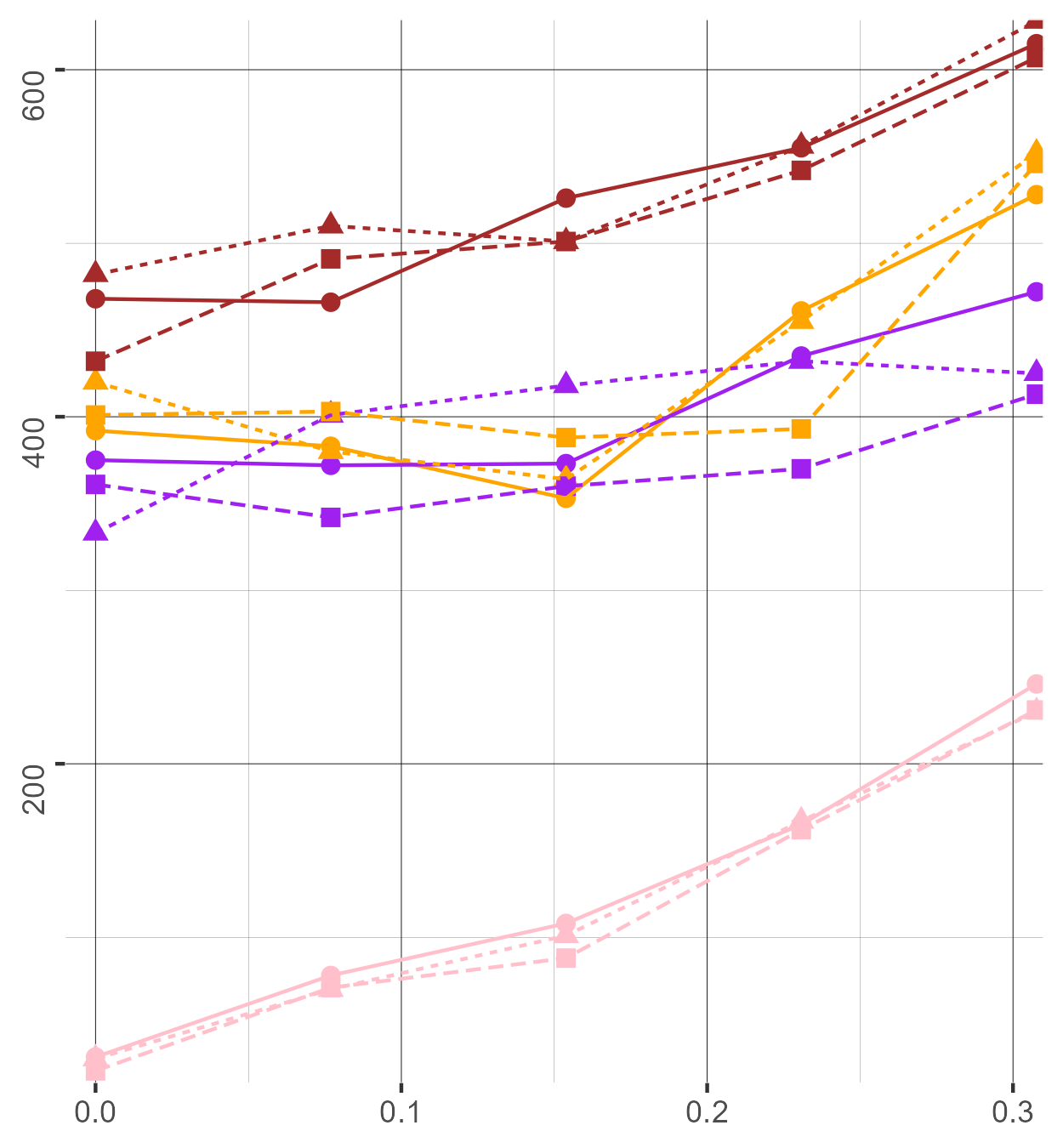}
&
\includegraphics[scale=.25]{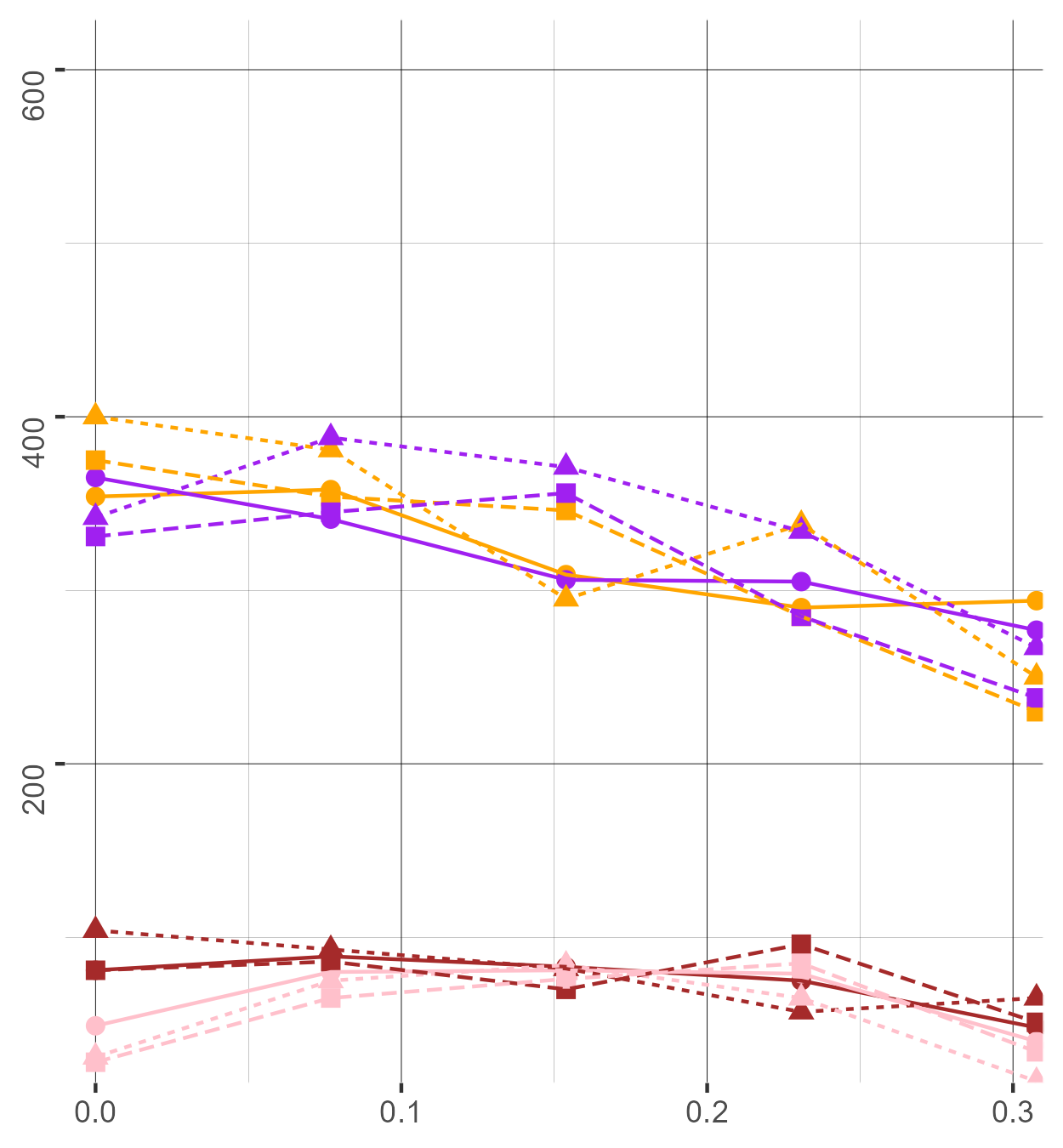}
\\
\hline 
\end{tabular}
\setlength\tabcolsep{6pt}
    
    \caption{Experiments w.r.t.~the fanout}
    \label{fig:exp2}
\end{figure}

The second experiments investigate the effect of parameter \textbf{(3)} i.e., the fanout from clients to nodes.
Parameter \textbf{(2)} is fixed as the slow network.
We consider $4$ possible values for parameter \textbf{(3)}: $1$, $f+1$, $2*f+1$ or $3*f+1$, the latter coinciding with the first set of experiments (clients broadcasting their transactions to all the nodes).
Fig.\ref{fig:exp2} summarizes experimental results.

The simulations highlight a tradeoff between robustness to transaction reordering and a higher rate of duplications in the DAG (i.e., the same transaction is found in multiple vertices).
Indeed, on the bottom left diagram, we can see, that the number of duplications increases with the fanout.
At higher fanouts, there is a slight decrease along the $x$ axis because Byzantine nodes do not include certain transactions.
On the top left diagram, we can see that, the lower the fanout is, the more vulnerable the system is (i.e., the decreases in the score becomes higher as the adversarial power increases).
For $OF_{S_{ND}}^{F_{IN}}$, we can see that when $b=0$, the lower the fanout, the higher the number of violations is.
However, while for fanout higher than $f$, this number increases with $b$, when it is $1$, it decreases. 
This is explained by the fact that in the former cases, all transactions are guaranteed to be included in the DAG (any transaction being send to at least one honest node) while it is not the case for the latter.
The transactions that disappear in the case of fanout $1$ are not taken into account when counting the pairs of transactions which violate $OF_{S_{ND}}^{F_{IN}}$, which explains the decrease.

\subsection{Varying the number of nodes\label{ssec:exp3}}

\definecolor{nodesnum10}{RGB}{248,118,109}
\definecolor{nodesnum13}{RGB}{124,174,0}
\definecolor{nodesnum16}{RGB}{0,191,196}
\definecolor{nodesnum19}{RGB}{199,124,255}

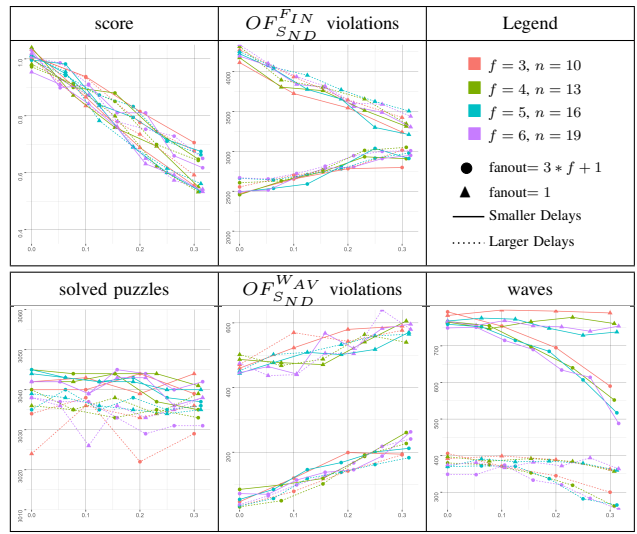
\begin{figure}[h]
    \centering

\setlength\tabcolsep{1.5pt}
\begin{tabular}{|c|c|c|}
\hline
{\scriptsize score}
&
{\scriptsize$OF_{S_{ND}}^{F_{IN}}$ violations}
&
{\scriptsize Legend}
\\
\hline 
\includegraphics[scale=.25]{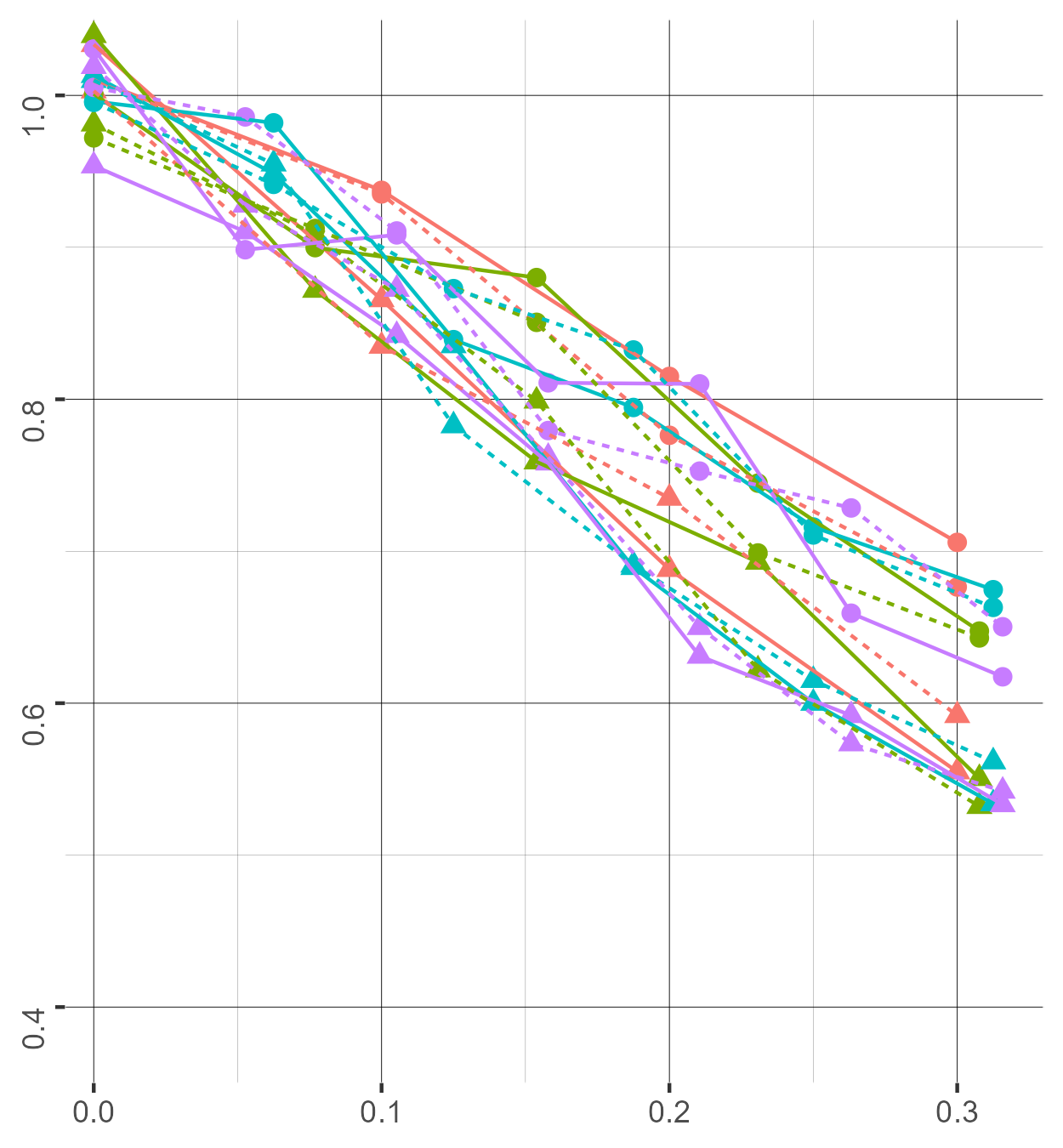}
&
\includegraphics[scale=.25]{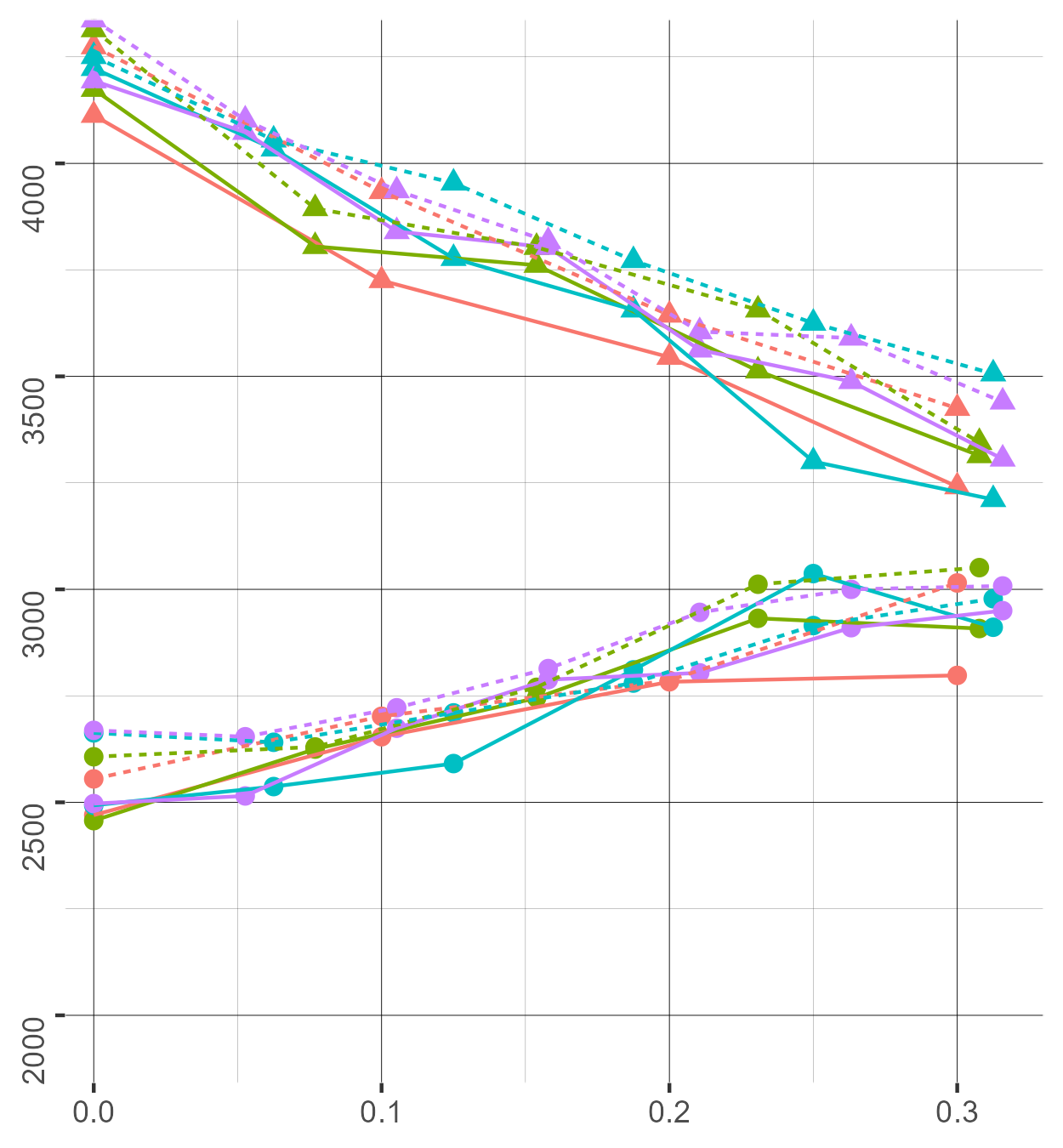}
&
\scalebox{.7}{
\begin{tikzpicture}
\node[align=center] (leg1) at (0,0) {\textcolor{nodesnum10}{$\blacksquare$} {\footnotesize $f=3$, $n=10$}};
\node[below=.175cm of leg1.south west,anchor=west,align=center] (leg2) {\textcolor{nodesnum13}{$\blacksquare$} {\footnotesize $f=4$, $n=13$}};
\node[below=.175cm of leg2.south west,anchor=west,align=center] (leg3) {\textcolor{nodesnum16}{$\blacksquare$} {\footnotesize $f=5$, $n=16$}};
\node[below=.175cm of leg3.south west,anchor=west,align=center] (leg4) {\textcolor{nodesnum19}{$\blacksquare$} {\footnotesize $f=6$, $n=19$}};
\node[below right=.1cm and -2cm of leg4] (leg5) {\footnotesize fanout$=3*f+1$};
\node[draw=black,fill=black,left=.2cm of leg5,circle,inner sep=2pt] (leg5s) {};
\node[below=.2cm of leg5.south west,anchor=west,align=center] (leg6) {\footnotesize fanout$=1$};
\node[left=.2cm of leg6,inner sep=0pt] (leg6s) {$\blacktriangle$};
\node[below=.25cm of leg6.south west,anchor=west,align=center] (leg7) {\footnotesize Smaller Delays};
\node[left=.2cm of leg7,inner sep=0pt] (leg7s) {$~$};
\draw[thick] ($(leg7s) + (-.3,0) $) -- ($ (leg7s) + (.3,0) $);
\node[below=.25cm of leg7.south west,anchor=west,align=center] (leg8) {\footnotesize Larger Delays};
\node[left=.2cm of leg8,inner sep=0pt] (leg8s) {$~$};
\draw[thick,dotted] ($(leg8s) + (-.3,0) $) -- ($ (leg8s) + (.3,0) $);
\end{tikzpicture}
}
\\
\hline 
\hline
{\scriptsize solved puzzles}
&
{\scriptsize $OF_{S_{ND}}^{W_{AV}}$ violations}
&
{\scriptsize waves}
\\
\hline 
\includegraphics[scale=.25]{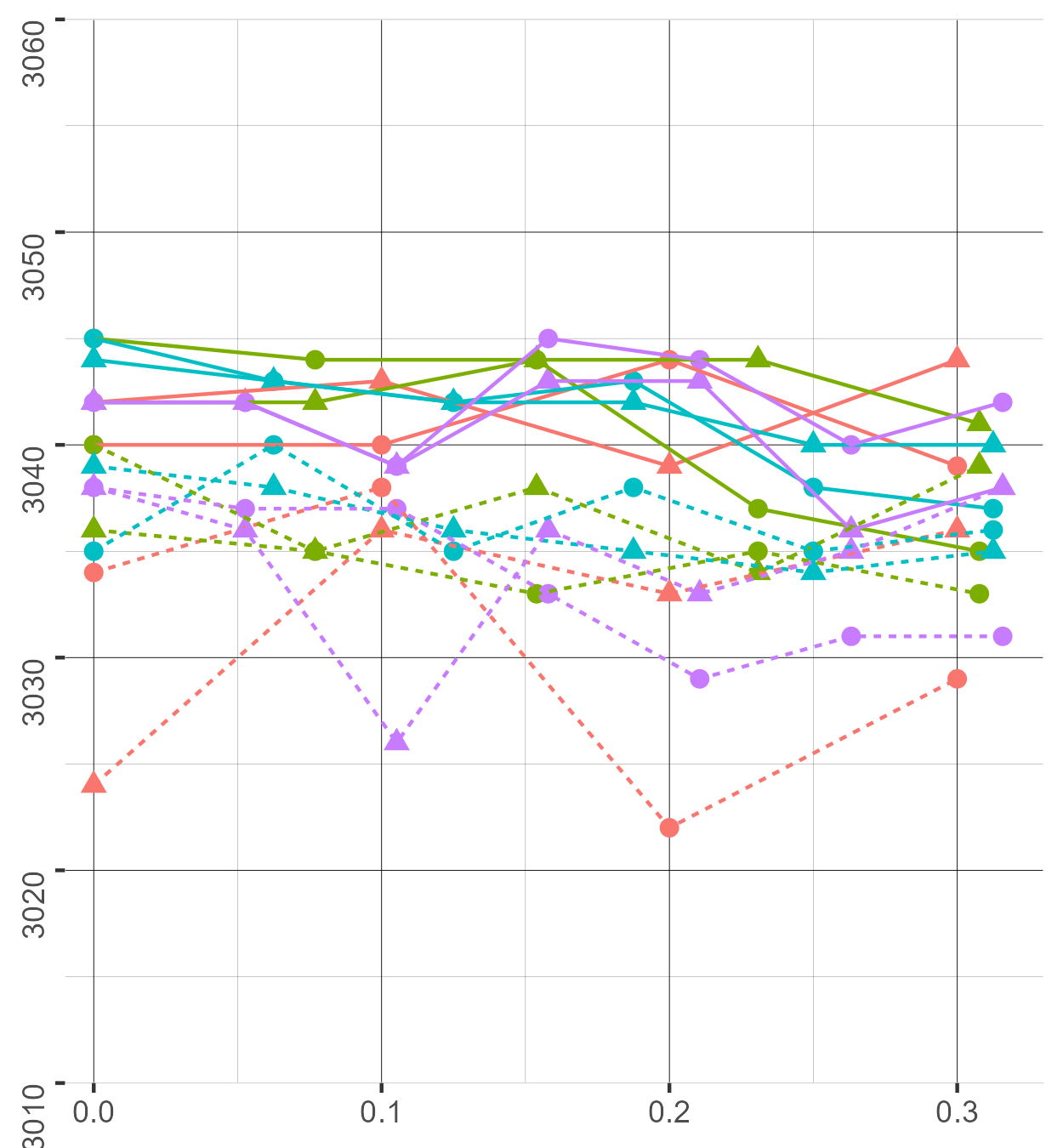}
&
\includegraphics[scale=.25]{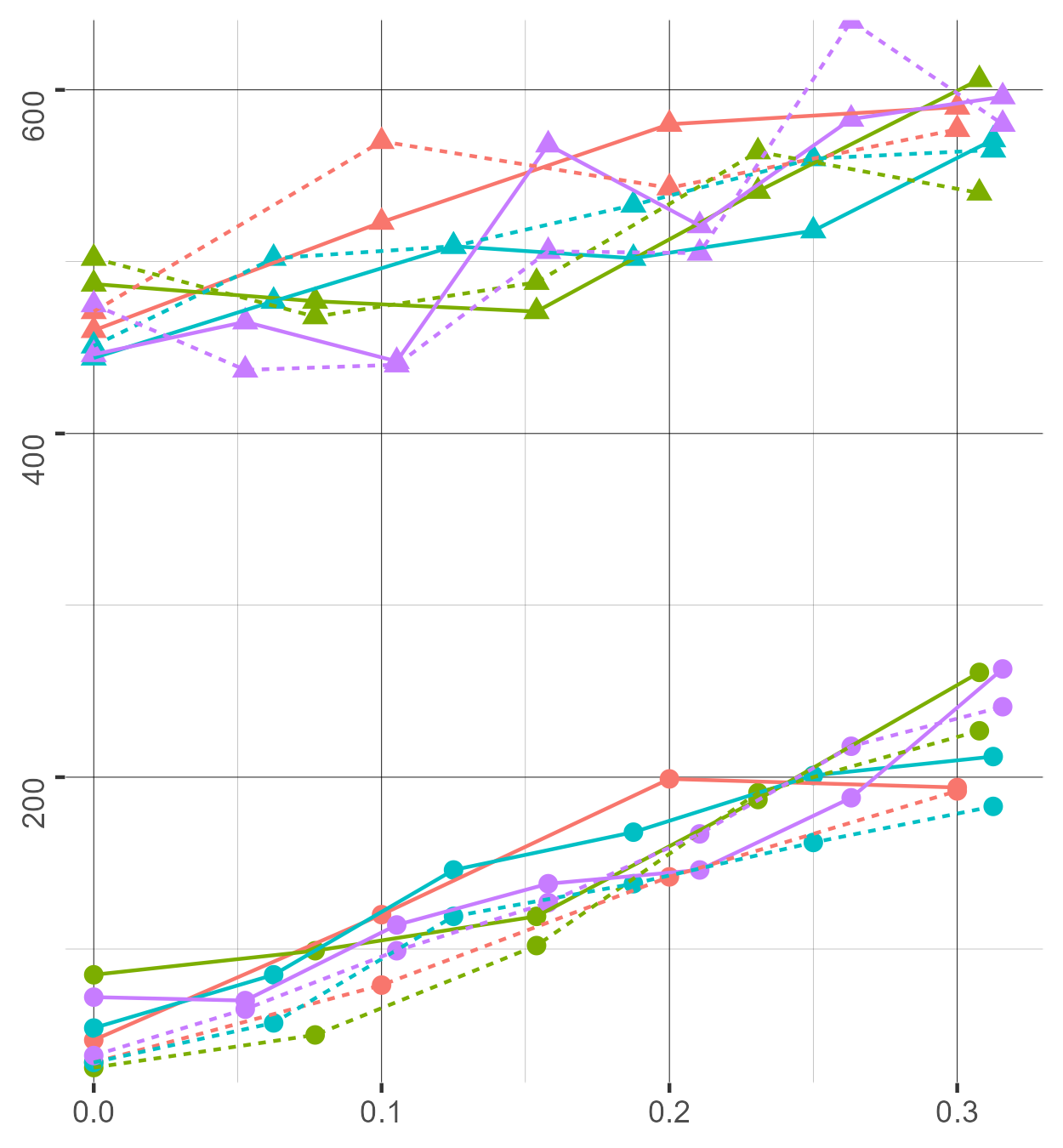}
&
\includegraphics[scale=.25]{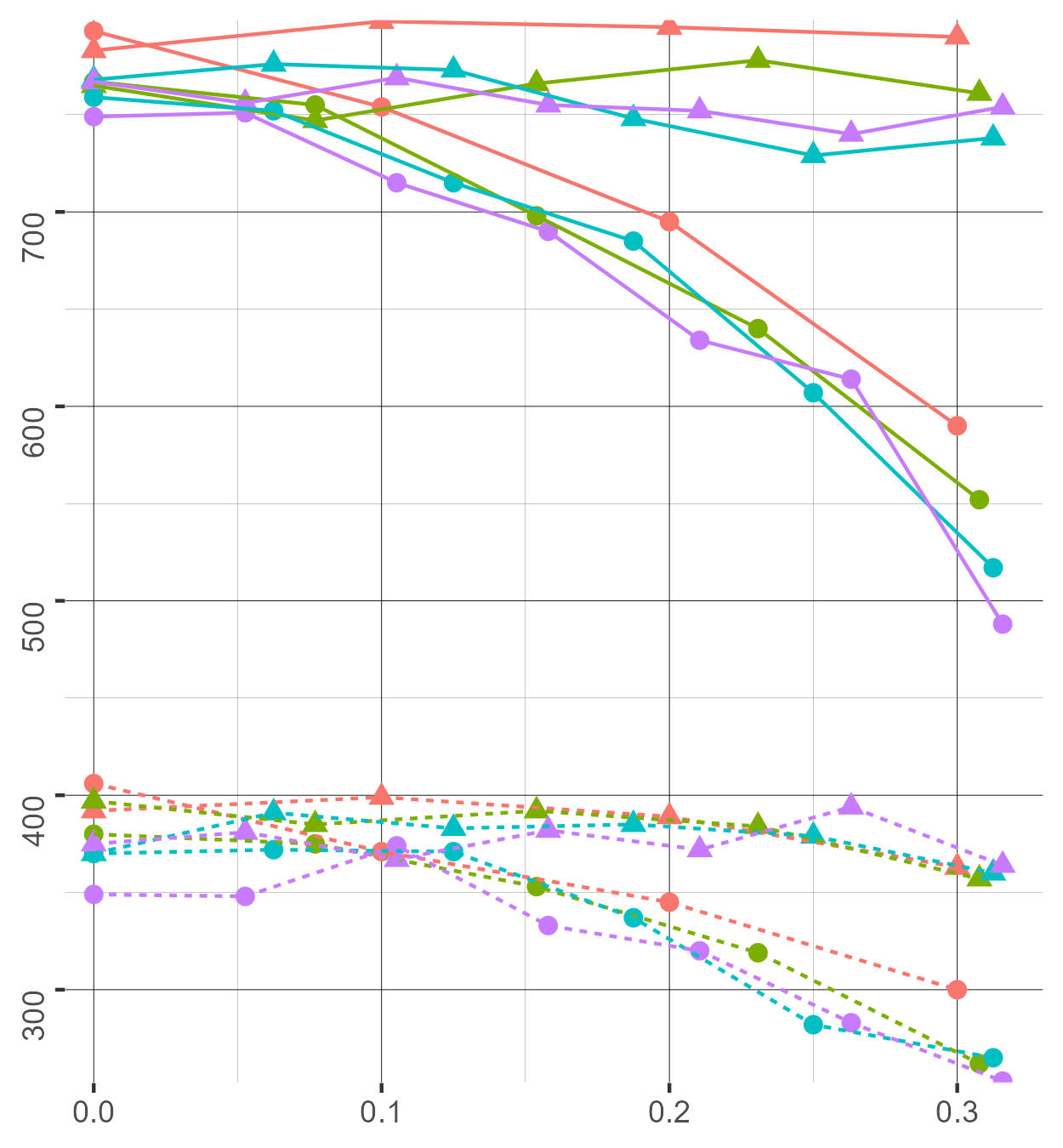}
\\
\hline 
\end{tabular}
\setlength\tabcolsep{6pt}
    
    \caption{Varying the number of nodes.}
    \label{fig:exp3}
\end{figure}

In Sec.\ref{ssec:exp1} and Sec.\ref{ssec:exp2}, we had fixed the number of nodes $n$ to $13 = 1 + 3*4$ and varied the number of Byzantine nodes $b$ between $0$ and $f=4$ (the proportion of Byzantine nodes thus varying from $0$ to $33\%$).
This arbitrary $n=13$ is a tradeoff w.r.t.~the computational costs of performing the simulations.
Indeed, in order to have statistically significant results, we need long simulations (in each, the system solves at least 3000 puzzles so that the score values converge, as per the law of large numbers). 
With $n=13$, at 2 puzzles per wave, we must simulate around $7$ million message exchanges on the Bracha layer.

In this third set of experiments, we show that our results for $n=13$ can be generalized to any value of $n$. To do so, we show that varying $n$ do not influence the impact of the other parameters on the robustness of the system w.r.t.~the proportion of Byzantine nodes $b/n$.

Fig.\ref{fig:exp3} summarizes our results.
Here, we consider the ``PerColumnShuffle'' deterministic order.
On the legend, in the top right, the colors correspond to different values of $n$ that we consider (10, 13, 16 and 19).
While the shape of the points (circles or triangles) correspond to the fanout value parameter,
the style of the lines (continuous or dotted) correspond to the \faClockO~delay distribution.
On the top of Fig.\ref{fig:exp3}, we can see that varying $n$ does not impact the effect of increasing the proportion $b/n$ of Byzantine nodes (on the $x$ axis) on the number of $OF_{S_{ND}}^{F_{IN}}$ violations (on the $y$ axis).
Indeed, for all values of $n$, at a fixed value of fanout, the number of violations is lower in the case of smaller delays and higher in the case of larger delays.
Reciprocally, at a fixed value of the delay distribution, the number of violations is lower at fanout $3*f+1$ and increases with $b/n$ while it is higher at fanout $1$ and decreases with $b/n$.
Similar observations can be made on all the other metrics.

\section{Conclusion\label{sec:conc}}

Our contribution concerns a theoretical and empirical security evaluation of DAG-based ledgers' robustness to reordering attacks (i.e., in which the adversary causes transactions to be delivered in a specific order) and therefore of order fairness.
We formalize novel order fairness properties that specifically describe DAG-based ledgers.
We demonstrate that, via targeting both the DAG construction logic and the underlying Byzantine Reliable Broadcast protocol, even a weak adversary can have a significant impact on the order with which transactions are delivered
(the adversary being weak as it does not control the network but only coordinates a minority of Byzantine nodes, below the fault tolerance thresholds of the involved algorithms).
Moreover, we highlight the tradeoff between transaction duplication and the robustness to such attacks.

Our study underlines the importance of considering order fairness as properties of interest when defining algorithms that implement distributed ledgers. 
It especially suggests that DAG-based ledgers are also vulnerable (in the same way as classical Blockchains) to transaction reordering attacks.

\bibliographystyle{IEEEtran}
\bibliography{
biblio/adversary,
biblio/blockchain,
biblio/conflictless,
biblio/dag,
biblio/generic_distributed_computing_literature,
biblio/mev,
biblio/multi_agent,
biblio/order_fairness,
biblio/other_fairness,
biblio/others,
biblio/own,
biblio/reliable_broadcast,
biblio/selfish_quasi_honest,
biblio/trust
}

\appendices

\section{Problem with weak edges in DagRider\label{anx:bug_dagrider}}

Let us remark that, thanks to our analysis and extensive simulation of DagRider, we have found an error in the definition of the algorithm in \cite{all_you_need_is_dag}.

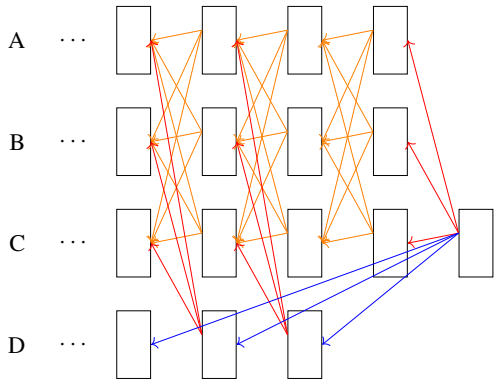
\begin{figure}[h]
    \centering
    \scalebox{.9}{\begin{tikzpicture}
\node (rowA) at (0,0) {$\cdots$};
\node[draw=black,minimum height=1cm, minimum width=.5cm,inner sep=0cm,right=.25cm of rowA] (v_0_A) {
};
\node[draw=black,minimum height=1cm, minimum width=.5cm,right=.75cm of v_0_A] (v_1_A) {
};
\node[draw=black,minimum height=1cm, minimum width=.5cm,right=.75cm of v_1_A] (v_2_A) {
};
\node[draw=black,minimum height=1cm, minimum width=.5cm,right=.75cm of v_2_A] (v_3_A) {
};
%
%
\node (rowB) at (0,-1.5) {$\cdots$};
\node[draw=black,minimum height=1cm, minimum width=.5cm,inner sep=0cm,right=.25cm of rowB] (v_0_B) {
};
\node[draw=black,minimum height=1cm, minimum width=.5cm,right=.75cm of v_0_B] (v_1_B) {
};
\node[draw=black,minimum height=1cm, minimum width=.5cm,right=.75cm of v_1_B] (v_2_B) {
};
\node[draw=black,minimum height=1cm, minimum width=.5cm,right=.75cm of v_2_B] (v_3_B) {
};
%
%
\node (rowC) at (0,-3) {$\cdots$};
\node[draw=black,minimum height=1cm, minimum width=.5cm,inner sep=0cm,right=.25cm of rowC] (v_0_C) {
};
\node[draw=black,minimum height=1cm, minimum width=.5cm,right=.75cm of v_0_C] (v_1_C) {
};
\node[draw=black,minimum height=1cm, minimum width=.5cm,right=.75cm of v_1_C] (v_2_C) {
};
\node[draw=black,minimum height=1cm, minimum width=.5cm,right=.75cm of v_2_C] (v_3_C) {
};
%
%
\node (rowD) at (0,-4.5) {$\cdots$};
\node[draw=black,minimum height=1cm, minimum width=.5cm,inner sep=0cm,right=.25cm of rowD] (v_0_D) {
};
\node[draw=black,minimum height=1cm, minimum width=.5cm,right=.75cm of v_0_D] (v_1_D) {
};
\node[draw=black,minimum height=1cm, minimum width=.5cm,right=.75cm of v_1_D] (v_2_D) {
};
%
%
\draw[->,orange] (v_1_A.150) -- (v_0_A.east);
\draw[->,orange] (v_1_A.150) -- (v_0_B.east);
\draw[->,orange] (v_1_A.150) -- (v_0_C.east);
\draw[->,orange] (v_2_A.150) -- (v_1_A.east);
\draw[->,orange] (v_2_A.150) -- (v_1_B.east);
\draw[->,orange] (v_2_A.150) -- (v_1_C.east);
\draw[->,orange] (v_3_A.150) -- (v_2_A.east);
\draw[->,orange] (v_3_A.150) -- (v_2_B.east);
\draw[->,orange] (v_3_A.150) -- (v_2_C.east);
\draw[->,orange] (v_1_B.150) -- (v_0_A.east);
\draw[->,orange] (v_1_B.150) -- (v_0_B.east);
\draw[->,orange] (v_1_B.150) -- (v_0_C.east);
\draw[->,orange] (v_2_B.150) -- (v_1_A.east);
\draw[->,orange] (v_2_B.150) -- (v_1_B.east);
\draw[->,orange] (v_2_B.150) -- (v_1_C.east);
\draw[->,orange] (v_3_B.150) -- (v_2_A.east);
\draw[->,orange] (v_3_B.150) -- (v_2_B.east);
\draw[->,orange] (v_3_B.150) -- (v_2_C.east);
\draw[->,orange] (v_1_C.150) -- (v_0_A.east);
\draw[->,orange] (v_1_C.150) -- (v_0_B.east);
\draw[->,orange] (v_1_C.150) -- (v_0_C.east);
\draw[->,orange] (v_2_C.150) -- (v_1_A.east);
\draw[->,orange] (v_2_C.150) -- (v_1_B.east);
\draw[->,orange] (v_2_C.150) -- (v_1_C.east);
\draw[->,orange] (v_3_C.150) -- (v_2_A.east);
\draw[->,orange] (v_3_C.150) -- (v_2_B.east);
\draw[->,orange] (v_3_C.150) -- (v_2_C.east);
\draw[->,red] (v_1_D.150) -- (v_0_A.east);
\draw[->,red] (v_1_D.150) -- (v_0_B.east);
\draw[->,red] (v_1_D.150) -- (v_0_C.east);
\draw[->,red] (v_2_D.150) -- (v_1_A.east);
\draw[->,red] (v_2_D.150) -- (v_1_B.east);
\draw[->,red] (v_2_D.150) -- (v_1_C.east);
\node[draw=black,minimum height=1cm, minimum width=.5cm,right=.75cm of v_3_C] (v_4_C) {
};
\draw[->,red] (v_4_C.150) -- (v_3_A.east);
\draw[->,red] (v_4_C.150) -- (v_3_B.east);
\draw[->,red] (v_4_C.150) -- (v_3_C.east);
\draw[->,blue] (v_4_C.150) -- (v_0_D.east);
\draw[->,blue] (v_4_C.150) -- (v_1_D.east);
\draw[->,blue] (v_4_C.150) -- (v_2_D.east);
\node[left=.25cm of rowA] {A};
\node[left=.25cm of rowB] {B};
\node[left=.25cm of rowC] {C};
\node[left=.25cm of rowD] {D};
\end{tikzpicture}}
    \caption{Bug in DagRider : more than $f$ weak edges}
    \label{fig:dagrider_bug}
\end{figure}

If a given node (at row $r$) begins the reliable broadcast of two successive vertices $v^r_c$ and $v^r_{c+1}$ (via emitting the corresponding INIT messages). Then, under an asynchronous communication model, and due to the non-determinism incurred by networked communications, there are no guarantee, for any node $r'$, that $r'$ can collect $f+1$ READY messages for $v^r_c$ before observing $f+1$ READY messages for $v^r_{c+1}$. This may result in $v^r_{c+1}$ being delivered before $v^r_{c}$, as illustrated on Fig.\ref{fig:dagrider_bug}.

In \cite{all_you_need_is_dag}, it is said that there can be up to $f$ weak edges.
However, these weak edges are programmatically defined in Algorithm 2 lines 27-31 in \cite{all_you_need_is_dag} by gathering all ``orphan'' vertices from columns before the previous column that have no path towards the new vertex proposal.
However, the creation of a new vertex ends with a call to $\mathtt{rbcast}$ (line 15 in Algorithm 2 of \cite{all_you_need_is_dag}) without waiting for the corresponding delivery ($\mathtt{rdlver}$) in the reliable broadcast layer. 
Because the subsequent creation of a new vertex does not depend on the reliable delivery of the previous one (it only depends on the delivery of $2*f+1$ vertices in the previous column, see lines 10 in Algorithm 2 of \cite{all_you_need_is_dag}), then it is possible for a node to emit several vertices with no edges between one another. 
This is exemplified on Fig.\ref{fig:dagrider_bug} with node ``D'' (on the bottom) which submits 3 vertices that are not linked with strong edges (in red and orange for better readability). This may occur if the delivery operation on the reliable broadcast layer is particularly slow for vertices submitted by ``D'' (this may be due to ``D'' having network issues, whether or not these are caused by malice).
In any case, supposing that node ``C'' (third row) then finally delivers the previous vertices from ``D'' and submits a new vertex proposal, by construction, this vertex will then includes at least the 3 weak edges drawn in blue on Fig.\ref{fig:dagrider_bug}.
In that case, because we have $n = 3*f+1$ and $n=4$, the limit is $f=1$ which contradicts the remark from \cite{all_you_need_is_dag}.

The fact that delivery operations on the reliable broadcast layer occur in a certain order is not guaranteed at all. In \cite{all_you_need_is_dag}, only three properties are given for the reliable broadcast abstraction: ``Agreement'', ``Integrity'' and ``Validity'' and none of them refer to ordering.
However, \cite{all_you_need_is_dag} seems to take for granted one such property, because in all the examples they propose there is always a strong edge between any two consecutive vertices submitted by the same node.
The aforementioned issue with the number of weak edges is related to this oversight w.r.t.~the ordering of delivery operation on the reliable broadcast layer. This underlines the importance of (fair) ordering properties in the definition of distributed protocols.

In order to correct the issue we could either \textbf{(1)} add additional properties on the reliable broadcast abstraction from \cite{all_you_need_is_dag} or \textbf{(2)} enforce the requirement that there is at most $f$ outgoing weak edges on a vertex.

Let us remark that \cite{reducing_latency_of_dag_based_consensus_in_the_asynchronous_setting_via_the_utxo_model} acknowledges that this problem may occur in Narwhal \& Tusk \cite{narwhal_and_tusk} and explicitly requires that a node's proposal at column $c$ always includes (i.e., has a strong edge towards) its proposal for column $c-1$ (i.e., solution \textbf{(1)}).

However, a problem with solution \textbf{(1)} is that it forces all nodes to wait for the delivery of their vertex at column $c$ before being able to initiate the broadcast of their vertex at column $c+1$.
Yet, as we discussed in Sec.\ref{ssec:dags_and_blockchains}, one of the advantages of DAG-based ledgers w.r.t.~Blockchain lies in the ability of each node to independently broadcast new vertices to be added to the DAG, instead of having to wait for the resolution of a consensus algorithm to select a new block.
Moreover, solution \textbf{(1)} may further increase the power of nuisance of an adversary that is able to delay the delivery of vertices (as described in Sec.\ref{ssec:attack_bracha}). Indeed, in addition of increasing the time between the moment of broadcast initiation and broadcast delivery for the vertex of column $c$, if solution \textbf{(1)} is implemented, the adversary will also delay the moment of the subsequent broadcast initiation (i.e., that of the vertex of column $c+1$).

For this reason, in this paper, we have considered solution \textbf{(2)}: limiting the number of weak edges to $f$ (randomly selecting those targeting the lower columns) even though more could be included.

\section{Bracha's Byzantine Reliable Broadcast\label{anx:bracha}}

Bracha's Byzantine Reliable Broadcast algorithm \cite{asynchronous_byzantine_agreement_protocols} relies on two ``echoing'' phases to ascertain that all nodes agree on the same message that is broadcast.
A node that wants to initiate the reliable broadcast start with a simple broadcast of an ``INIT'' message carrying the message it wants to reliably broadcast (this corresponds to the call of $\mathtt{rbcast}$ on Fig.\ref{fig:layers_dagrider}).
The first echoing phase then consists in nodes emitting and collecting ``ECHO'' messages that answer the first ``INIT'' message.
The second echoing phase likewise consists in collecting ``READY'' messages.

Once enough ``READY'' messages have been received locally by a given node, that node can deliver the reliably broadcast message (this corresponds to the call of $\mathtt{rdlver}$ on Fig.\ref{fig:layers_dagrider}) because it is then certain that all the other nodes will also eventually deliver that same message.

The thresholds for the collection of ``ECHO'' and ``READY'' messages depend on the maximum number of byzantine nodes $f$ as follows: $\lfloor \frac{n+f}{2} \rfloor + 1$ ``ECHO'' or $f+1$ ``READY'' messages from distinct nodes are required to broadcast a new ``READY'' message and $2*f + 1$ ``READY'' messages are required for delivery ($\mathtt{rdlver}$).

\section{Problem with the order of Board\&Clerk\label{anx:bug_board_and_clerk}}

In \cite{reducing_latency_of_dag_based_consensus_in_the_asynchronous_setting_via_the_utxo_model}, the deterministic order $<_{\text{\faEnvelopeO}}$ that is proposed is such that $x <_{\text{\faEnvelopeO}} x'$ iff $|\{\text{nodes vote}~x~\text{first}\}| > |\{\text{nodes vote}~x'~\text{first}\}|$.

However, as illustrated on Fig.\ref{fig:do_counter_ex_board_and_clerk_votes}, this definition may yield cases in which we have $x_1 <_{\text{\faEnvelopeO}} x_2$, $x_2 <_{\text{\faEnvelopeO}} x_3$ and $x_3 <_{\text{\faEnvelopeO}} x_1$ i.e., a Condorcet cycle \cite{condorcet_attack_against_fair_transaction_ordering}. 
Indeed, on Fig.\ref{fig:do_counter_ex_board_and_clerk}, one can see that $n_0$ and $n_3$ vote for $v_4^0$ before $v_4^2$ but only $n_2$ vote for $v_4^2$ before $v_4^0$. Thus, $v_4^0 <_{\text{\faEnvelopeO}} v_4^2$.
Similarly, both $n_0$ and $n_2$ vote for $v_4^2$ before $v_4^3$ while only $n_3$ does the opposite. Hence $v_4^2 <_{\text{\faEnvelopeO}} v_4^3$.
However, $n_2$ and $n_3$ vote for $v_4^3$ before they do so for $v_4^0$, only $n_0$ doing the opposite.
Thus $v_4^3 <_{\text{\faEnvelopeO}} v_4^0$.

\begin{figure}[h]
    \centering

\begin{subfigure}{.15\textwidth}
    \centering
    \scalebox{.8}{\begin{tikzpicture}
\node[draw] at (0,0)                  (v04) {$v^0_{4}$};
\node[draw,        below=.3cm of v04] (v14) {$v^1_{4}$};
\node[draw,        below=.3cm of v14] (v24) {$v^2_{4}$};
\node[draw,        below=.3cm of v24] (v34) {$v^3_{4}$};
\node[draw,red,        right=.6cm of v04] (v05) {$v^0_{5}$};
\node[draw,blue,        right=.6cm of v24] (v25) {$v^2_{5}$};
\node[draw,darkspringgreen,        right=.6cm of v34] (v35) {$v^3_{5}$};
\node[draw,        right=.6cm of v25] (v26) {$v^2_{6}$};
\draw[->,red] (v05) -- (v04);
\draw[->,red] (v05) -- (v14);
\draw[->,red] (v05) -- (v24);
\draw[->,blue] (v25) -- (v14);
\draw[->,blue] (v25) -- (v24);
\draw[->,blue] (v25) -- (v34);
\draw[->,darkspringgreen] (v35) -- (v04);
\draw[->,darkspringgreen] (v35) -- (v14);
\draw[->,darkspringgreen] (v35) -- (v34);
\draw[->] (v26) -- (v05);
\draw[->] (v26) -- (v25);
\draw[->] (v26) -- (v35);
\end{tikzpicture}}
    \caption{DAG}
    \label{fig:do_counter_ex_board_and_clerk_dag}
\end{subfigure}
\begin{subfigure}{.3\textwidth}
    \centering
\begin{tabular}{|c|c|c|c|c|}
\hline
\multirow{2}{*}{} & \multicolumn{4}{c|}{earliest vote by} \\
\cline{2-5} 
     & $n_0$ & $n_1$ & $n_2$ & $n_3$ \\
     \hline 
$v_4^0$ & $4$ & $\infty$ & $6$ & $5$ \\
\hline 
$v_4^2$ & $5$ & $\infty$ & $4$ & $\infty$ \\
\hline 
$v_4^3$ & $\infty$ & $\infty$ & $5$ & $4$ \\
\hline 
\end{tabular}
    \caption{Vote table}
    \label{fig:do_counter_ex_board_and_clerk_votes}
\end{subfigure}

    \caption{Example where $<_{\text{\faEnvelopeO}}$ is not transitive}
    \label{fig:do_counter_ex_board_and_clerk}
\end{figure}

\cite{reducing_latency_of_dag_based_consensus_in_the_asynchronous_setting_via_the_utxo_model} does not seem to acknowledge that this problem may occur.
Indeed, \cite{reducing_latency_of_dag_based_consensus_in_the_asynchronous_setting_via_the_utxo_model} proposes to sort the vertices using a stable sorting algorithm and the $<_{\text{\faEnvelopeO}}$ relation as a comparator.
However, it is impossible to do it reliably if $<_{\text{\faEnvelopeO}}$ is not transitive.

\section{Our solution for the ``VoteCount'' mechanism\label{anx:vote_count_mechanism}}

\begin{figure}[h]
    \centering
    \includegraphics[scale=.275]{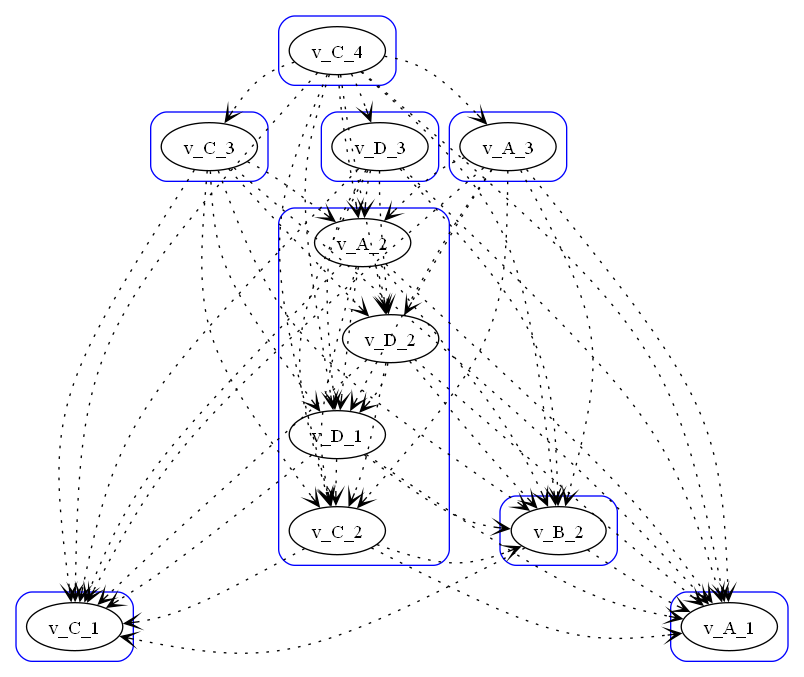}
    \caption{$\leq_{\text{\faEnvelopeO}}$ while highlighting SCCs}
    \label{fig:ord_graph_condensation}
\end{figure}

\begin{figure}[h]
    \centering
    \includegraphics[scale=.275]{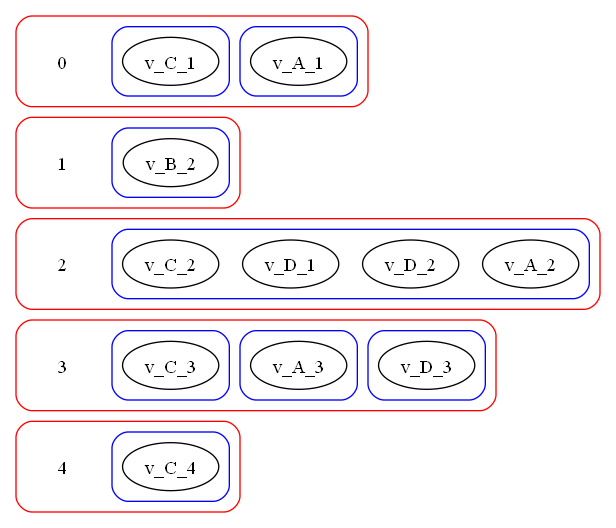}
    \caption{Topological ranks}
    \label{fig:ord_topological_ranks}
\end{figure}

In this appendix,
we detail the ``VoteCount'' mechanism used to order vertices in a wave of the DAG.
It is inspired by 
\textbf{(1)}
the initial idea from \cite{reducing_latency_of_dag_based_consensus_in_the_asynchronous_setting_via_the_utxo_model}
of dressing a table of votes from the structure of the DAG wave and use it to define a precedence relation $\leq_{\text{\faEnvelopeO}}$,
\textbf{(2)} the idea from \cite{order_fairness_for_byzantine_consensus,themis_fast_strong_order_fairness_in_byzantine_consensus} to solve the issue caused by Condorcet cycles that could exist in the precedence relation 
via identifying them as Strongly Connected Components of the corresponding graph representation 
and \textbf{(3)} the topological ranking notion from
\cite{diversified_top_k_graph_pattern_matching} as an alternative to the Hamiltonian path method used in \cite{themis_fast_strong_order_fairness_in_byzantine_consensus}.

We illustrate this mechanism's use on Fig.\ref{fig:ord_graph_condensation} and Fig.\ref{fig:ord_topological_ranks}, applying it to the example from Fig.\ref{fig:do_example_1}.

Representing $\geq_{\text{\faEnvelopeO}}$ as a directed graph, the first step is to compute its Strongly Connected Components (SCCs) which we highlight in blue on Fig.\ref{fig:ord_graph_condensation}.
This allows identifying Condorcet cycles (on Fig.\ref{fig:ord_graph_condensation} there is a cycle with $4$ vertices).
The condensation graph \cite{order_fairness_for_byzantine_consensus,themis_fast_strong_order_fairness_in_byzantine_consensus}, obtained via merging vertices in the same SCC, is guaranteed to be acyclic.
This allows to use the topological rank notion from \cite{diversified_top_k_graph_pattern_matching} which is defined as follows:
leaves of the condensation graph have rank 0, while the rank of the other vertices correspond to their shortest distance to any of the leaves.
On Fig.\ref{fig:ord_topological_ranks} we give the topological rank of each vertex (the rank is written on the left, from 0 at the top to 4 at the bottom). 
This allows sorting the vertices in a manner that is consistent with $\geq_{\text{\faEnvelopeO}}$.
The order between vertices with the same topological rank can be resolved in any deterministic manner (e.g., lexicographic order on their hash value).

\section{Details on DagRider-layer Byzantine behavior\label{anx:detail_dag_rider_layer_attack}}

\begin{figure}[h]
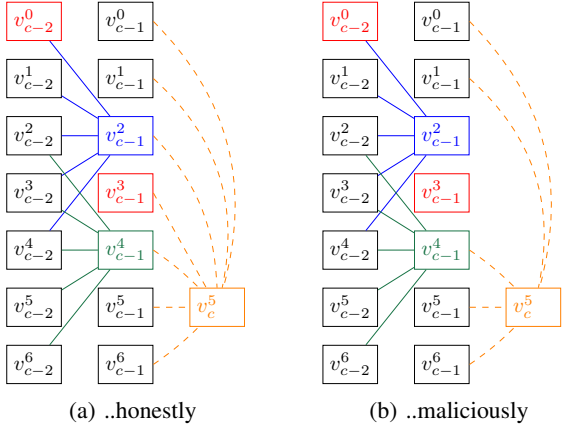

    \centering

\begin{subfigure}{.225\textwidth}
    \centering
\scalebox{.8}{\begin{tikzpicture}
\input{figures/pariah/pariah_dag_base}
\node[draw,orange, right=.6cm of n52] (n51) {$v^5_{c\phantom{-1}}$};
\draw[orange,dashed] (n02.east) edge[bend left=30] (n51);
\draw[orange,dashed] (n12.east) edge[bend left=30] (n51);
\draw[orange,dashed] (n22.east) edge[bend left=20] (n51);
\draw[orange,dashed] (n32.east) -- (n51);
\draw[orange,dashed] (n42.east) edge[bend left=5] (n51);
\draw[orange,dashed] (n52.east) -- (n51);
\draw[orange,dashed] (n62.east) edge[bend right=5] (n51);
\end{tikzpicture}}
    \caption{..honestly}
    \label{fig:pariah_attack_concept_honest}
\end{subfigure}
\begin{subfigure}{.225\textwidth}
    \centering
\scalebox{.8}{\begin{tikzpicture}
\input{figures/pariah/pariah_dag_base}
\node[draw,orange, right=.6cm of n52] (n51) {$v^5_{c\phantom{-1}}$};
\draw[orange,dashed] (n02.east) edge[bend left=30] (n51);
\draw[orange,dashed] (n12.east) edge[bend left=30] (n51);
\draw[orange,dashed] (n42.east) edge[bend left=5] (n51);
\draw[orange,dashed] (n52.east) -- (n51);
\draw[orange,dashed] (n62.east) edge[bend right=5] (n51);
\end{tikzpicture}}
    \caption{..maliciously}
    \label{fig:pariah_attack_concept_depth1}
\end{subfigure}
    
    \caption{Vertex proposal reasoning..}
    \label{fig:pariah_attack_concept}
\end{figure}

Fig.\ref{fig:pariah_attack_concept} provides a more detailed example for illustrating the Byzantine behavior described in Sec.\ref{ssec:byz_dagrider_layer}.

We consider $7$ nodes ($N_0$ to $N_6$) from the perspective of node $N_5$ which adds its vertices to the sixth row.
For the sake of clarity, not all edges are represented.
We consider which edges (in orange and dashed) should be included when node $N_5$ defines vertex \textcolor{orange}{$v^5_c$}.
Here, we suppose that the two vertices in red contain transactions of the target client.
If node $N_5$ is honest, it will choose at least $2*f+1$ strong edges among all the 7 possibilities (see Fig.\ref{fig:pariah_attack_concept_honest}).
If however, $N_5$ acts maliciously as per our attack scenario, it will not include the strong edge towards \textcolor{red}{$v^3_{c-1}$} because \textcolor{red}{$v^3_{c-1}$} contains transactions from the target client. Additionally, the proposed \textcolor{orange}{$v^5_c$} may also not point towards \textcolor{blue}{$v^2_{c-1}$} (in blue) because \textcolor{blue}{$v^2_{c-1}$} itself points towards \textcolor{red}{$v^0_{c-2}$} which contains unwanted transactions.

It is clear here that performing such ``pariah'' attacks (ostracizing specific vertices) may yield the desired effect of excluding said vertices from the next wave. 
Their exclusion is not guaranteed but manipulating the edges as described above makes it more likely. 

More generally, an infected node may define a lexicographic rank for every strong edge candidate according to the number of transactions from the target client that each column in the associated causal sub-graph (up to a certain depth) contains.
For instance, in the example from Fig.\ref{fig:pariah_attack_concept}, vertex \textcolor{darkspringgreen}{$v^4_{c-1}$} has a depth 2 rank of $(0,0)$ because no such transactions are present at column $c-1$ and $c-2$.
On the other hand, \textcolor{red}{$v^3_{c-1}$} has a rank $(1,0)$ and \textcolor{blue}{$v^2_{c-1}$} has $(0,1)$.
Supposing that vertices $v^0_{c-1}$, $v^1_{c-1}$, $v^5_{c-1}$ and $v^6_{c-1}$ have a rank $(0,0)$, the infected node $N_5$ may sort them as follows : [\textcolor{darkspringgreen}{$v^4_{c-1}$}, $v^0_{c-1}$, $v^1_{c-1}$, $v^5_{c-1}$, $v^6_{c-1}$,\textcolor{blue}{$v^2_{c-1}$},\textcolor{red}{$v^3_{c-1}$}] and select the $2*f+1$ first so as to define the strong edges of \textcolor{orange}{$v^5_c$}.

\section{Details on Bracha-layer Byzantine behavior\label{anx:detail_bracha_layer_attack}}

Because Bracha's Byzantine Reliable Broadcast algorithm \cite{asynchronous_byzantine_agreement_protocols} relies on collecting (and thus waiting for) a certain number of ECHO and READY messages from other nodes, the adversary can statistically delay this $\mathtt{rdlver}$ operation on all nodes via infecting a minority of nodes.
In this scenario, infected nodes will not emit ECHO and READY messages for vertices that contain transactions from the target client.

Let us indeed suppose that the probability to receive an emitted ECHO before timestamp $z$ is given by variable $X$.
Among $n$ trials, the probability of having collected exactly $k \leq n$ such messages before $z$ is:
\[
\binom{n}{k} * X^k * (1-X)^{n - k}
\]

For vertices that contain transactions from the target client, sabotaged nodes never send these ECHO messages (we always have $X=0$ for any timestamp $z$) and can therefore be ignored when counting the numbers of messages.
Therefore, given $b \leq f$ the number of sabotaged nodes, the probability of having collected at least $\lfloor \frac{n+f}{2} \rfloor + 1$ ECHO messages from distinct nodes before $z$ is:
\[
Y = \sum_{k = \lfloor \frac{n+f}{2} \rfloor + 1}^{n-b} \binom{n - b}{k} * X^k * (1-X)^{n - b - k}
\]

$\mathtt{rdlver}$ then requires collecting at least $2*f+1$ READY messages, which probability we can approximate as follows:
\[
Z = \sum_{k = 2*f + 1}^{n-b} \binom{n - b}{k} * Y^k * (1-Y)^{n - b - k}
\]

\begin{figure}[h!]
\vspace*{-.25cm}
    \centering

\scalebox{.85}{
\input{figures/bracha_sabotage/theory_plot}
}

    \caption{Theoretical effect of Bracha sabotage for $n=25$}
    \label{fig:bracha_sabotage_theory}
\vspace*{-.25cm}
\end{figure}

Via plotting this probability $Z$ w.r.t.~$X$ on Fig.\ref{fig:bracha_sabotage_theory} (with $n=25$ and $f=8$), we observe that the more nodes are sabotaged, the less likely is the occurrence of the delivery $\mathtt{rdlver}$ operation before timestamp $z$.
Consequently, even if we stay below the maximum number $f$ of Byzantine nodes, sabotage can still result in the delivery of vertices to be delayed (statistically).
In turn, if the delivery of vertices that contain transactions from the target client is delayed, this makes them less likely to be the target of strong edges (which may result in situations such as the one described in Sec.\ref{ssec:attack_bracha} and Fig.\ref{fig:dag_attacked_bracha_layer}).
This can be partly mitigated by DagRider supporting weak edges.
The mitigation is only partial because for the delayed vertex to be included in the next wave $w$, it still requires to be targeted (via a weak edge) by a vertex in the causal subgraph of $w$'s leader. Otherwise it won't be included until at least wave $w+1$.

\section{Details of the 10 DAG-specific OF properties\label{anx:enumeration_order_fairness_props}}

The 8 properties, covering all $OF^\beta_\alpha$ cases of Sec.\ref{ssec:metrics} are:\\
\noindent$\bullet$ $OF^{F_{IN}}_{S_{ND}}$ as ``finalization-order-fairness w.r.t.~transaction emission'': if the initial emission of $x$ by a certain client precedes that of $x'$ then all honest nodes must finalize $x$ before $x'$.\\
\noindent$\bullet$ $OF^{W_{AV}}_{S_{ND}}$ as ``wave-order-fairness w.r.t.~transaction emission'': if the initial emission of $x$ by a certain client precedes that of $x'$ then no honest node can finalize $x'$ in a wave before that in which $x$ is finalized.\\
\noindent$\bullet$ $OF^{F_{IN}}_{R_{EC}}$ as ``finalization-order-fairness w.r.t.~reception from client'': if a majority of nodes receive $x$ from a client before they do so for $x'$ then all honest nodes must finalize $x$ before $x'$.\\
\noindent$\bullet$ $OF^{W_{AV}}_{R_{EC}}$ as ``wave-order-fairness w.r.t.~reception from client'': if a majority of nodes receive $x$ from a client before they do so for $x'$ then no honest node can finalize $x'$ in a wave before that in which $x$ is finalized.\\
\noindent$\bullet$ $OF^{F_{IN}}_{I_{NI}}$ as ``finalization-order-fairness w.r.t.~broadcast initiation'': if a majority of nodes begin their participation in the reliable broadcast of a vertex that contains $x$ before they do so for $x'$ then all honest nodes must finalize $x$ before $x'$.\\
\noindent$\bullet$ $OF^{W_{AV}}_{I_{NI}}$ as ``wave-order-fairness w.r.t.~broadcast initiation'': if a majority of nodes begin their participation in the reliable broadcast of a vertex that contains $x$ before they do so for $x'$ then no honest node can finalize $x'$ in a wave before that in which $x$ is finalized.\\
\noindent$\bullet$ $OF^{F_{IN}}_{D_{LV}}$ as ``finalization-order-fairness w.r.t.~broadcast delivery'': if a majority of nodes deliver a vertex containing $x$ before they do so for $x'$ then all honest nodes must finalize $x$ before $x'$.\\
\noindent$\bullet$ $OF^{W_{AV}}_{D_{LV}}$ as ``wave-order-fairness w.r.t.~broadcast delivery'': if a majority of nodes deliver a vertex containing $x$ before they do so for $x'$ then no honest node can finalize $x'$ in a wave before that in which $x$ is finalized.

\section{Details on simulation parameterization\label{anx:max_sim}}

We use the MAX Multi-Agent eXperimenter tool \cite{max_tool}, which is based on the Agent-Group-Role \cite{from_agents_to_organizations_an_organizational_view_of_multi_agent_systems} Multi-Agent System \cite{an_introduction_to_multiagent_systems} formalism.
It allows discrete time simulation of complex distributed systems.
Our implementation of DagRider \cite{all_you_need_is_dag} is available in \cite{max_dagrider} and that of the underlying Bracha BRB algorithm \cite{on_the_versatility_of_bracha_byzantine_reliable_broadcast_algorithm} in \cite{max_bracha}.
We use the adversary model defined in \cite{adversary_augmented_simulation_to_evaluate_client_fairness_on_hyperledger_fabric} and implemented in \cite{max_p2p_adversarial_model}. 

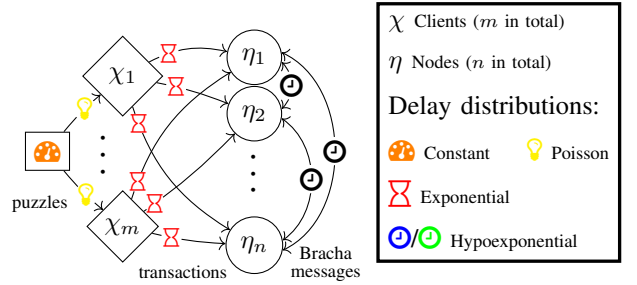
\begin{figure}[h]
\vspace*{-.25cm}
    \centering
    \scalebox{1}{\begin{tikzpicture}
%
%
%
\node[draw,diamond] at (0,1) (client1) {$\chi_1$};
\node at (-.25,0) {\rotatebox{90}{\Large$\cdots$}};
\node[draw,diamond,inner sep=1.75] at (0,-1) (clientm) {$\chi_{m}$};
\node[draw,circle] at (1.75,1.25) (n1) {$\eta_1$};
\node[draw,circle] at (1.75,.5) (n2) {$\eta_2$};
\node at (1.75,-.25) {\rotatebox{90}{\Large$\cdots$}};
\node[draw,circle] at (1.75,-1.25) (nn) {$\eta_n$};
\node[draw,rectangle] at (-1,0) (puzzler) {\textcolor{orange}{\faDashboard}};
%
%
%
%
\draw (puzzler) edge[->,bend left=5] node[pos=.6,circle,fill=white,inner sep=-.5] {\textcolor{yellow}{\faLightbulbO}} (client1);
\draw (puzzler) edge[->,bend right=5] node[pos=.6,circle,fill=white,inner sep=-.5] {\textcolor{yellow}{\faLightbulbO}} (clientm);
\node[align=center] at (-1.1,-.7) {\scriptsize puzzles};
\draw (client1) edge[->,bend left=20] node[pos=.25,circle,fill=white,inner sep=-.5] {\scriptsize\textcolor{red}{\faHourglassO}} (n1);
\draw (client1) edge[->,bend left=5] node[pos=.25,circle,fill=white,inner sep=-.5] {\scriptsize\textcolor{red}{\faHourglassO}} (n2);
\draw (client1) edge[->,bend right=20] node[pos=.1,circle,fill=white,inner sep=-.5] {\scriptsize\textcolor{red}{\faHourglassO}} (nn);
\draw (clientm) edge[->,bend left=20] node[pos=.1,circle,fill=white,inner sep=-.5] {\scriptsize\textcolor{red}{\faHourglassO}} (n1);
\draw (clientm) edge[->,bend right=5] node[pos=.1,circle,fill=white,inner sep=-.5] {\scriptsize\textcolor{red}{\faHourglassO}} (n2);
\draw (clientm) edge[->,bend right=5] node[pos=.25,circle,fill=white,inner sep=-.5] {\scriptsize\textcolor{red}{\faHourglassO}} (nn);
\node[align=center] at (.8,-1.6) {\scriptsize transactions};
\draw (n1.350) edge[<->,bend left=45] node[midway,circle,fill=white,inner sep=-.5] {\textcolor{black}{\faClockO}} (n2.10);
\draw (n2.350) edge[<->,bend left=55] node[midway,circle,fill=white,inner sep=-.5] {\textcolor{black}{\faClockO}} (nn.10);
\draw (nn.350) edge[<->,bend right=65] node[midway,circle,fill=white,inner sep=-.5] {\textcolor{black}{\faClockO}} (n1.10);
\node[align=center] at (2.7,-1.35) {\scriptsize Bracha};
\node[align=center] at (2.7,-1.65) {\scriptsize messages};
%
%
%
\node[draw,line width=1.25,inner sep=2] at (5,0.25) {
\begin{tikzpicture}
\node (leg1) at (0,0) {$\chi$ {\scriptsize Clients ($m$ in total)}};
\node[below=0.15cm of leg1.south west, anchor=north west] (leg2) {$\eta$ {\scriptsize Nodes ($n$ in total)}};
\node[below=0.15cm of leg2.south west, anchor=north west] (leg3) {Delay distributions:};
\node[below=0.1cm of leg3.south west, anchor=north west] (leg4) {\textcolor{orange}{\faDashboard} {\scriptsize Constant} ~~ \textcolor{yellow}{\faLightbulbO} {\scriptsize Poisson}};
\node[below=0.1cm of leg4.south west, anchor=north west] (leg5) {\textcolor{red}{\faHourglassO} {\scriptsize Exponential}};
\node[below=0.1cm of leg5.south west, anchor=north west] (leg6) {\textcolor{blue}{\faClockO}/\textcolor{green}{\faClockO} {\scriptsize Hypoexponential}};
\end{tikzpicture}
};
\end{tikzpicture}}
    \caption{Network parameterization}
    \label{fig:network_param}
\vspace*{-.25cm}
\end{figure}

\begin{figure}[h]
\vspace*{-.25cm}
    \centering
    \includegraphics[width=.475\textwidth]{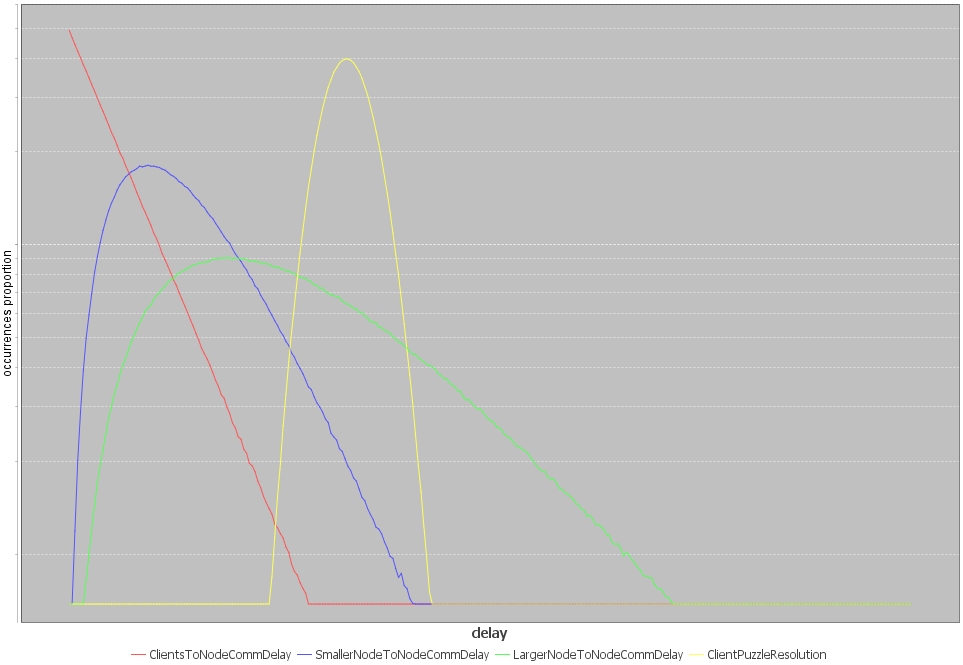}
    \caption{Distributions of delays (log scale)}
    \label{fig:delays_distros}
\vspace*{-.25cm}
\end{figure}

Fig.\ref{fig:network_param} describes the parameterization of the system, with the colors of the delay distributions corresponding to the curves on Fig.\ref{fig:delays_distros} (except from the constant rate \textcolor{orange}{\faDashboard} at which new puzzles are revealed).

We use an arbitrary unit of time denoted as ``tick'' in our discrete time simulations.
We consider that a new puzzle is revealed every 200 ticks \textcolor{orange}{\faDashboard}.
Any client can solve it after a given time that is modeled by a Poisson distribution \textcolor{yellow}{\faLightbulbO} of mean 100 (in ticks).
Once a client solves a puzzle, it creates a transaction and broadcasts it to nodes.
We consider the \textcolor{red}{\faHourglassO} distribution of delays taken by such transactions to reach any given node to be an Exponential distribution of mean 20.

As for the delay \textcolor{black}{\faClockO} of node to node communications, we consider 2 different cases in order to model quicker and slower P2P networks:
\begin{itemize}
    \item a hypoexponential distribution \textcolor{blue}{\faClockO} with rates (1/10, 1/15 and 1/20) modelling an average P2P network with smaller delays
    \item a hypoexponential distribution \textcolor{green}{\faClockO} with rates (1/20, 1/30 and 1/40) modelling a slow P2P network with larger delays
\end{itemize}

\section{Details on simulation execution\label{anx:max_sim_exec}}

\begin{figure}[h]
    \centering

    \begin{subfigure}{.475\textwidth}
        \includegraphics[scale=.28]{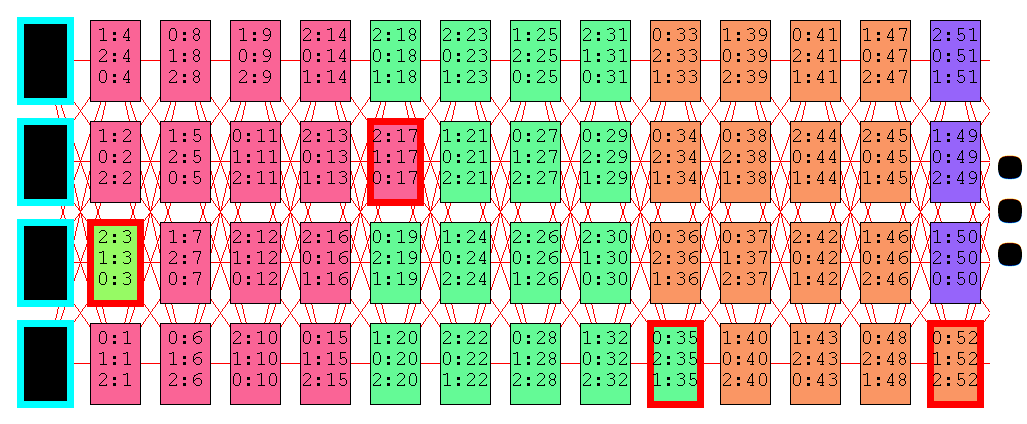}
        \caption{simulation with a perfect network}
        \label{fig:dag_from_simu_perfect}
    \end{subfigure}
    
    \begin{subfigure}{.475\textwidth}
        \includegraphics[scale=.28]{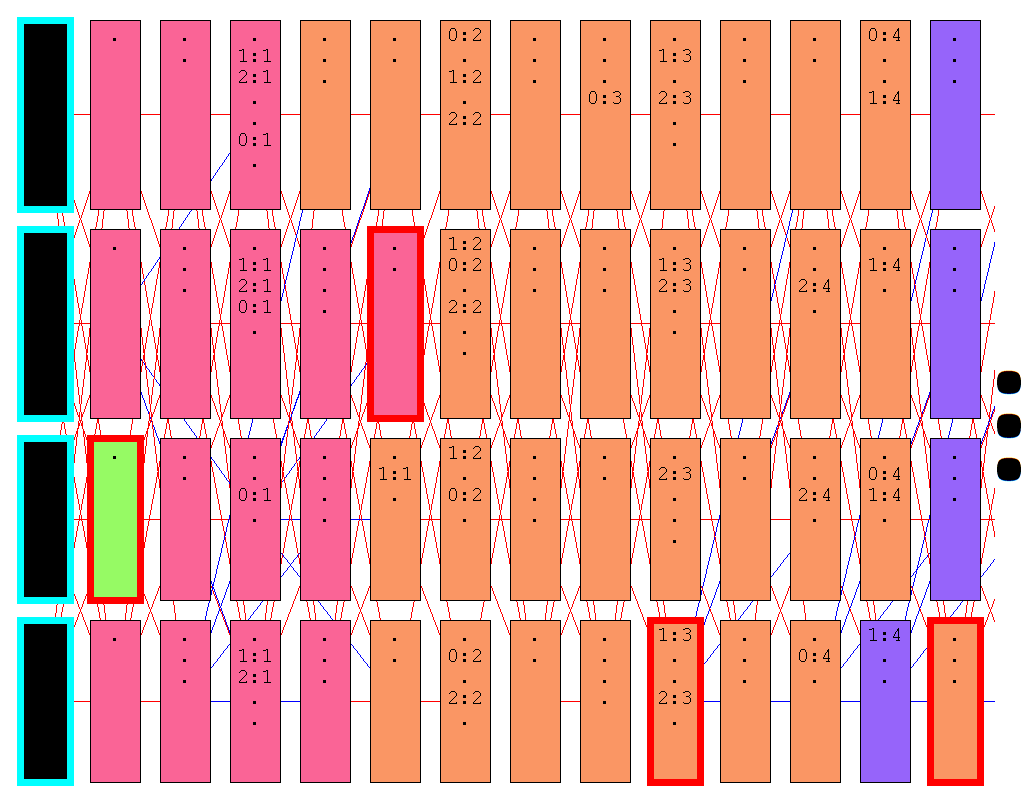}
        \caption{simulation to illustrate leader skipping}
        \label{fig:dag_from_simu_quickest}
    \end{subfigure}
    
    \caption{Simulation examples}
    \label{fig:dag_simu_further_discussion}
\end{figure}

\begin{figure*}
    \centering

\begin{minipage}{.65\textwidth}
    
    \begin{subfigure}{\textwidth}
        \includegraphics[scale=.27]{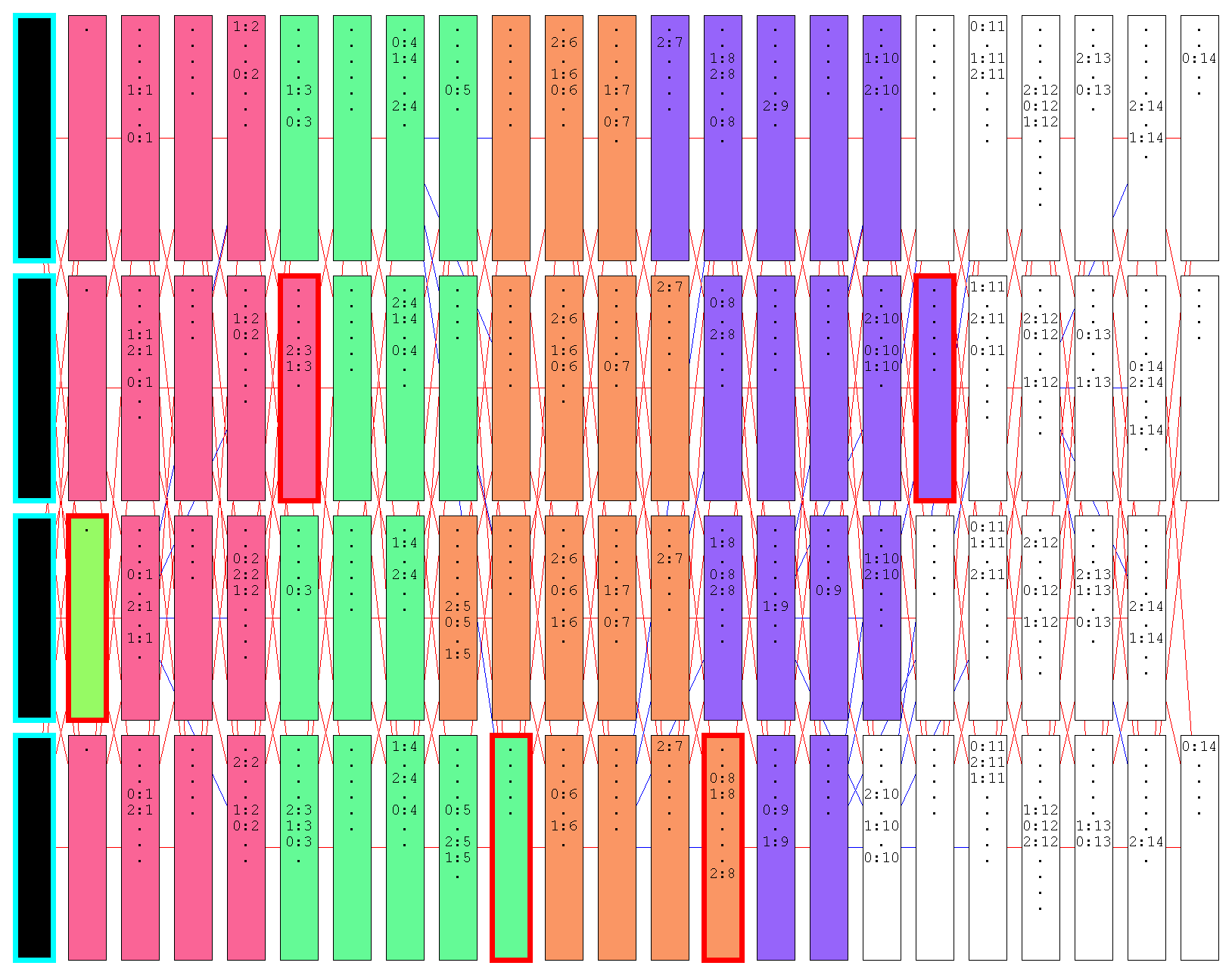}
        \caption{...the \textcolor{blue}{\faClockO} (quicker network)}
        \label{fig:dag_from_simu_quick}
    \end{subfigure}
    
\end{minipage}
\begin{minipage}{.325\textwidth}
    \begin{subfigure}{\textwidth}
        \includegraphics[scale=.27]{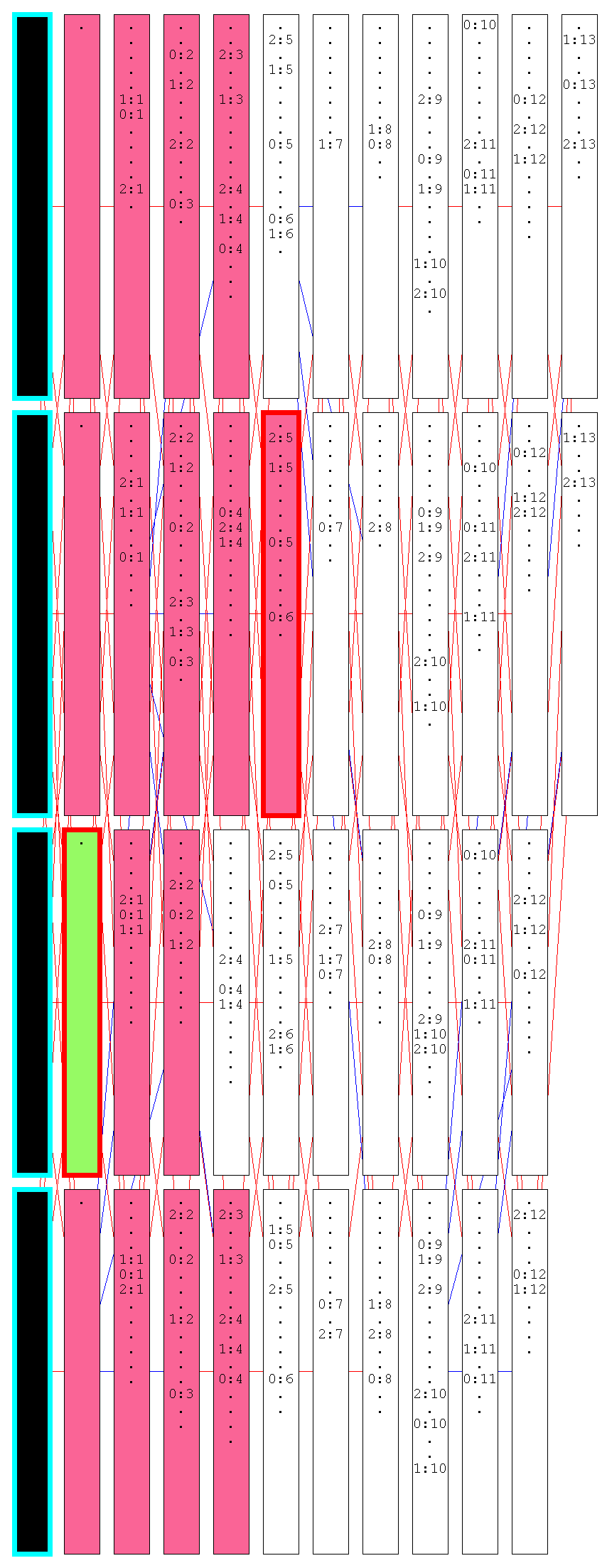}
        \caption{...the \textcolor{green}{\faClockO} (slower network)}
        \label{fig:dag_from_simu_slow}
    \end{subfigure}
\end{minipage}
    
    \caption{DAG content excerpts at the end of a $n=4$ simulation with ...}
    \label{fig:dag_from_simu}
\end{figure*}

Running a simulation yields client puzzle solutions being included in the local copies of the DAG hosted on each node.
In the following, we consider a simple case with $3$ clients and $4$ nodes which are all honest.
The diagrams on Fig.\ref{fig:dag_simu_further_discussion} and Fig.\ref{fig:dag_from_simu} are produced by our tool. 
Each corresponds to a representation of the local copy of the DAG at a certain node at the end of the simulation. 
Here, strong edges are drawn in red and weak edges in blue.
Vertices framed in red are leaders.
The colors of the vertices correspond to the wave they are finalized in (white vertices are not yet included in any wave).

Fig.\ref{fig:dag_from_simu_perfect} results from a simulation in which there are no third party transactions and in which both \textcolor{red}{\faHourglassO} and \textcolor{black}{\faClockO} are s.t.~there is always the same fixed delay of $1$ simulation tick between the moment an emission or reception is scheduled and executed in the reliable broadcast layer and there is always a delay of $1$ between the moment a new puzzle is revealed and the moment a node receives its solution from its corresponding client.
This allows every node to produce a new vertex as soon as it receives the solutions (which all arrive at the same time) from the clients at exactly the rate in which puzzles are revealed.
Here, it is as if the construction of the DAG is executed in lock-step synchrony across all nodes.
As a result, every vertex has exactly $n = 4$ strong edges and no weak edge.
In these conditions, the adversary cannot delay the delivery of vertices i.e., the attack on the reliable broadcast layer from Sec.\ref{ssec:attack_bracha} has no effect.
The DagRider layer attack from Sec.\ref{ssec:byz_dagrider_layer} is also ineffective because, even if a Byzantine node at column $c$ decides not to include strong edges to certain vertices at column $c-1$, it must include $2*f+1$ strong edges and at least $f+1$ of these targeted vertices will, in any case, include all the pending puzzle solutions.
Here, the adversary can only manipulate the content of the DAG via Byzantine nodes not including certain transactions in their own vertices proposal.

In contrast, on Fig.\ref{fig:dag_from_simu_quickest}, we use a hypoexponential distribution of delays for node to node communications. One can see that the regularity of the DAG is broken : weak edges appear and the shape of the waves vary.

This comparison with a perfect network shows that it is the non-determinism in network communication delays that leave Byzantine nodes with ample leeway to change the outcome of an execution via manipulating these delays.
This is true, even when staying within the $\Delta$ bounds in a synchronous or partially synchronous communication model \cite{consensus_in_the_presence_of_partial_synchrony,impossibility_of_distributed_consensus_with_one_faulty_process}.

The reader may notice that the situation described on Fig.\ref{fig:dagrider_bug} in appendix \ref{anx:bug_dagrider} effectively occurs on Fig.\ref{fig:dag_from_simu_quickest} as it is possible for a vertex at column $c$ and row $r$ to be broadcast ($\mathtt{rbcast}$ of Fig.\ref{fig:layers_dagrider}) before the vertex at column $c-1$ and row $r$ is delivered ($\mathtt{rdlver}$ of Fig.\ref{fig:layers_dagrider}) due to the random delays in the reliable broadcast layer. As a result, there is no strong edge between $(c,r)$ and $(c-1,r)$ (see e.g., the absence of edge between the leader vertex of the orange wave and its immediate predecessor on the same row, which is colored in purple).

Fig.\ref{fig:dag_from_simu_quickest} also illustrates the fact that certain leader vertices can be skipped over.
Indeed, given a leader at column $c$, if the condition that there are at least $2*f+1$ strong paths between it and vertices at column $c + 4$ is not met, it is not included in the leader stack.
In that case, the wave is ignored, and its content may eventually be included in the next wave.
On Fig.\ref{fig:dag_from_simu_quickest}, wave 3 is ignored and its content is included in wave 4 (the orange wave).

In Sec.\ref{ssec:network_param}, we use two specific \faClockO~distributions of delays for our experiments.
Fig.\ref{fig:dag_from_simu} illustrate the use of these two specific distributions.
Fig.\ref{fig:dag_from_simu_quick} results from a simulation with the node to node delay being modeled using the \textcolor{blue}{\faClockO} distribution (smaller delays).
One can see that there are roughly 2 to 3 puzzles per wave of the DAG and 5 to 10 transactions per vertex of the DAG.
On Fig.\ref{fig:dag_from_simu_slow} there rather are roughly 5 to 6 puzzles per wave and 10 to 20 transactions per vertex.

\clearpage 

\section{Additional metrics on the first experiments\label{anx:exp1}}

\begin{figure}[h]
    \centering

\setlength\tabcolsep{1.5pt}
\begin{tabular}{|c|c|}
\hline
{\scriptsize $\#$ of solved puzzles}
&
{\scriptsize $\#$ of waves}
\\
\hline 
\includegraphics[scale=.3]{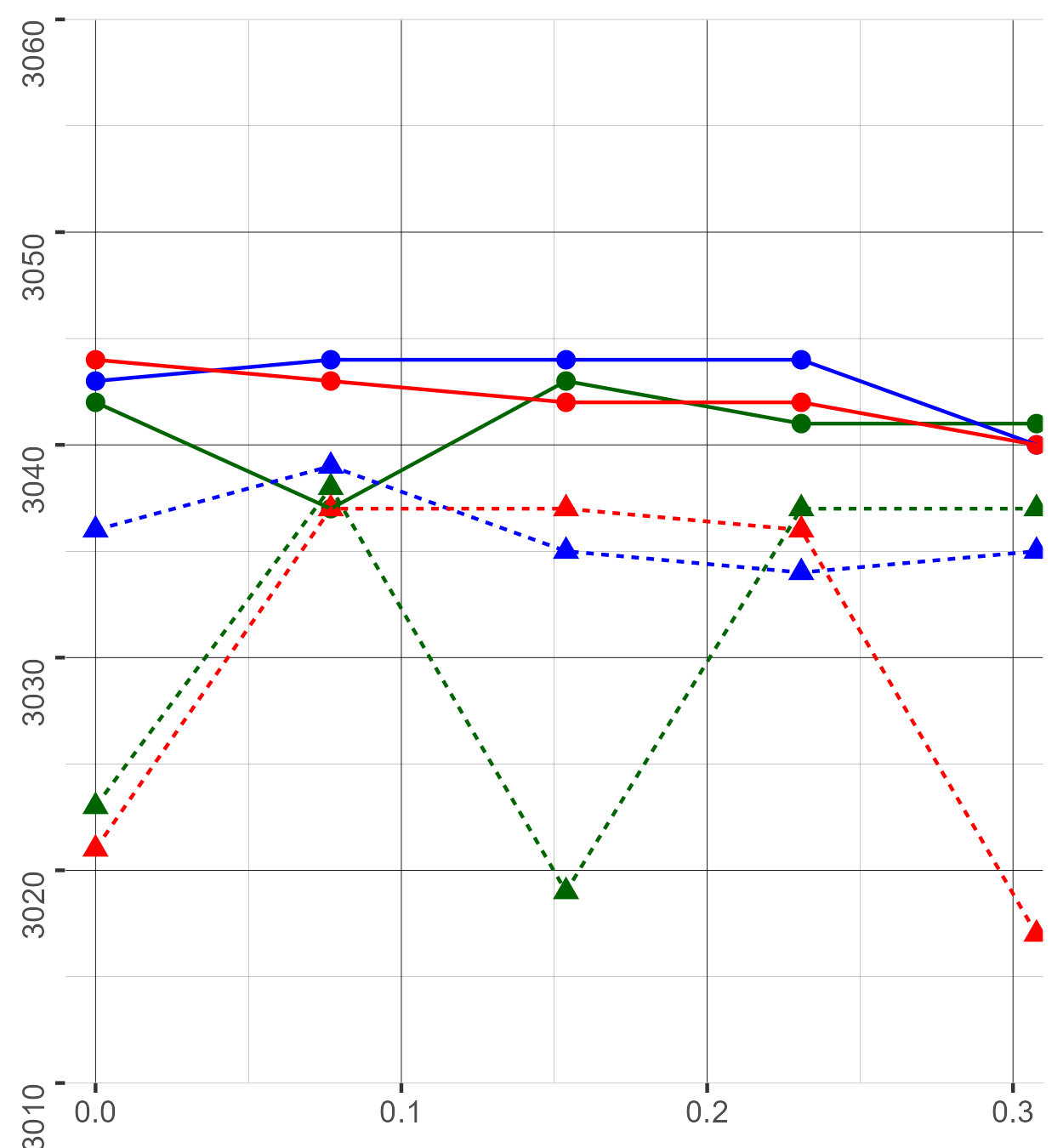}
&
\includegraphics[scale=.3]{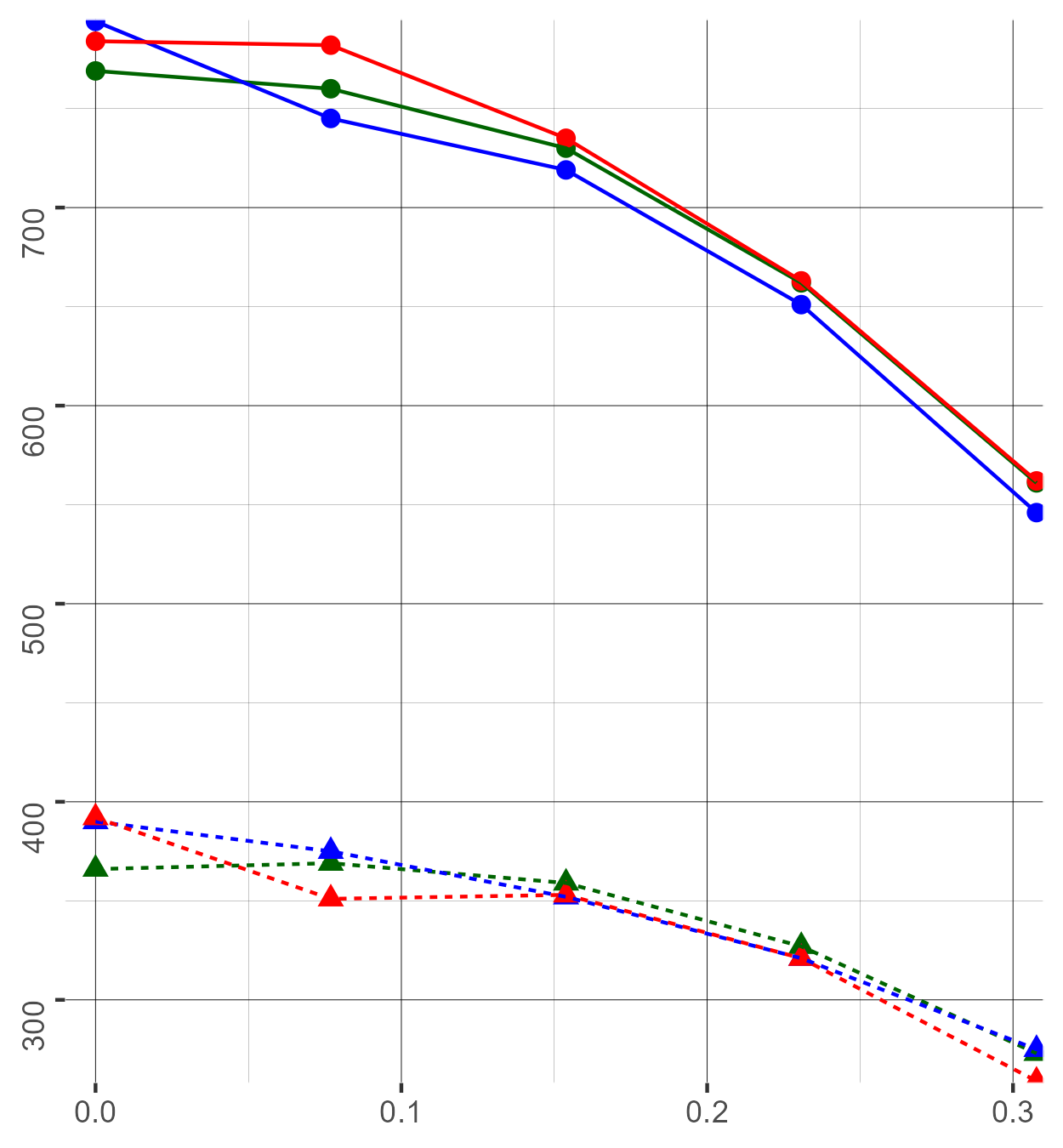}
\\
\hline 
\hline
{\scriptsize $\#$ of transactions per wave}
&
{\scriptsize$\#$ of $OF_{I_{NI}}^{W_{AV}}$ violations}
\\
\hline 
\includegraphics[scale=.3]{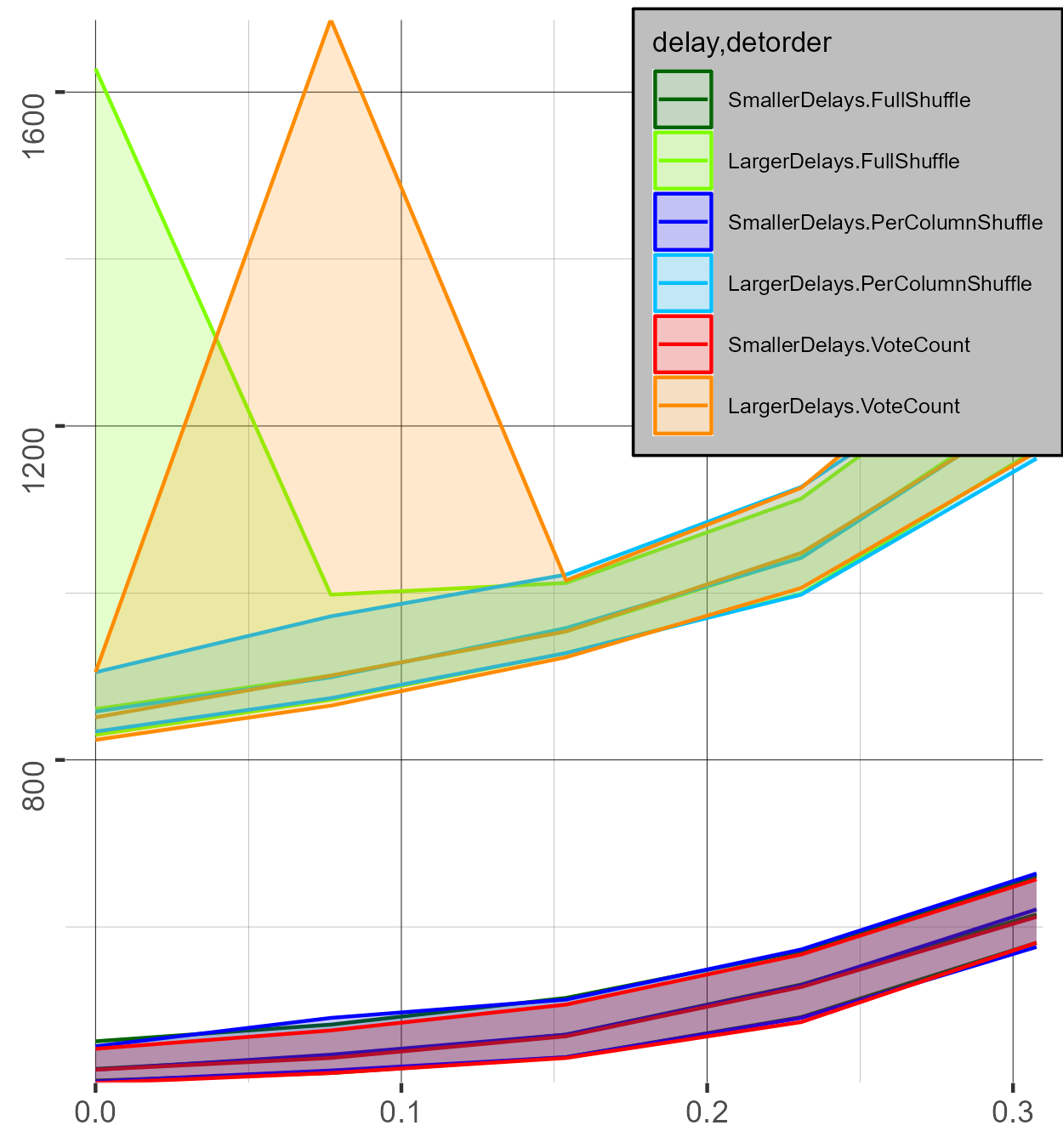}
&
\includegraphics[scale=.3]{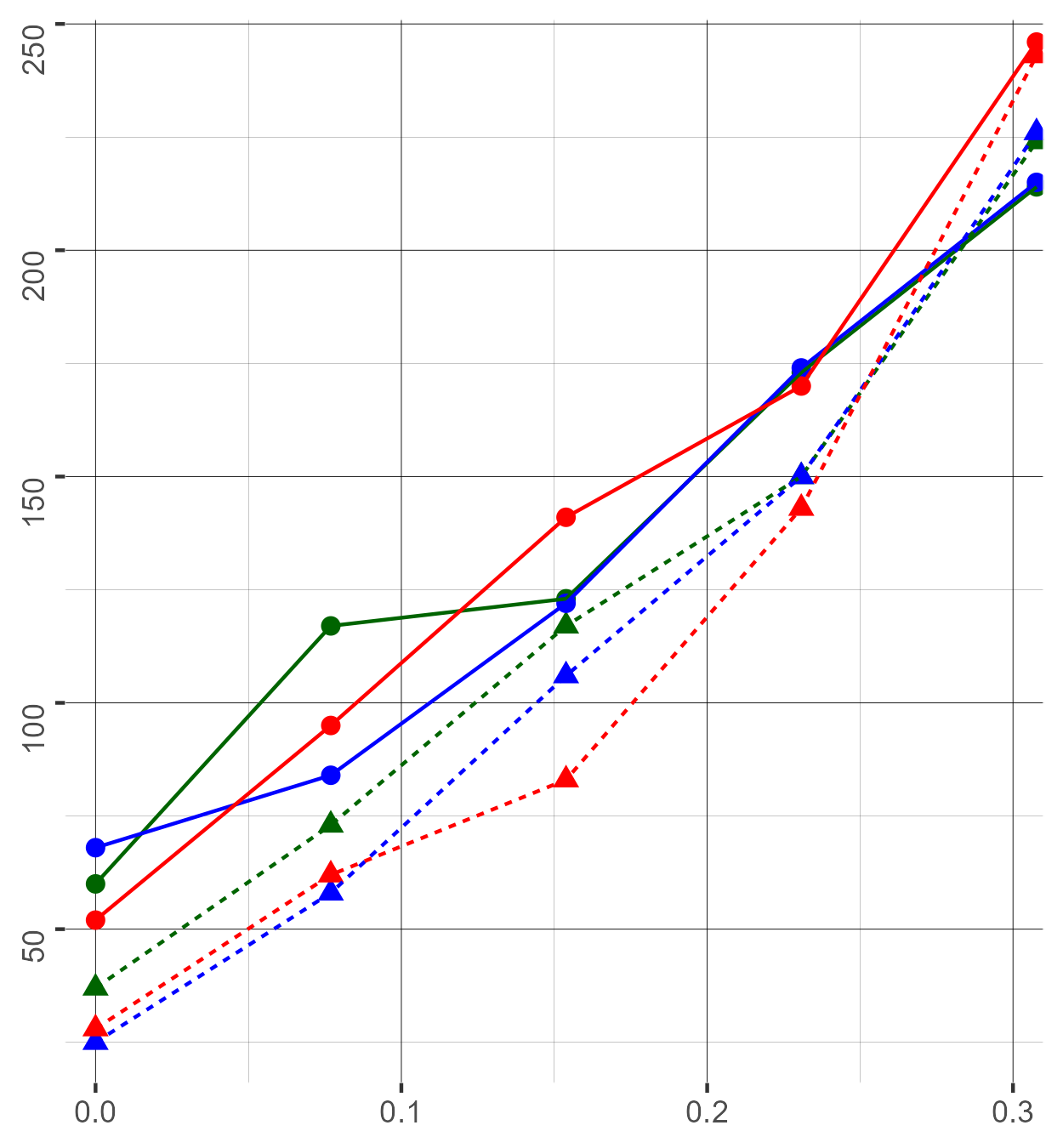}
\\
\hline 
\hline
{\scriptsize$\#$ of $OF_{R_{EC}}^{F_{IN}}$ violations}
&
{\scriptsize$\#$ of $OF_{I_{NI}}^{F_{IN}}$ violations}
\\
\hline 
\includegraphics[scale=.3]{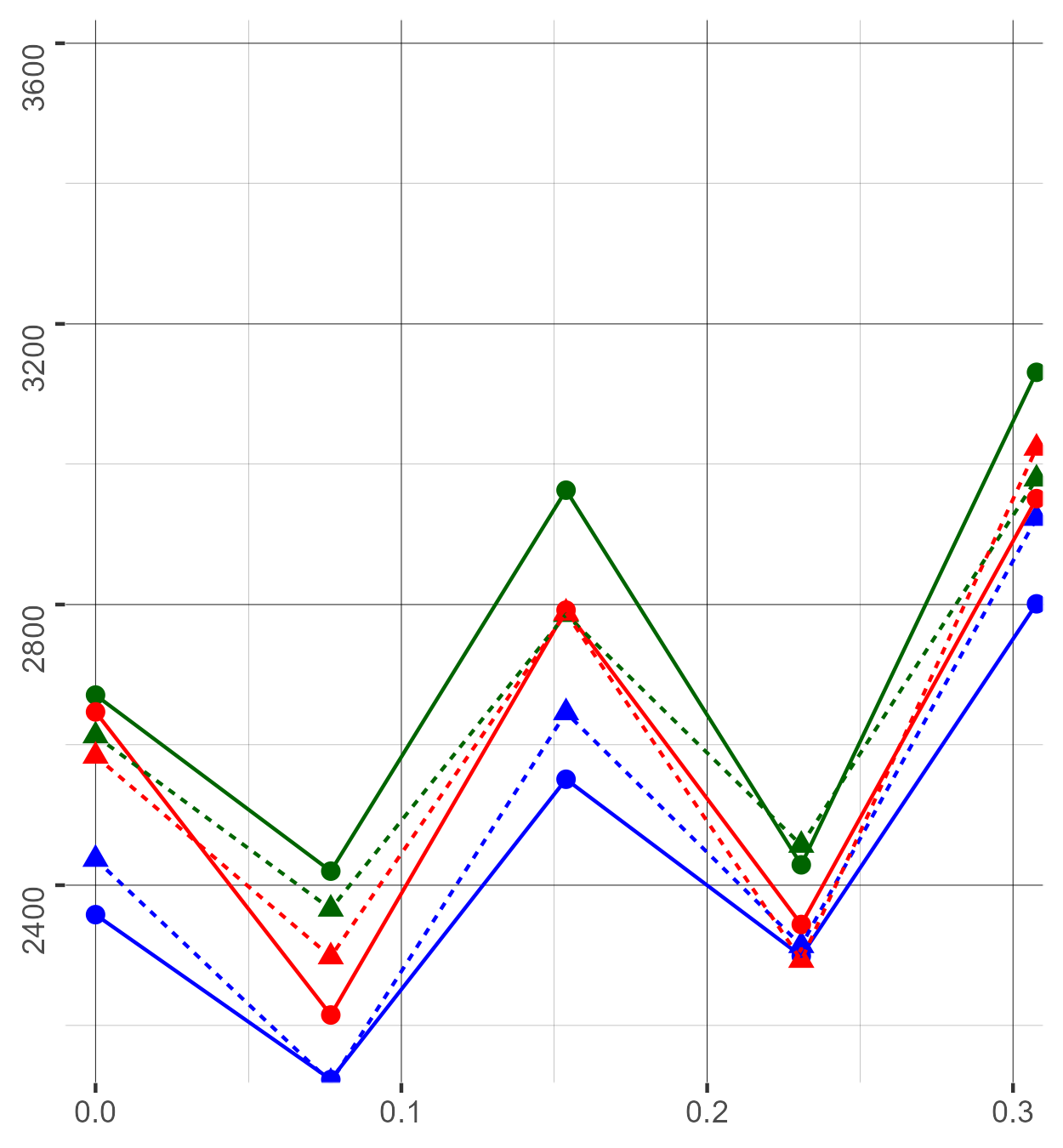}
&
\includegraphics[scale=.3]{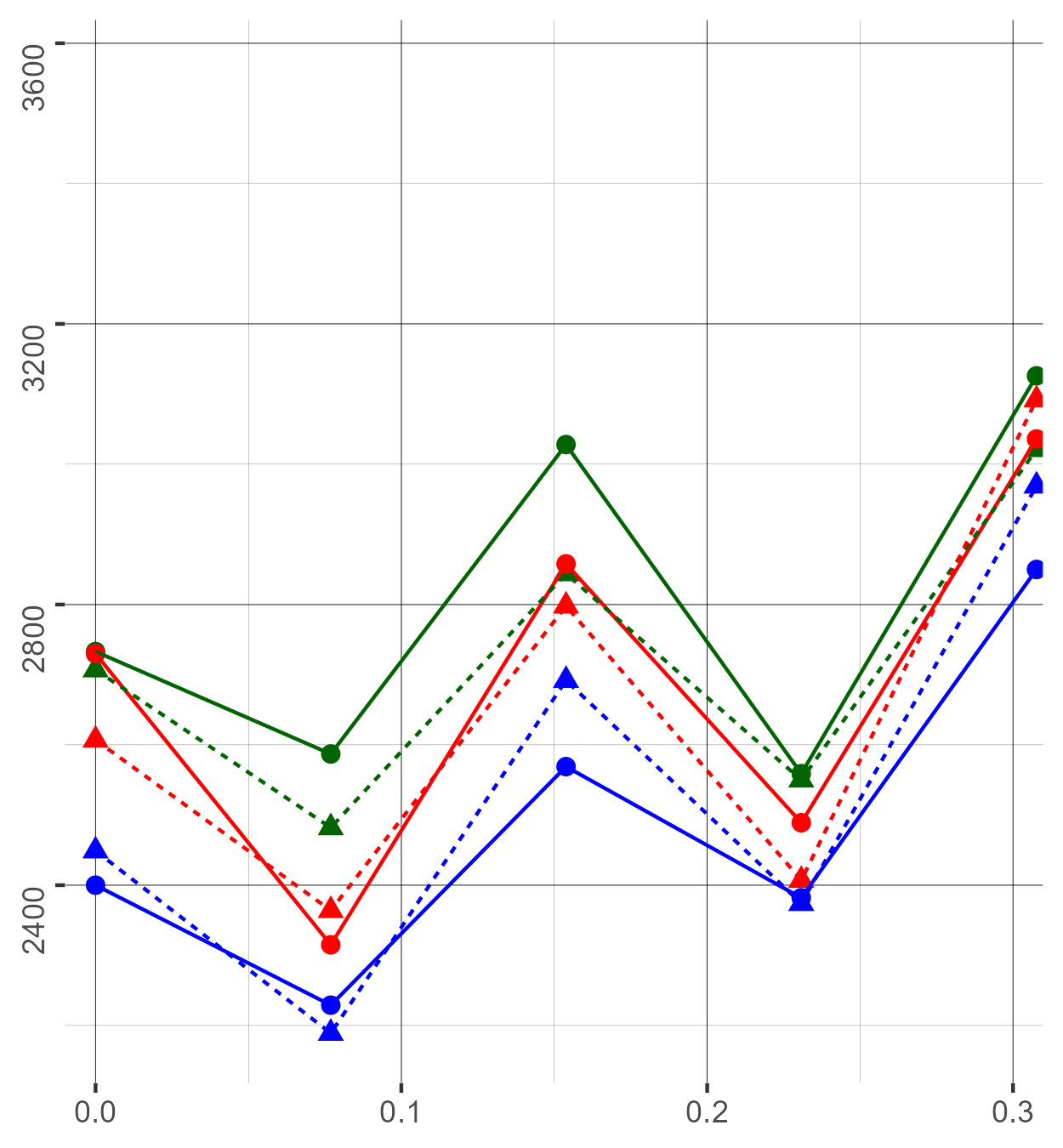}
\\
\hline 
\hline 
{\scriptsize$\#$ of $OF_{D_{LV}}^{F_{IN}}$ violations}
&
\\
\hline 
\includegraphics[scale=.3]{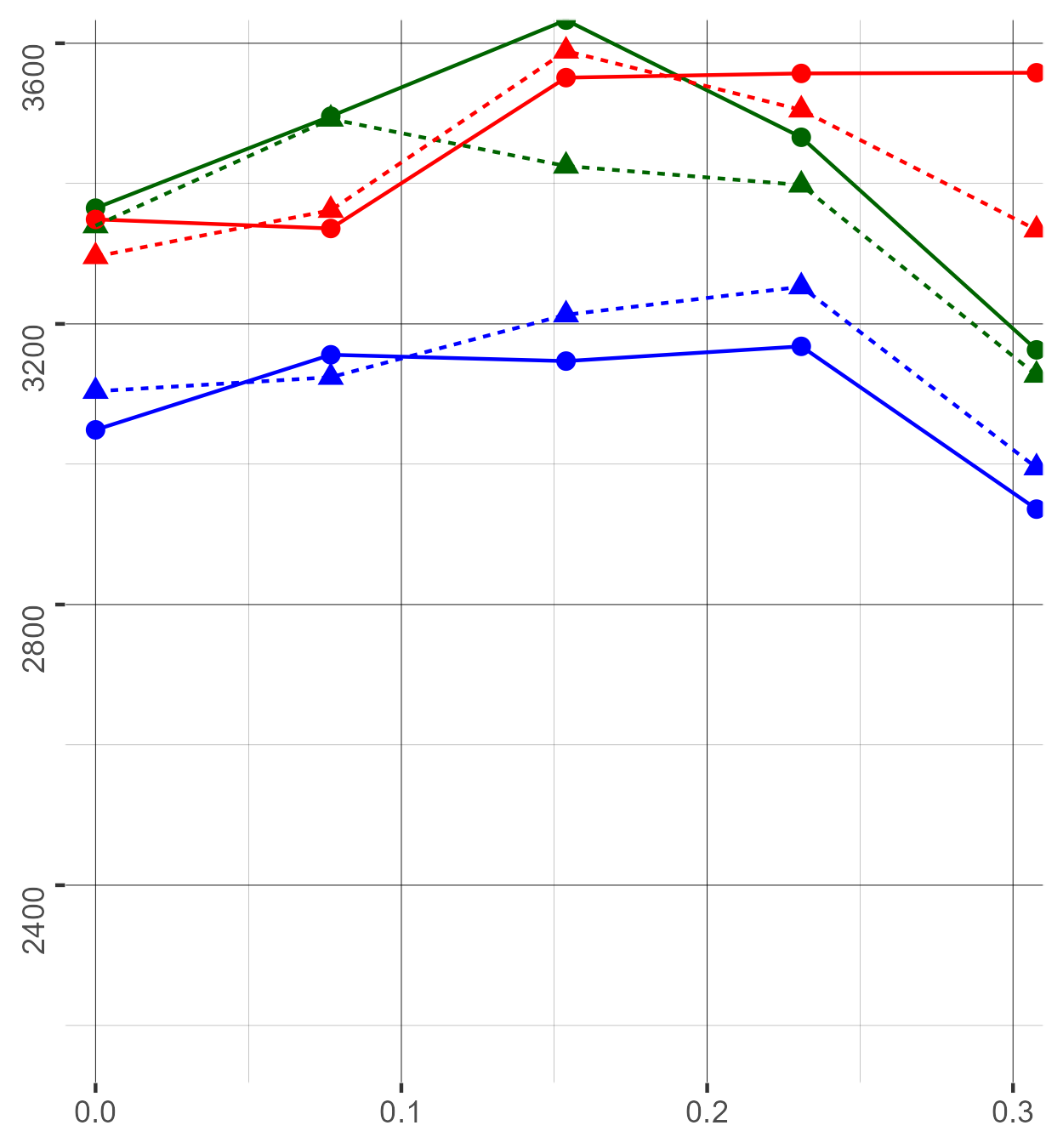}
&
\\
\hline 
\end{tabular}
\setlength\tabcolsep{6pt}
    
    \caption{Other metrics for the first experiment}
    \label{fig:exp1_other}
\end{figure}

Fig.\ref{fig:exp1_other} provide additional metrics (in addition of those of Fig.\ref{fig:exp1}) that characterize the simulations performed in the first set of experiments.

The total number of solved puzzles (top left of Fig.\ref{fig:exp1_other}) indicates that in all simulations, all metrics are measured after having solved at least 3010 puzzles.
Thus, we have statistically significant results in regards to fairness.

Let us then consider two next diagrams. 
The top right diagram gives the total number of waves at the end of the simulation.
The left diagram on the the second row gives, in the form of ribbon plots, the 1st quartile, median value and 3rd quartile of the distribution of the number of transactions per wave (across the 300 to 800 waves in the simulation).
Let us also recall that the duration of the simulations are configured so that at least 3000 puzzles are solved and the frequency at which new puzzles are revealed does not change.
In the slower network with higher delays, the number of waves taken to solve these 3000 puzzles is smaller (dotted lines with triangles on the top right diagram), and each such wave contains more transactions (light green, orange and cyan ribbons on the left diagram of the second row).

The high values on the number of transactions per wave (left diagram of the second row) are explained by the fact that certain leaders may be skipped over (see discussion in Appendix \ref{anx:max_sim}). Whenever a leader of wave $w$ is skipped, the content of wave $w$ is eventually included in that of wave $w+1$, leading to larger waves that contain more transactions.
Because there are less waves in the simulations on the slower network (around 350 compared to the 700 waves in the quicker network) it is also more likely that the 3rd quartile corresponds to such enlarged waves.
The fact that there are larger delays (light green, orange and cyan ribbons) may also increase the risk that certain leaders are skipped over (because of the unreliable delivery, the strong path condition is less likely to be met).

These diagrams also highlight a side effect of the impact of the adversary (the $x$-axis corresponding to the power of the adversary).
Indeed, as the adversary causes the delivery of certain vertices to be delayed, it slows down the overall throughput (indeed, in addition of these vertices themselves being delayed, as nodes require at least $2*f+1$ strong edges to vertices at column $c$ to propose a new vertex at column $c+1$, the production of new vertices may also be slowed down).
In turn, this causes delayed vertices (and thus also waves) to contain more transactions (increase in the number of transactions per wave) and the total number of waves to decrease.

The number of violations of $OF_{I_{NI}}^{W_{AV}}$ and $OF_{D_{LV}}^{F_{IN}}$ follow the same trends as discussed in Sec.\ref{ssec:exp1}.
As for $OF_{R_{EC}}^{F_{IN}}$ and $OF_{I_{NI}}^{F_{IN}}$, we observe that the statistical noise is greater than for $OF_{S_{ND}}^{F_{IN}}$ and prevents reaching conclusions.

\section{Additional metrics on the second experiments\label{anx:exp2}}

\begin{figure}[h]
    \centering

\setlength\tabcolsep{1.5pt}
\begin{tabular}{|c|c|}
\hline
{\scriptsize $\#$ of solved puzzles}
&
{\scriptsize $\#$ of waves}
\\
\hline 
\includegraphics[scale=.3]{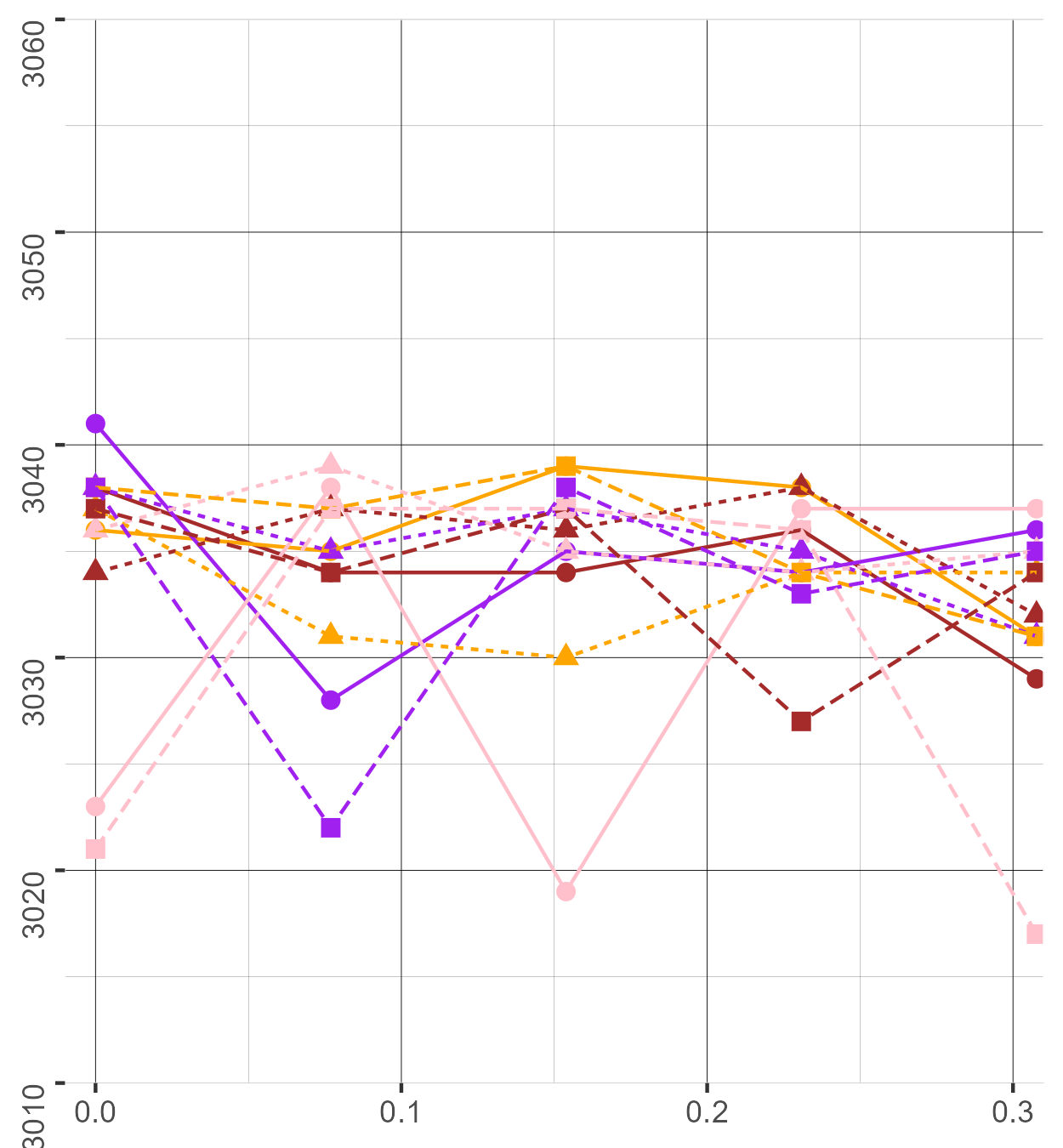}
&
\includegraphics[scale=.3]{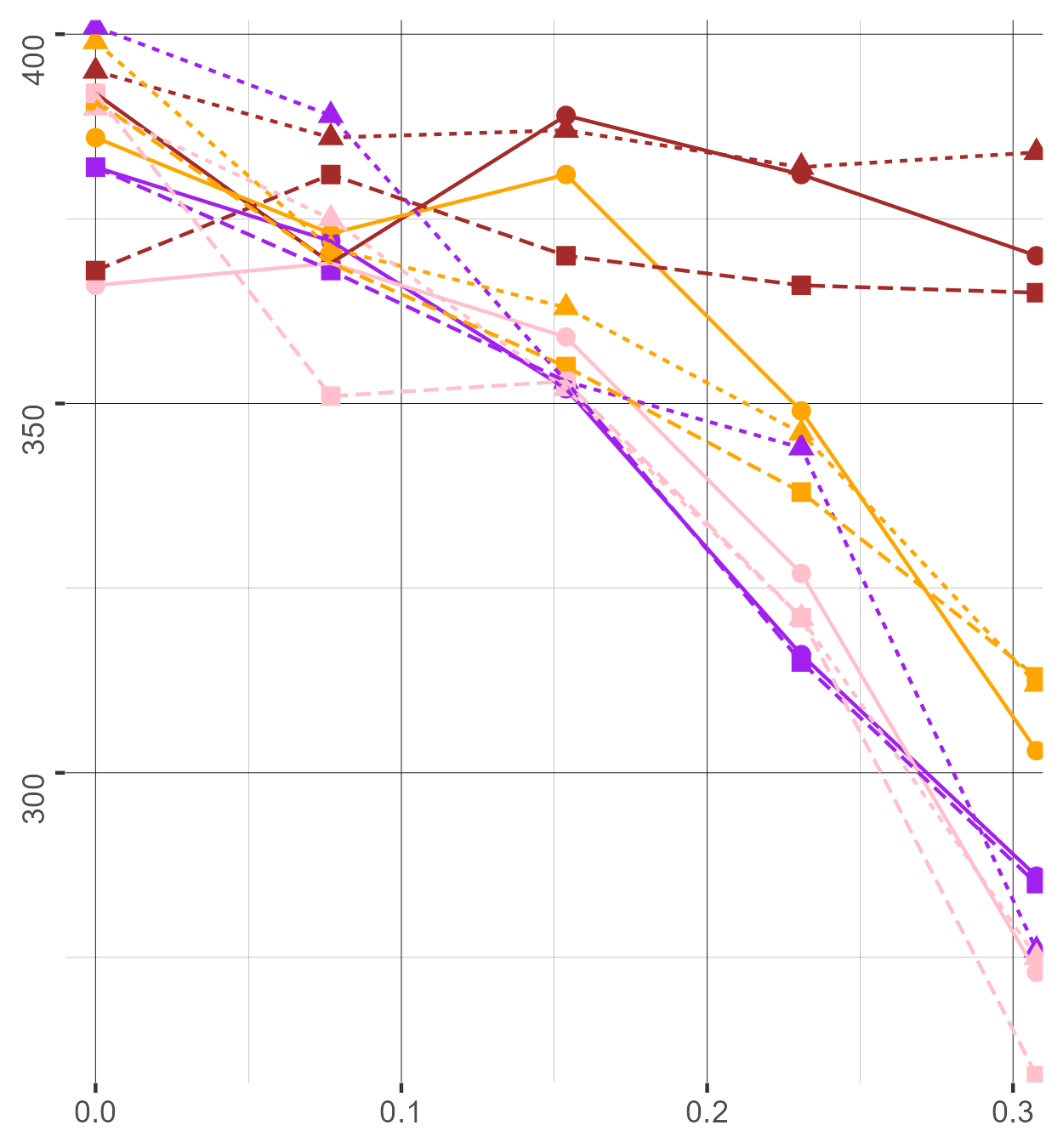}
\\
\hline 
\hline
{\scriptsize$\#$ of $OF_{I_{NI}}^{W_{AV}}$ violations}
&
{\scriptsize$\#$ of $OF_{I_{NI}}^{F_{IN}}$ violations}
\\
\hline 
\includegraphics[scale=.3]{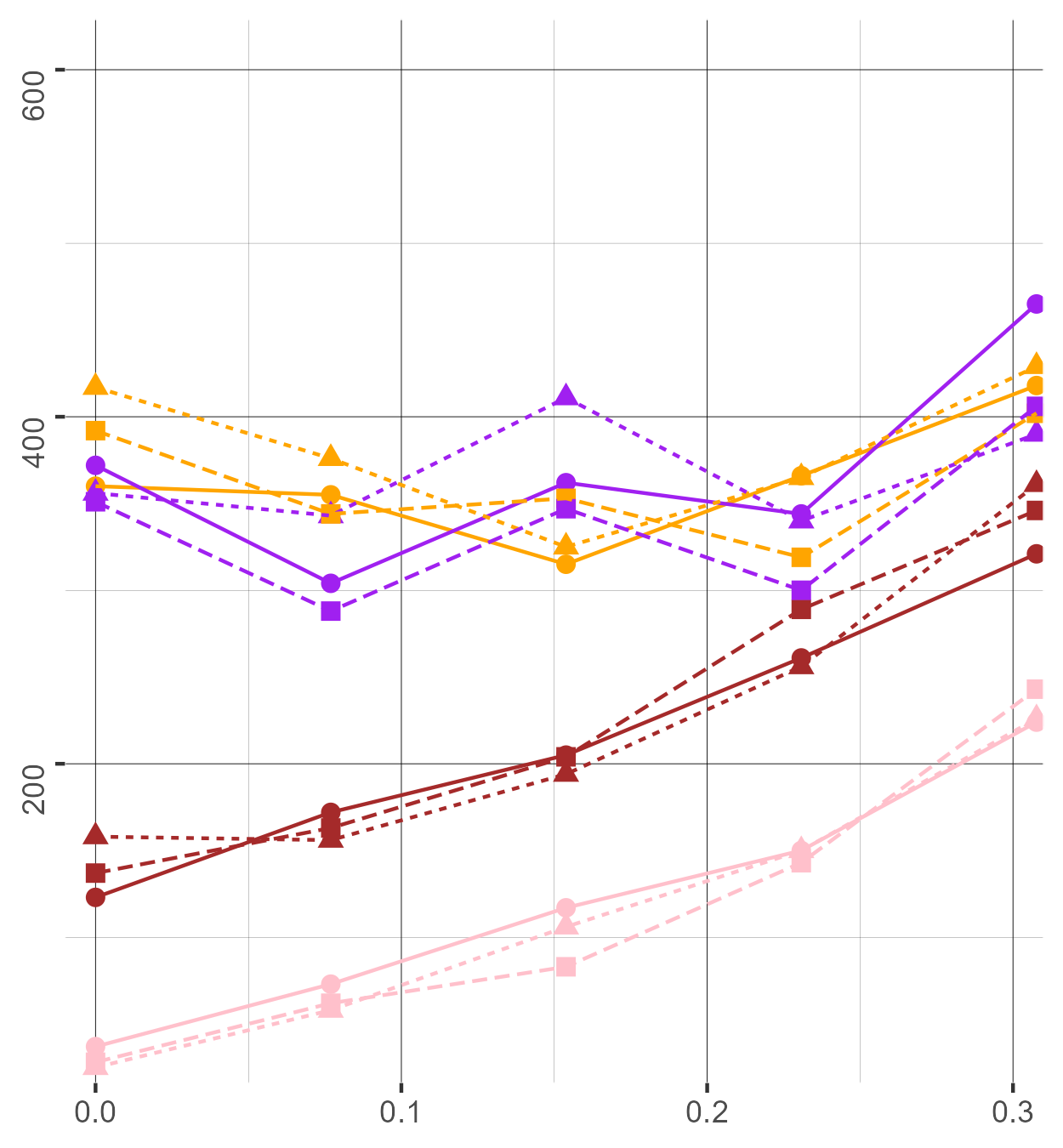}
&
\includegraphics[scale=.3]{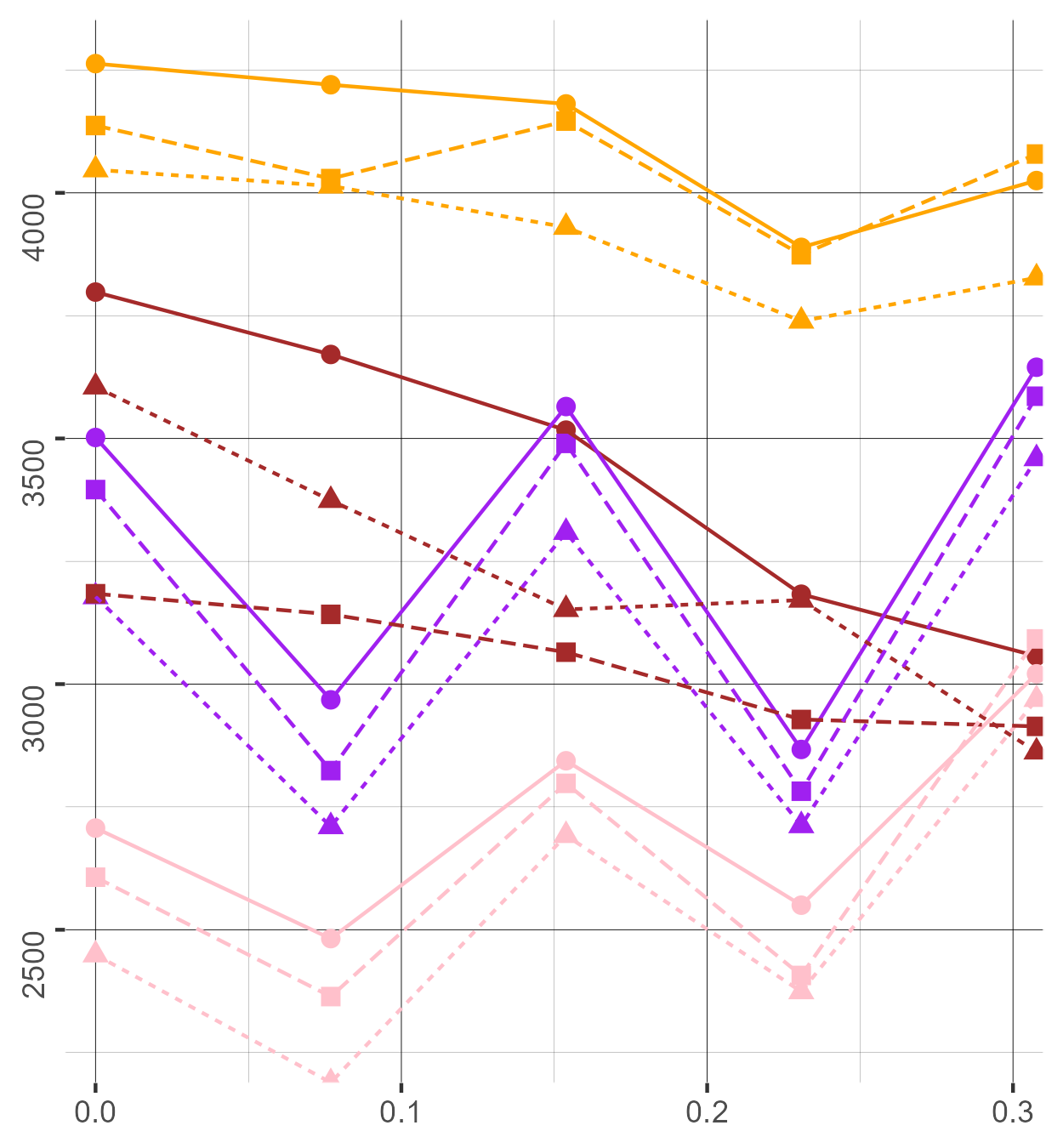}
\\
\hline 
\hline
{\scriptsize$\#$ of $OF_{D_{LV}}^{F_{IN}}$ violations}
&

\\
\hline 
\includegraphics[scale=.3]{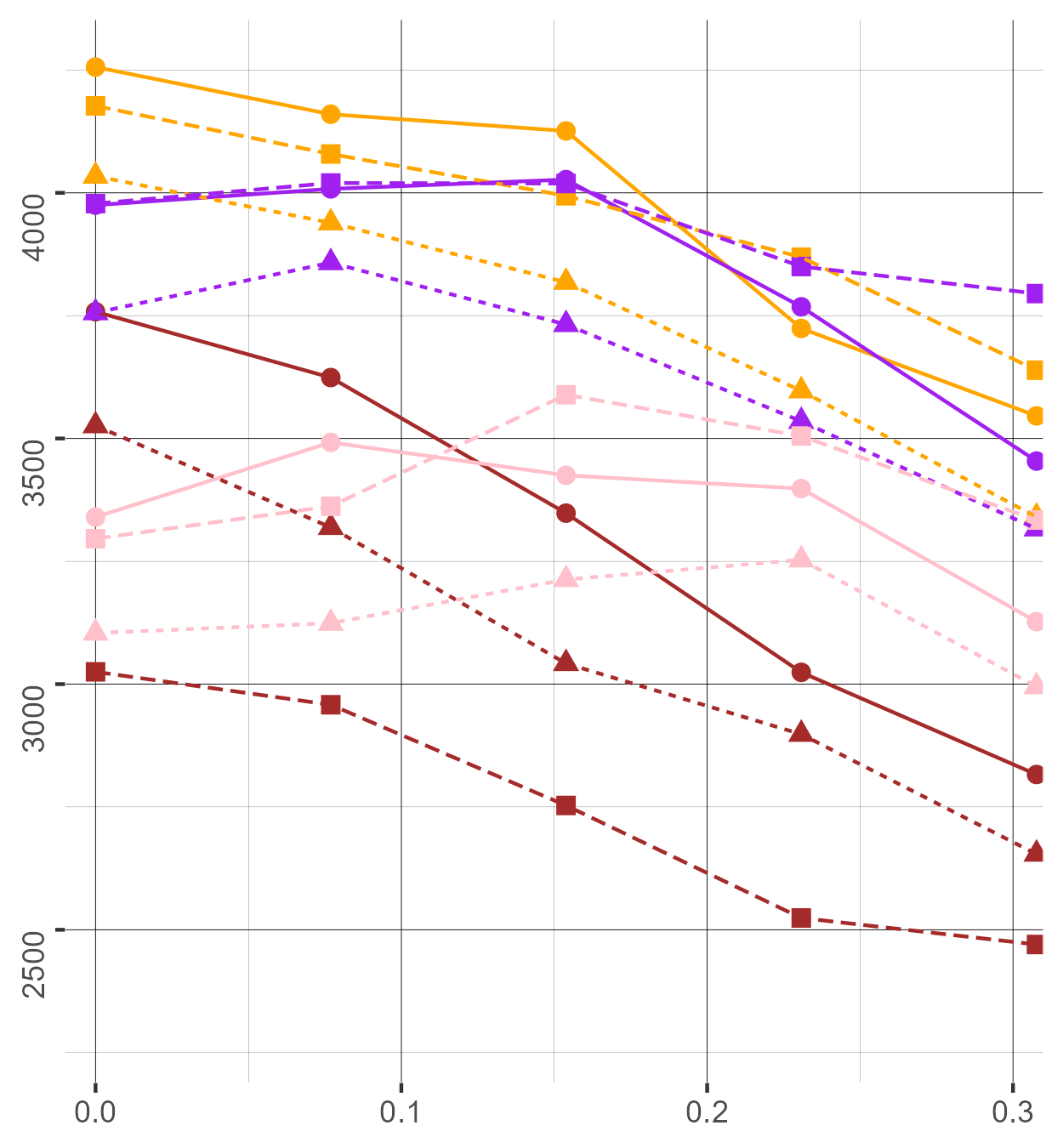}
&

\\
\hline 
\end{tabular}
\setlength\tabcolsep{6pt}
    
    \caption{Other metrics for the second experiment}
    \label{fig:exp2_other}
\end{figure}

Fig.\ref{fig:exp2_other} provide additional metrics (in addition of those of Fig.\ref{fig:exp2}) that characterize the simulations performed in the second set of experiments.

\section{Additional metrics on the third experiments\label{anx:exp3}}

\begin{figure}[h]
    \centering

\setlength\tabcolsep{1.5pt}
\begin{tabular}{|c|c|}
\hline
{\scriptsize$\#$ of $OF_{I_{NI}}^{W_{AV}}$ violations}
&
{\scriptsize$\#$ of $OF_{D_{LV}}^{W_{AV}}$ violations}
\\
\hline 
\includegraphics[scale=.3]{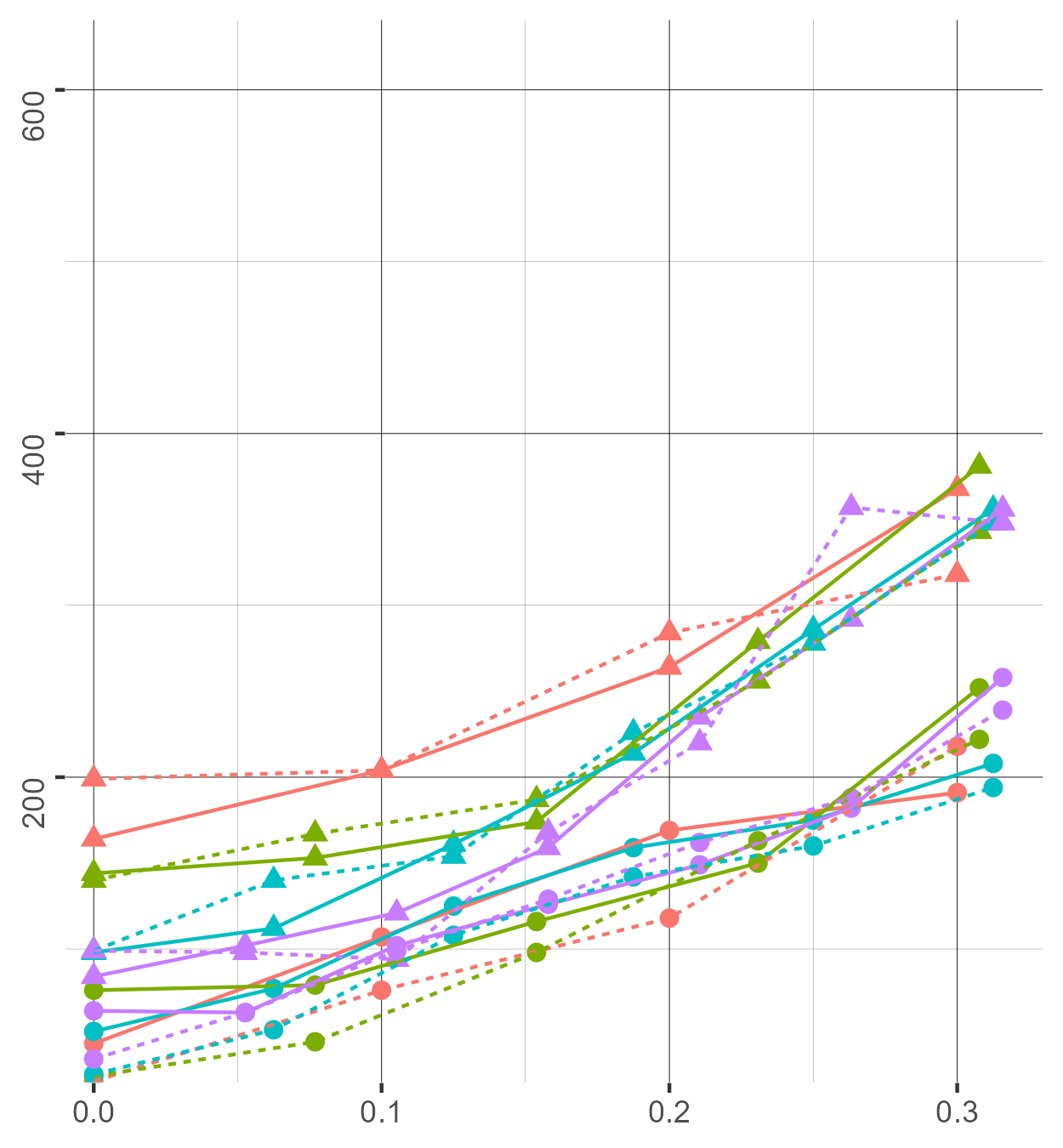}
&
\includegraphics[scale=.3]{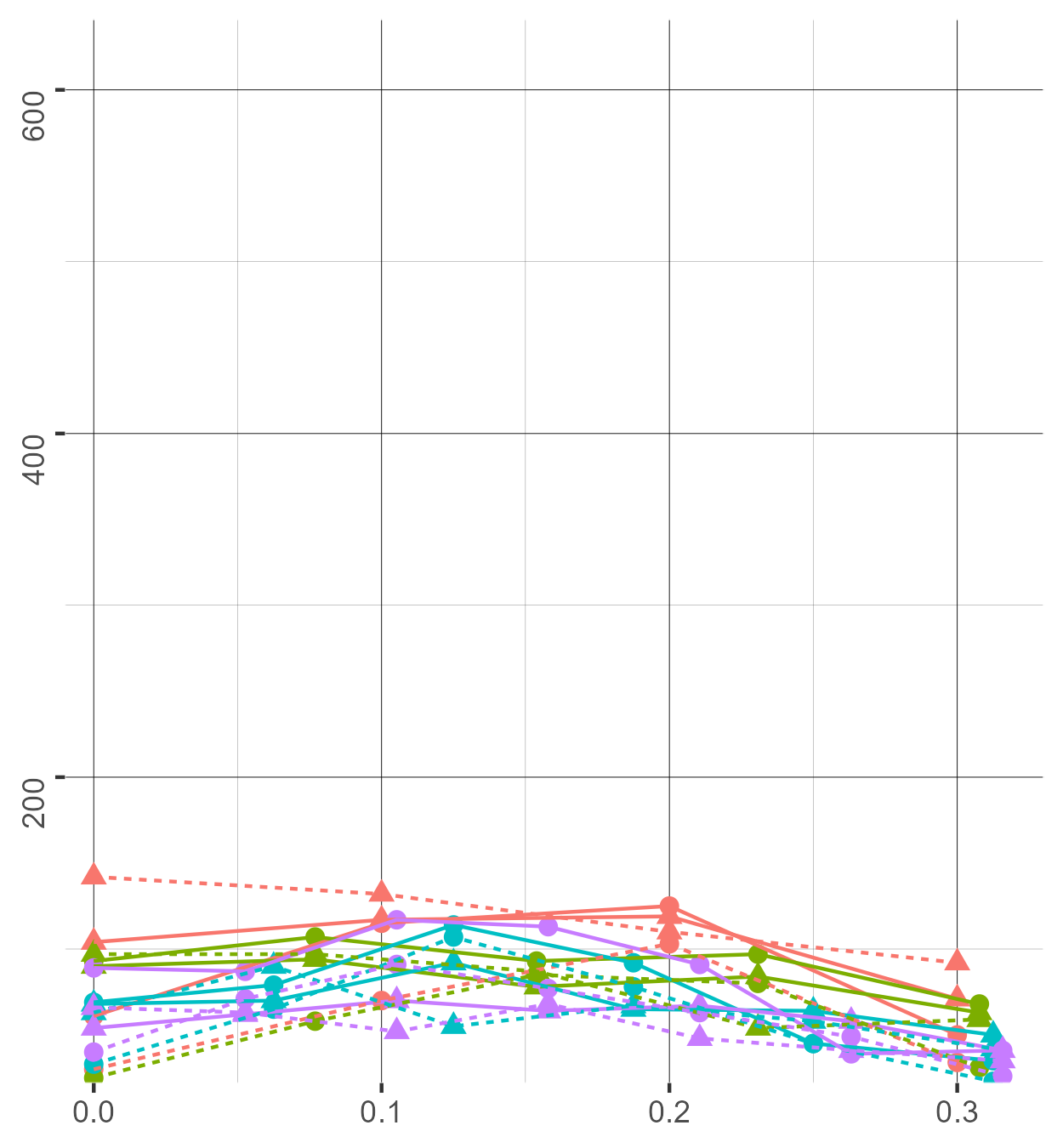}
\\
\hline 
\hline
{\scriptsize$\#$ of $OF_{I_{NI}}^{F_{IN}}$ violations}
&
{\scriptsize$\#$ of $OF_{D_{LV}}^{F_{IN}}$ violations}
\\
\hline 
\includegraphics[scale=.3]{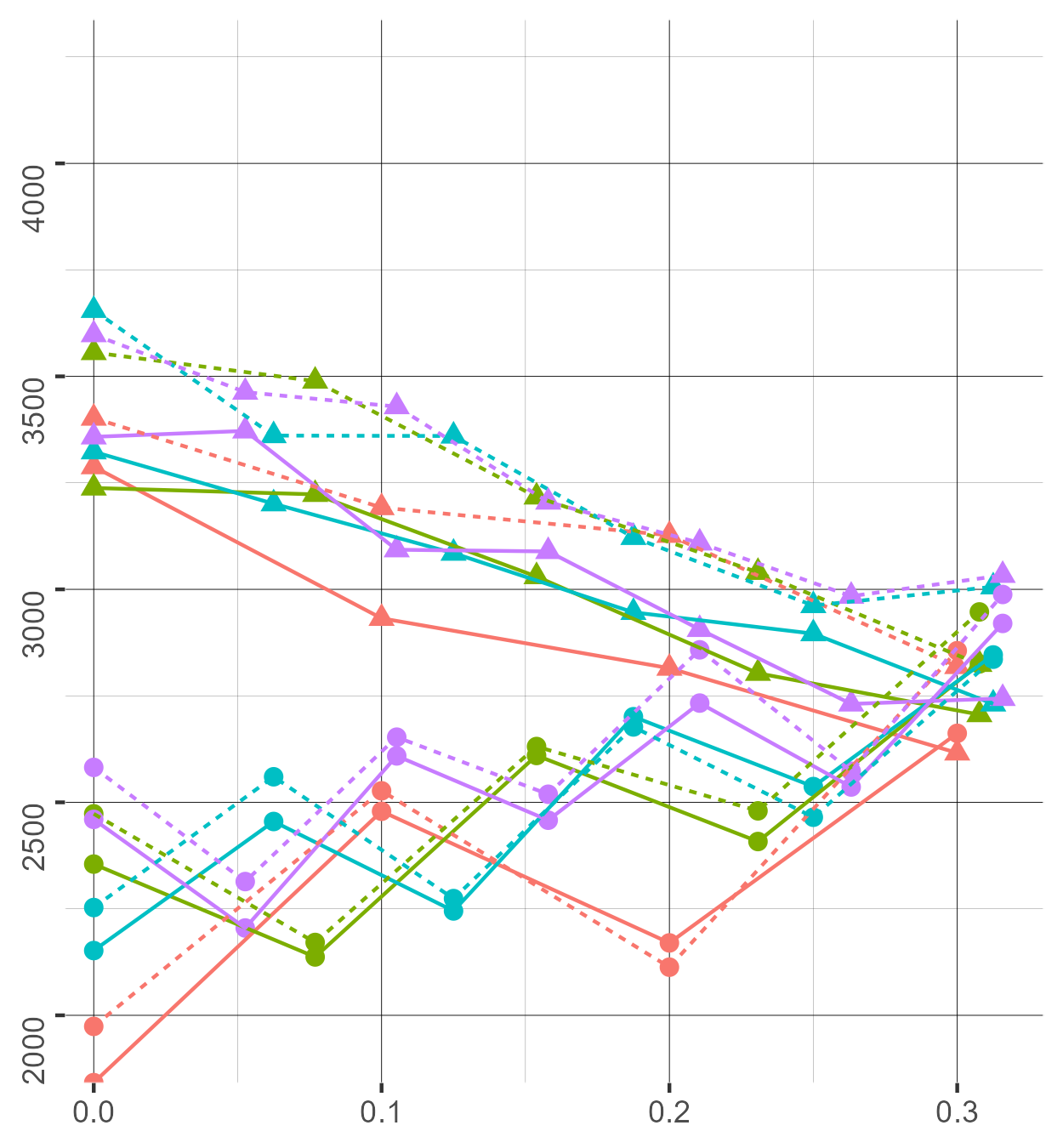}
&
\includegraphics[scale=.3]{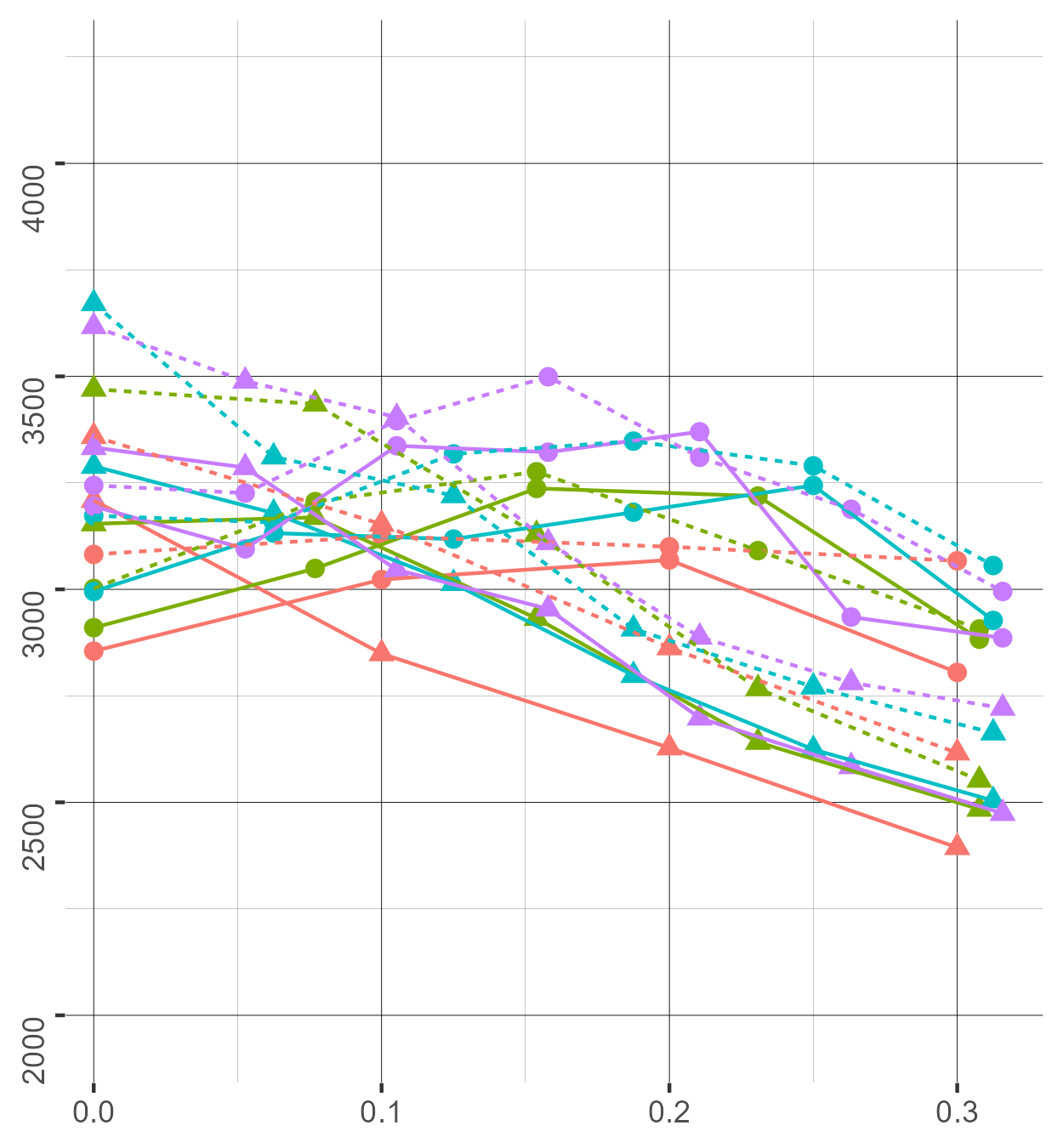}
\\
\hline 
\end{tabular}
\setlength\tabcolsep{6pt}
    
    \caption{Other metrics for the third experiment}
    \label{fig:exp3_other}
\end{figure}

Fig.\ref{fig:exp3_other} provide additional metrics (in addition of those of Fig.\ref{fig:exp3}) that characterize the simulations performed in the third set of experiments.

\clearpage

\end{document}